\documentclass[11pt,reqno]{amsproc}
\allowdisplaybreaks
\usepackage{cite}
\usepackage{color,soul}
\usepackage[dvipsnames]{xcolor}
\usepackage{float}
\usepackage{psfrag}
\usepackage{fullpage}
\usepackage{rotating}
\usepackage{stmaryrd}
\usepackage{subfigure}
\usepackage{listings}
\usepackage{mcode}
\usepackage{enumerate}
\usepackage[small]{caption}
\usepackage{morefloats}
\numberwithin{equation}{section}
\usepackage{graphicx}
\DeclareGraphicsExtensions{.eps}
\usepackage{epstopdf}
\usepackage{array,multirow,graphicx}
\usepackage{rotating}
\usepackage{booktabs,tabularx}
\usepackage{adjustbox}
\usepackage[font=small,labelfont=bf,tableposition=top]{caption}
\usepackage[utf8x]{inputenc}
\usepackage{makecell}
\usepackage{multicol}
\usepackage[flushleft]{threeparttable}
\usepackage[customcolors,shade]{hf-tikz}
\usepackage{tikz}
\usetikzlibrary{shapes,snakes}
\usepackage[semicolon,square,authoryear]{natbib}
\usepackage[debug=false, colorlinks=true, pdfstartview=FitV, 
linkcolor=blue, citecolor=blue, urlcolor=blue]{hyperref}
\definecolor{orange}{RGB}{255,127,0}
\newcolumntype{R}[2]{%
    >{\adjustbox{angle=#1,lap=\width-(#2)}\bgroup}%
    l%
    <{\egroup}%
}
\usepackage{enumitem,booktabs}
\usepackage[referable]{threeparttablex}
\renewlist{tablenotes}{enumerate}{1}
\makeatletter
\setlist[tablenotes]{label=\tnote{\alph*},ref=\alph*,itemsep=\z@,topsep=\z@skip,partopsep=\z@skip,parsep=\z@,itemindent=\z@,labelindent=\tabcolsep,labelsep=.2em,leftmargin=*,align=left,before={\footnotesize}}
\makeatother
\makeatletter
\def\maketag@@@#1{\hbox{\m@th\normalfont\normalsize#1}}
\makeatother
\usepackage{morefloats}
\newlength{\drop}
\definecolor{amethyst}{rgb}{0.6, 0.4, 0.8}
\definecolor{burgundy}{rgb}{0.5, 0.0, 0.13}

\title{\textbf{Composable block solvers for the 
four-field double porosity/permeability model}}

\author{
\textbf{M.~S.~Joshaghani}, 
\textbf{J.~Chang}, 
\textbf{K.~B.~Nakshatrala}
and
\textbf{M.~G.~Knepley}
 \\
  {\small 
    \textbf{Correspondence to:}~\textsf{knakshatrala@uh.edu}}}


\newsavebox{\measurebox}
\begin{document}

\date{\today}

	 \begin{titlepage}
		 \drop=0.1\textheight
		 \centering
		 \vspace*{\baselineskip}
		 \rule{\textwidth}{1.6pt}\vspace*{-\baselineskip}\vspace*{2pt}
		 \rule{\textwidth}{0.4pt}\\[0.25\baselineskip]
				  {\Large \textbf{\color{burgundy}
	 Composable block solvers for the four-field double porosity/permeability model}}
	 \\[0.3\baselineskip]
				  \rule{\textwidth}{0.4pt}\vspace*{-\baselineskip}\vspace{2pt}
				  \rule{\textwidth}{1.6pt}\\[0.25\baselineskip]
				  \scshape
				  An e-print of the paper will be made available on arXiv. \par 
				  \vspace*{0.3\baselineskip}
				  Authored by \\[0.3\baselineskip]

		 {\Large M.~S.~Joshaghani\par}
		 {\itshape Graduate Student, University of Houston}\\[0.25\baselineskip]

		 {\Large J.~Chang\par}
		 {\itshape Postdoctoral Research Associate, Rice University}\\[0.25\baselineskip]
						 
		 {\Large K.~B.~Nakshatrala\par}
		 {\itshape Department of Civil \& Environmental Engineering \\
		 University of Houston, Houston, Texas 77204--4003 \\ 
		 \textbf{phone:} +1-713-743-4418, \textbf{e-mail:} knakshatrala@uh.edu \\
		 \textbf{website:} http://www.cive.uh.edu/faculty/nakshatrala}\\[0.25\baselineskip]
	 
	 	 {\Large M.~G.~Knepley\par}
	 	{\itshape Associate Professor, State University of New York at Buffalo}
		 \vspace{-3mm} 
		\begin{figure}[h]
			\centering
			\includegraphics[clip,trim=0 0 0 1.5cm,width=0.5\linewidth]{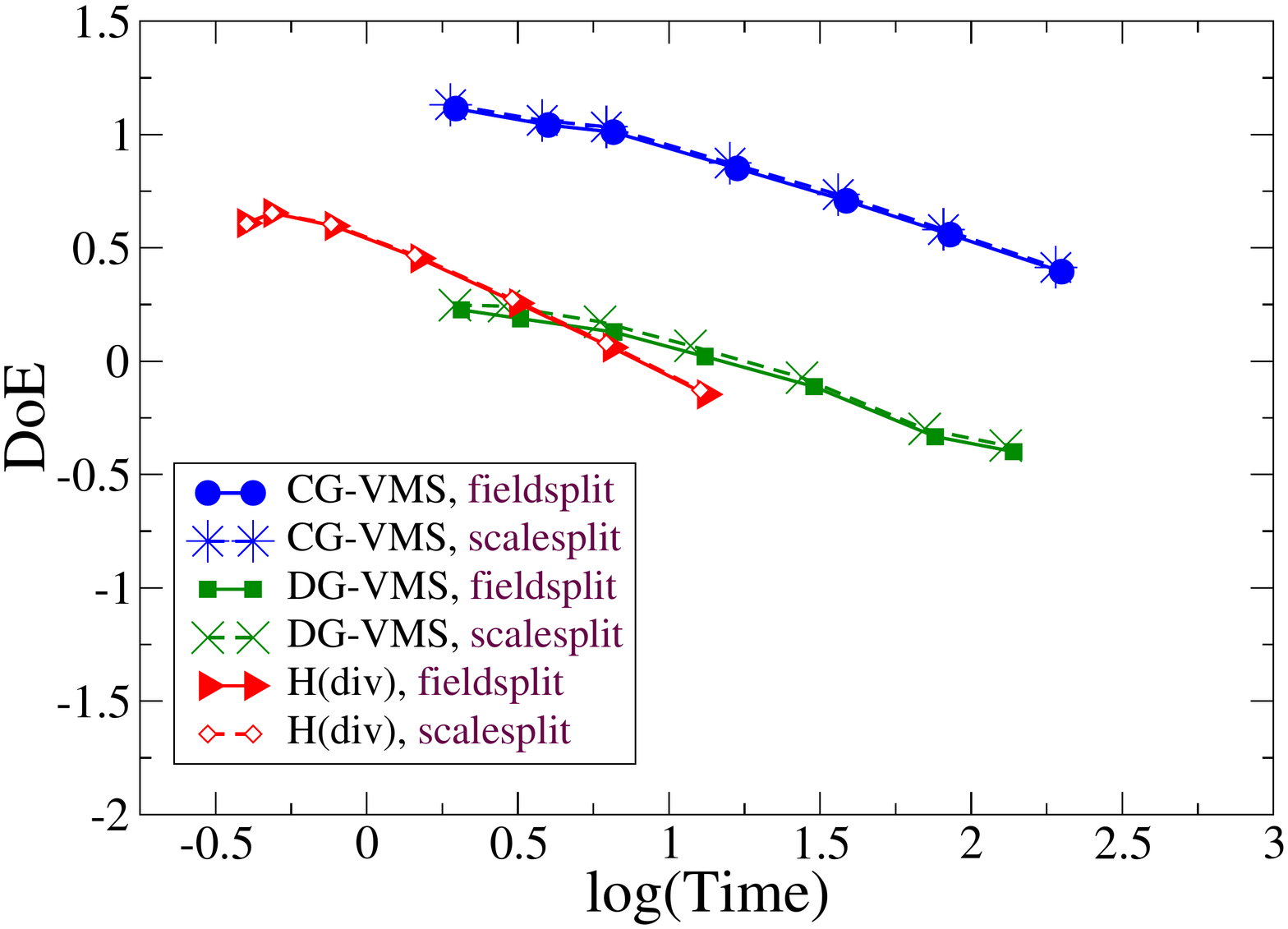}
		\vspace{-0.2in}
		\caption*{
			\emph{{\small{One of the main contributions 
			of this paper is performing Time-Accuracy-Size 
			spectrum analysis of various finite element 
			discretizations for the DPP mathematical 
			model. One of the main findings 
			under this analysis is that the continuous stabilized 
			Galerkin (CG-VMS) formulation has a higher 
			Digits of Efficacy (DoE), which measures 
			the combined effect of numerical accuracy 
			and computational time, 
			than the H(div) formulation. 
			The graphical abstract illustrates this finding 
			for the case of four-node tetrahedron element 
			and for the macro-pressure under the DPP model.}}}}
			\end{figure}
		  \vfill
		 {\scshape 2018} \\
		 {\small Computational \& Applied Mechanics Laboratory} \par
	 \end{titlepage}

\begin{abstract}
The objective of this paper is twofold. 
\emph{First}, we propose two composable block solver 
methodologies to solve the discrete systems that arise from 
finite element discretizations of the double porosity/permeability
(DPP) model. The DPP model, which is a four-field mathematical 
model, describes the flow of a single-phase incompressible fluid 
in a porous medium with two distinct pore-networks and 
with a possibility of mass transfer between them. 
Using the composable solvers feature available in 
PETSc and the finite element libraries available under 
the Firedrake Project, we illustrate two different ways 
by which one can effectively precondition these large 
systems of equations. 
\emph{Second}, we employ the recently developed
performance model called the Time-Accuracy-Size (TAS) spectrum 
to demonstrate that the proposed composable
block solvers are scalable in both the parallel
and algorithmic sense. Moreover, we utilize this 
spectrum analysis to compare the performance
of three different finite element discretizations 
(classical mixed formulation with H(div) elements, 
stabilized continuous Galerkin  mixed formulation, 
and stabilized discontinuous Galerkin mixed formulation) 
for the DPP model. 
Our performance spectrum analysis demonstrates that
the composable block solvers are fine choices for
any of these three finite element discretizations.
Sample computer codes are provided to illustrate 
how one can easily implement the proposed block 
solver methodologies through PETSc command 
line options.
\end{abstract}

\maketitle


\section{INTRODUCTION}
Due to recent growth in the exploration of hydrocarbons from 
unconventional sources (i.e., oil and gas from tight shale), 
there has been a growing interest to understand and to
model flows in porous media with complex pore-networks
\citep{straughan2017mathematical}. This interest has
been amplified due to recent advances in additive
manufacturing, which allow for creating materials
with complex pore-networks for various applications
ranging from water purification filters to composite
manufacturing.
The porous materials in the aforementioned applications
typically exhibit two or more dominant pore-networks with
each pore-network displaying distinctive hydro-mechanical
properties \citep{Schmidt_Mcdonald_1979,delage1996microstructure}.
Moreover, due to presence of fissures, there could be mass
transfer across the pore-networks
\citep{Barenblatt_Zheltov_Kochina_v24_p1286_1960_ZAMM}.

To address flows in these types of porous materials,
several mathematical models have been proposed in
the literature (see \citep{straughan2017mathematical}
and references therein). A class of models, which is
commonly referred to as double porosity/permeability
(DPP) models, have been found to be particularly attractive
in modeling flows in porous media with two pore-networks
(e.g., see \citep{Barenblatt_Zheltov_Kochina_v24_p1286_1960_ZAMM,warren1963behavior,Dykhuizen_v26_p351_1990_WRR,Boutin_Royer_2015,
  Nakshatrala_Joodat_Ballarini,Choo_White_Borja_2016_IJG}).
Recently, a DPP mathematical model with strong
continuum thermomechanics underpinning has been
derived in \citep{Nakshatrala_Joodat_Ballarini}.
This model, which will be central to this paper
and will be referred to as \emph{the DPP model}
from here on, describes the flow of a single-phase
incompressible fluid in a rigid porous medium with
two distinct pore-networks, with possible mass
transfer across the pore-networks.
The governing equations form a boundary value
problem in terms of four-fields and the nature
of the PDE is elliptic under steady-state
responses. 

Except for some academic problems, it is not
possible to obtain analytical solutions for
the governing equations under the DPP model.
Hence, there is a need to resort to numerical
solutions. Recently, several numerical formulations
have been developed to solve the governing equations
under the DPP model; which include 
\citep{choo2015stabilized,joodat2018modeling,joshaghani2018stabilized}.
However, these works addressed small-scale problems. 

The problems that arise in subsurface modeling and
other applications involving flow through porous
media are typically large-scale in nature. These
large-scale problems cannot be solved on a standard
desktop or by employing direct solvers; as such a
computation will be prohibitively expensive. But
large-scale problems from subsurface modeling are
routinely tackled using parallel computing tools
and by employing iterative linear solvers.
The current iterative solver methodologies have
been developed and successfully employed for
either single-field problems (e.g., Poisson's
equation, linear elasticity) or for two-field
problems (e.g., Darcy equations, Stokes equations)
using two-field composable solvers \citep{Saad,
  rathgeber2016firedrake,brown2012composable}.  
However, there is a gap in knowledge when
one wants to solve large-scale problems
under the DPP model, which involves four
independent field variables. Unlike Darcy
equations, the governing equations under the DPP model
cannot
be written as a single-field Poisson
equation solely in terms of pressures
\citep{joodat2018modeling}, or even 
as a two-field problem.

To facilitate solving large-scale problems
under the DPP model, we present two four-field
composable block solver methodologies.
Appealing to PETSc's composable solver 
features \citep{petsc-user-ref,brown2012composable}
and Firedrake Project's finite element libraries
\citep{rathgeber2016firedrake}, we will show that
the proposed composable block solvers can be
effectively implemented in a parallel setting.
The two salient features of the proposed block
solvers are: they are scalable in both the algorithmic
and parallel senses. They can be employed under
a wide variety of finite element discretizations.
Both these features will be illustrated in this
paper using representative two- and three-dimensional
problems. 

In order to illustrate that the proposed
composable solvers can be used under a wide
variety of finite element discretizations,
we will employ three popular finite element
discretizations -- the classical mixed
formulation (which is
based on the Galerkin formalism) using H(div)
elements, the CG-VMS stabilized formulation
\citep{joodat2018modeling} and the DG-VMS stabilized
formulation \citep{joshaghani2018stabilized}.
We will consider H(div) discretizations
for simplicial elements (triangle [TRI] and
tetrahedron [TET]) and non-simplicial elements
(quadrilateral [QUAD] and hexahedron [HEX]).
In particular, we employ the lowest-order
Raviart-Thomas spaces for simplicial
elements \citep{raviart1977mixed,boffi2013mixed}
\footnote{The classical mixed formulation
  using the lowest-order Raviart-Thomas
  spaces for simplicial elements is
  commonly referred to as the RT0 formulation.}.
For non-simplicial elements, the velocity
spaces for QUAD and HEX elements are,
respectively, $\mathrm{RCTF_1}$ and
$\mathrm{NCF_1}$
\citep{McRae2016,arnold2014periodic}
\footnote{See Figure \ref{Fig:elements}
  and Table \ref{tab:ELE_level_discretization}
  for a description of these discretizations.}. 
The CG-VMS formulation is based on the
variational multi-scale (VMS) formalism
\citep{hughes1995multiscale} and is stable
under any arbitrary interpolation order
for velocity and pressure fields.
The DG-VMS formulation is a discontinuous
version of the CG-VMS formulation and is
built by combining the VMS formalism and
discontinuous Galerkin techniques.
The DG-VMS formulation has been shown to
accurately capture physical jumps in flow
profiles across highly heterogeneous porous
media \citep{joshaghani2018stabilized}.

Recently, \citep{chang2018comparative} have proposed
the Time-Accuracy-Size (TAS) performance spectrum
model, which is an enhanced version of the original
spectrum model proposed in \citep{chang2018performance}
obtained by incorporating accuracy into the spectrum
model. The TAS spectrum model can be used to
study performance of numerical formulations in
a parallel setting. Herein, we will utilize the
TAS model specifically to achieve the following:
(i) We show that the proposed composable
solvers are algorithmically scalable.
(ii) We compare the performance of the two
proposed composable solvers on a particular
hardware.
(iii) We discuss how the choice of finite
element mesh type could affect the solver
performance.
(iv) We compare
  the performance of the chosen three finite element
  discretizations (the classical mixed formulation
  with H(div) elements,
  the CG-VMS stabilized formulation and
  the DG-VMS stabilized formulation) for
  solving the governing equations under
  the DPP model.    

The work reported in this paper will
be valuable to subsurface modelers on
three fronts. First and the obvious
one is that the proposed composable 
block solver methodologies facilitate
solving large-scale problems involving
flow through porous media with multiple
pore-networks. Second, our work can guide
an application scientist to choose a
finite element discretization among
several choices. Third, our work illustrates
on how to utilize performance metrics other than the commonly
used metric -- the total time to solution
-- in subsurface modeling. A couple  of these 
other metrics include Digits of Efficacy (DoE)
and the total Degrees-of-Freedom (DoF) processed per 
second (DoF/s). 

The rest of this paper is organized as
follows. The governing equations under
the DPP model and convenient grouping
of the field variables are presented
in Section \ref{Sec:S2_Comparison_GE}. 
The weak forms of the three finite
element formulations that are employed
in this paper are presented in Section
\ref{Sec:S3_Comparison_Weak}.
The proposed two block solver
methodologies are discussed
in detail in Section
\ref{Sec:S4_Comparison_Solver}.
The framework of performance spectrum
model along with the guidelines on how
to interpret the resulting diagrams are
presented in Section
\ref{Sec:S5_Comparison_Performance}.
The performance of the proposed block
solvers is illustrated using numerical
simulations in Section \ref{Sec:S6_Comparison_NR}.
In the same section, we also compare the
performance of the chosen finite element
discretizations using the TAS performance
spectrum model, which provides guidance
to an application scientist with respect
to several metrics (e.g., time-to-solution,
digits-of-efficacy).
Finally, conclusions are drawn in 
Section \ref{Sec:S7_Comparison_Conclusion}.

\section{GOVERNING EQUATIONS AND GROUPING OF FIELD VARIABLES}
\label{Sec:S2_Comparison_GE}
We now document the most important equations under
the DPP mathematical model. In this paper, we restrict 
our treatment of the model to a steady-state response;
however, the proposed composable block solvers are
equally applicable in a transient setting.
We refer
the two dominant
pore-networks as \emph{macro-pore} and \emph{micro-pore},
and the quantities associated with the two pore-networks
are, respectively, identified by subscripts $1$ and $2$.
The porous medium is denoted by $\Omega$. Mathematically, 
$\Omega \subset \mathbb{R}^{nd}$ is assumed to be open
and bounded, where $nd$ denotes the number of spatial
dimensions. In this paper, $nd = 2 \; \mathrm{or} \; 3$. 
The gradient and divergence operators with respect
to a spatial point $\mathbf{x} \in \Omega$ are,
respectively, denoted by $\mathrm{grad}[\cdot]$
and $\mathrm{div}[\cdot]$. 
The pressure and the discharge
(or Darcy) velocity fields in the macro-pore network
are, respectively, denoted by $p_{1}(\mathbf{x})$ and
$\mathbf{u}_{1}(\mathbf{x})$, and the corresponding
fields in the micro-pore network are denoted by
$p_{2}(\mathbf{x})$ and $\mathbf{u}_{2}(\mathbf{x})$.
The viscosity and true density of the
fluid are denoted by $\mu$ and $\gamma$, respectively.

The governing equations for a steady-response
under the DPP mathematical model take the
following form:
\begin{subequations}
  \label{Eqn:DG_GE_Darcy_BLM}
  \begin{alignat}{2}
    \label{Eqn:DG_GE_Darcy_BLM_1}
    &\mu k_{1}^{-1} \mathbf{u}_1(\mathbf{x})
    + \mathrm{grad}[p_1] = \gamma \mathbf{b}(\mathbf{x})
    &&\quad \mathrm{in} \; \Omega \\
    \label{Eqn:DG_GE_Darcy_BLM_2}
    &\mu k_{2}^{-1} \mathbf{u}_2(\mathbf{x})
    + \mathrm{grad}[p_2] = \gamma \mathbf{b}(\mathbf{x})
    &&\quad \mathrm{in} \; \Omega \\
    \label{Eqn:DG_GE_Darcy_mass_balance_1}
    &\mathrm{div}[\mathbf{u}_1] = -\frac{\beta}{\mu}(p_1(\mathbf{x}) - p_2(\mathbf{x}))
    &&\quad \mathrm{in} \; \Omega \\
    \label{Eqn:DG_GE_Darcy_mass_balance_2}
    &\mathrm{div}[\mathbf{u}_2] = +\frac{\beta}{\mu}(p_1(\mathbf{x}) - p_2(\mathbf{x}))
    &&\quad \mathrm{in} \; \Omega 
  \end{alignat}
\end{subequations}
where $k_1(\mathbf{x})$ and $k_2(\mathbf{x})$,
respectively, denote the (isotropic)
permeabilities of the macro-pore and
micro-pore networks, $\beta$ is a dimensionless
characteristic of the porous medium,
and $\mathbf{b}(\mathbf{x})$ denotes
the specific body force.
In the above equations, the mass transfer
per unit volume, $\chi(\mathbf{x})$, from
the macro-pore network to the micro-pore
network is modeled as follows:
\begin{align}
  \label{Eqn:Composable_mass_transfer}
  \chi(\mathbf{x}) = -\frac{\beta}{\mu}(p_1 - p_2)
\end{align}
As one can see from equations
\eqref{Eqn:DG_GE_Darcy_BLM_1}--\eqref{Eqn:DG_GE_Darcy_mass_balance_2},
the flow in one pore-network is coupled with its counterpart
in the other pore-network through the inter-pore mass
transfer. Unlike Darcy equations, it is \emph{not}
possible to rewrite the governing equations under the
DPP model in form of a single-field Poisson's equation
in terms of pressures. One has to deal with the governing
equations in their mixed form. 

\subsection{Grouping of field variables in continuum setting}
We now discuss two ways of grouping the field
variables, which form the basis for the proposed
composable block solvers.
Under the first approach, the field variables
are grouped based on the scale of the pore-network.
That is, all the field variables (i.e., velocity and
pressure) pertaining to the macro-pore network
are placed in one group, and the field variables
of the micro-pore network are placed into another.
We refer to this splitting of field variables as the
\emph{scale-split} and the associated grouping
takes the following form: 
\begin{align}
  \label{Eqn:Composable_group_scale_split}
  \boldsymbol{\Upsilon}_1 = \left\{\begin{array}{c}
  \mathbf{u}_1(\mathbf{x}) \\
  p_1(\mathbf{x})          \\ \hline
  \mathbf{u}_2(\mathbf{x}) \\
  p_2(\mathbf{x})   
  \end{array}\right\}
\end{align}
The governing equations of the DPP model
under the scale-split can be compactly
written as follows:
\begin{align}
  \mathcal{L}_{1}[\boldsymbol{\Upsilon}_{1}]
  = \mathcal{F}_{1}
\end{align}
In the above equation, the differential
operator takes the following form:
\begin{align}
  \mathcal{L}_{1} := \left[\begin{array}{cc|cc}
      \mu k_1^{-1} \mathbf{I} & \mathrm{grad}[\cdot] & \mathbf{O} & 0 \\
      \mathrm{div}[\cdot] & \frac{\beta}{\mu} & \mathbf{O} & -\frac{\beta}{\mu} \\ \hline
      \mathbf{O} & 0 & \mu k_2^{-1} \mathbf{I} & \mathrm{grad}[\cdot] \\
      \mathbf{O} & -\frac{\beta}{\mu} & \mathrm{div}[\cdot] & \frac{\beta}{\mu}
    \end{array}\right]
\end{align}
where $\mathbf{I}$ denotes the identity
tensor, $\mathbf{O}$ denotes the zero
tensor, and the forcing function takes
the following form:
\begin{align}
  \mathcal{F}_1 = \left\{\begin{array}{c}
  \gamma \mathbf{b}(\mathbf{x}) \\
  0          \\ \hline
  \gamma \mathbf{b}(\mathbf{x}) \\
  0
  \end{array}\right\}
\end{align}

Under the second approach, the field variables
are grouped based on the nature of the fields.
That is, field variables of a similar kind are
placed in the same group. We refer to this splitting
of field variables as the \emph{field-split}
and the associated grouping takes the following
form: 
\begin{align}
  \label{Eqn:Composable_group_field_split}
  \boldsymbol{\Upsilon}_2 = \left\{\begin{array}{c}
  \mathbf{u}_1(\mathbf{x})   \\
  \mathbf{u}_2(\mathbf{x})   \\ \hline
  p_1(\mathbf{x})            \\ 
  p_2(\mathbf{x})   
  \end{array}\right\}
\end{align}
The governing equations of the DPP model
under the field-split can be compactly
written as follows:
\begin{align}
  \mathcal{L}_{2}[\boldsymbol{\Upsilon}_{2}]
  = \mathcal{F}_{2}
\end{align}
where the differential operator
takes the following form:
\begin{align}
  \mathcal{L}_{2} := \left[\begin{array}{cc|cc}
      \mu k_1^{-1} \mathbf{I} & \mathbf{O} & \mathrm{grad}[\cdot] & 0 \\
      \mathbf{O} & \mu k_2^{-1} \mathbf{I} & 0 & \mathrm{grad}[\cdot] \\ \hline 
      \mathrm{div}[\cdot] & \mathbf{O} & \frac{\beta}{\mu} & -\frac{\beta}{\mu} \\ 
      \mathbf{O} & \mathrm{div}[\cdot] & -\frac{\beta}{\mu} & \frac{\beta}{\mu}
    \end{array}\right]
\end{align}
and the forcing function can
be written as follows:
\begin{align}
  \mathcal{F}_2 = \left\{\begin{array}{c}
  \gamma \mathbf{b}(\mathbf{x}) \\
  \gamma \mathbf{b}(\mathbf{x}) \\ \hline 
  0 \\ 
  0
  \end{array}\right\}
\end{align}

\subsection{Boundary conditions}
The boundary of the domain will be denoted
by $\partial \Omega$. The unit outward normal
to the boundary at $\mathbf{x} \in \partial \Omega$
is denoted by $\widehat{\mathbf{n}}(\mathbf{x})$.
The velocity boundary condition (i.e., no
penetration boundary condition) for each
pore-network takes the following form:
\begin{subequations}
	\label{Eqn:DG_GE_vBC}
  \begin{alignat}{2}
    \label{Eqn:DG_GE_vBC_1}
    &\mathbf{u}_1(\mathbf{x}) \cdot
    \widehat{\mathbf{n}}(\mathbf{x})
    = u_{n1}(\mathbf{x})
    &&\quad \mathrm{on} \; \Gamma^{u}_{1} \\
    \label{Eqn:DG_GE_vBC_2}
    &\mathbf{u}_2(\mathbf{x}) \cdot
    \widehat{\mathbf{n}}(\mathbf{x})
    = u_{n2}(\mathbf{x}) 
    &&\quad \mathrm{on} \; \Gamma^{u}_{2} 
  \end{alignat}
\end{subequations}
where $\Gamma_{i}^{u}$ denotes that part of the boundary on
which the normal component of the velocity is prescribed
in the macro-pore ($i=1$) and micro-pore ($i=2$) networks,
and $u_{n1}(\mathbf{x})$ and $u_{n2}(\mathbf{x})$ denote the
prescribed normal components of the velocities on $\Gamma^{u}_{1}$ and $\Gamma^{u}_{2}$, respectively.  
The pressure boundary condition for each
pore-network reads:
\begin{subequations}
	\label{Eqn:DG_GE_Darcy_pBC}
  \begin{alignat}{2}
    \label{Eqn:DG_GE_Darcy_pBC_1}
    &p_1(\mathbf{x}) = p_{01} (\mathbf{x})
    &&\quad \mathrm{on} \; \Gamma^{p}_{1} \\
    \label{Eqn:DG_GE_Darcy_pBC_2}
    &p_2(\mathbf{x}) = p_{02} (\mathbf{x})
    &&\quad \mathrm{on} \; \Gamma^{p}_{2} 
  \end{alignat}
\end{subequations}
in which $\Gamma_{i}^{p}$ is that part of the boundary on
which the pressure is prescribed in the macro-pore ($i=1$)
and micro-pore ($i=2$) networks, and $p_{01}(\mathbf{x})$
and $p_{02}(\mathbf{x})$ denote the prescribed pressures
on $\Gamma_{1}^{p}$ and $\Gamma_{2}^{p}$, respectively.
For mathematical well-posedness,
we assume that 
\begin{align}
  \Gamma_{1}^{u} \cup \Gamma_{1}^{p} =
  \partial \Omega, \quad
  \Gamma_{1}^{u} \cap \Gamma_{1}^{p} = \emptyset, \quad 
  \Gamma_{2}^{u} \cup \Gamma_{2}^{p} = \partial \Omega 
  \quad \mathrm{and} \quad
  \Gamma_{2}^{u} \cap \Gamma_{2}^{p} = \emptyset
\end{align}

\section{CLASSICAL AND STABILIZED MIXED WEAK FORMULATIONS}
\label{Sec:S3_Comparison_Weak}
As mentioned earlier, we will employ three
different mixed formulations -- the classical
mixed formulation using H(div) discretizations,
a continuous stabilized
mixed formulation and a discontinuous
stabilized mixed formulation.
Under a mixed formulation, velocities
and pressures are taken to be the primary
variables. However, for numerical stability,
a mixed formulation should either satisfy or
circumvent the Ladyzhenskaya-Babu\v{s}ka-Brezzi
(LBB) \emph{inf-sup} stability condition
\citep{brezzi2012mixed}. This naturally
places all the mixed formulations into
either of two categories.
A mixed formulation in the first category
is built on the classical mixed formulation
(which is based on the Galerkin formalism)
but places restrictions on the interpolation
functions for the independent field variables
to satisfy the LBB condition. To put it
differently, not all combinations of
interpolation functions for the field
variables satisfy the LBB condition
under the classical mixed formulation.
A mixed formulation in the second category
augments the classical mixed formulation
with stabilization terms so as to circumvent
the LBB condition and to render a stable
formulation. In this paper, we consider
one mixed formulation from the first
category and two from the second category. 

We denote the set of all square-integrable
functions on $\Omega$ by $L_{2}(\Omega)$.
The standard $L_2$ inner-product over a
set $K$ is denoted as follows:
\begin{align}
  (\mathbf{a};\mathbf{b})_{K} \equiv \int_{K}
  \mathbf{a}(\mathbf{x})\cdot \mathbf{b}
  (\mathbf{x}) \; \mathrm{d} K 
\end{align}
For convenience, the subscript $K$
will be dropped if $K = \Omega$ in
the case of classical mixed formulation
and the stabilized mixed continuous
Galerkin formulation. In the case of
the stabilized mixed discontinuous
Galerkin formulation, the subscript
$K$ will be dropped if $K = \widetilde{\Omega}$,
which will be defined later in equation
\eqref{Eqn:DG_Omega_tilde}.
We now provide details of the three
chosen mixed finite element discretizations.

Let us define the following function spaces
for the velocities and pressures fields as
follows:
\begin{subequations}
	\begin{align}
	\label{Eqn:VMS_Function_space_U1}
	\mathcal{U}_{1} &:= 
	\left\{\mathbf{u}_{1}(\mathbf{x}) \in 
	\left(L_{2}(\Omega)\right)^{nd} 
	\; \Big\vert \;
	\mathrm{div}[\mathbf{u}_{1}] \in L_{2}(\Omega), 
	\mathbf{u}_{1}(\mathbf{x}) \cdot \widehat{\mathbf{n}}(\mathbf{x}) 
	= u_{n1}(\mathbf{x}) \in H^{-1/2}(\Gamma_{1}^{u})\right\} \\
	\mathcal{U}_{2} &:= 
	\left\{\mathbf{u}_{2}(\mathbf{x}) \in 
	\left(L_{2}(\Omega)\right)^{nd} 
	\; \Big\vert \;
	\mathrm{div}[\mathbf{u}_{2}] \in L_{2}(\Omega), 
	\mathbf{u}_{2}(\mathbf{x}) \cdot \widehat{\mathbf{n}}(\mathbf{x}) 
	= u_{n2}(\mathbf{x}) \in H^{-1/2}(\Gamma_{2}^{u})\right\} \\
	\mathcal{W}_{1} &:= 
	\left\{\mathbf{w}_{1}(\mathbf{x}) \in 
	\left(L_{2}(\Omega)\right)^{nd} 
	\; \Big\vert \;
	\mathrm{div}[\mathbf{w}_{1}] \in L_{2}(\Omega), 
	\mathbf{w}_{1}(\mathbf{x}) \cdot \widehat{\mathbf{n}}(\mathbf{x}) 
	= 0 \; \mathrm{on} \; \Gamma_{1}^{u} \right\} \\
	\mathcal{W}_{2} &:= 
	\left\{\mathbf{w}_{2}(\mathbf{x}) \in 
	\left(L_{2}(\Omega)\right)^{nd} 
	\; \Big\vert \;
	\mathrm{div}[\mathbf{w}_{2}] \in L_{2}(\Omega), 
	\mathbf{w}_{2}(\mathbf{x}) \cdot \widehat{\mathbf{n}}(\mathbf{x}) 
	= 0 \; \mathrm{on} \; \Gamma_{2}^{u}\right\} \\
        \mathcal{P} &:= 
		\left\{(p_1(\mathbf{x}),p_{2}(\mathbf{x})) 
		\in L_{2}(\Omega) \times L_{2}(\Omega)
		\; \Big\vert \;
		\left(\int_{\Omega} p_{1}(\mathbf{x}) \mathrm{d} \Omega \right) 
		\left(\int_{\Omega} p_{2}(\mathbf{x}) \mathrm{d} \Omega \right) 
		= 0 \right\} \\
	\label{Eqn:VMS_Function_space_Q}
	\mathcal{Q} &:= 
	\left\{(p_1(\mathbf{x}),p_{2}(\mathbf{x})) 
	\in H^{1}(\Omega) \times H^{1}(\Omega)
	\; \Big\vert \;
	\left(\int_{\Omega} p_{1}(\mathbf{x}) \mathrm{d} \Omega \right) 
	\left(\int_{\Omega} p_{2}(\mathbf{x}) \mathrm{d} \Omega \right) 
	= 0 \right\}
	\end{align}
\end{subequations}
where $H^{1}(\Omega)$ is a standard Sobolev space,
and $H^{-1/2}(\cdot)$ is the dual space corresponding
to $H^{1/2}(\cdot)$. Rigorous discussion of Sobolev spaces are accessible in \citep{krylov2008lectures}; and further discussion of function spaces are provided by \citep{brezzi2012mixed}.

\subsection{Classical mixed formulation
  using H(div) elements}
The classical mixed formulation
can be written as follows:~Find
$\left(\mathbf{u}_1(\mathbf{x}),
\mathbf{u}_2(\mathbf{x})\right) \in
\mathcal{U}_1 \times \mathcal{U}_2$
and $\left(p_1(\mathbf{x}),p_2(\mathbf{x})
\right) \in \mathcal{P}$ such that we have
\begin{align}
  \label{Eqn:VMS_classical_mixed_formulation}
  \mathcal{B}_{\mathrm{Gal}}(\mathbf{w}_1,\mathbf{w}_2,q_1,q_2;
  \mathbf{u}_1,\mathbf{u}_2,p_1,p_2) = \mathcal{L}_{\mathrm{Gal}}
  (\mathbf{w}_1,\mathbf{w}_2,q_1,q_2) \nonumber \\
  \quad \forall \left(\mathbf{w}_1(\mathbf{x}),
  \mathbf{w}_{2}(\mathbf{x})\right) \in
  \mathcal{W}_1 \times \mathcal{W}_2,~
  \left(q_1(\mathbf{x}),~q_2(\mathbf{x})\right) \in \mathcal{P} 
\end{align}
where the bilinear form and the linear
functional are, respectively, defined
as follows:
\begin{subequations}
  \begin{align}
    \mathcal{B}_{\mathrm{Gal}} 
    &:= (\mathbf{w}_1;\mu k_{1}^{-1}\mathbf{u}_1)
    - (\mathrm{div}[\mathbf{w}_1];p_1)
    + (q_1;\mathrm{div}[\mathbf{u}_1]) 
    + (\mathbf{w}_2;\mu k_{2}^{-1}\mathbf{u}_2) \notag \\
    &- (\mathrm{div}[\mathbf{w}_2];p_2)
    + (q_2;\mathrm{div}[\mathbf{u}_2]) 
    + (q_1 - q_2;\beta/\mu(p_1 - p_2)) \\
    \mathcal{L}_{\mathrm{Gal}}
    &:= (\mathbf{w}_1;\gamma \mathbf{b})
    - (\mathbf{w}_1 \cdot \widehat{\mathbf{n}};p_{01}
    )_{\Gamma^{\mathrm{p}}_{1}}~+ (\mathbf{w}_2;\gamma \mathbf{b})
    - (\mathbf{w}_2 \cdot \widehat{\mathbf{n}};p_{02}
    )_{\Gamma^{\mathrm{p}}_{2}}
  \end{align}
\end{subequations}

\subsubsection{H(div) elements}
Classes of H(div) finite element discretizations
such as Raviart-Thomas (RT) \citep{raviart1977mixed},
generalized RTN \citep{nedelec1980mixed}, BDM
\citep{brezzi1985two}, and BDFM \citep{brezzi1987efficient}
have been shown to satisfy the LBB condition.
Moreover, these finite element discretizations
satisfy element-wise mass balance property
\citep{brezzi2012mixed}. 

The classical mixed formulation based
on discretizations from the lowest-order
Raviart-Thomas spaces is commonly referred
to as the RT0 formulation; which is
frequently used in subsurface modeling 
\citep{chen2006computational}. The unknowns
under the RT0 formulation on a triangle
are fluxes at the midpoints of edges
of the element and element-wise constant
pressures. 
The finite dimensional subspaces for
each velocity and pressure fields under
the lowest-order Raviart-Thomas discretization on
a triangle, which are collectively denoted by
$\mathrm{RTF}_1
\oplus \mathrm{DP}_0$, take the following form: 
\begin{subequations}
  \begin{align}
    &\mathcal{U}^{h}:=\{\mathbf{u} =
    (u,v)~|~u_{K} = a_{K}+b_{K}x,v_{K}
    = c_{K}+b_{K}y;~a_{K}, b_{K}, c_{K}
    \in \mathbb{R}; K \in \mathcal{T}_h\} \\
    &\mathcal{P}^{h}:=
    \{
    p~|~p = \mbox{constant on each
      triangle}~K \in \mathcal{T}_{h}
    \}
\end{align}
\end{subequations}
where $\mathcal{T}_{h}$ is a triangulation on $\Omega$. 
These subspaces on a tetrahedron,
which are denoted by $\mathrm{N1F}_1
\oplus \mathrm{DP}_0$, take the
following form:
\begin{subequations}
\begin{align}
&\mathcal{U}^{h}:=\{\mathbf{u} =
(u,v,w)~|~
u_{K} = a_{K} + b_{K} x,
v_{K} = c_{K} + b_{K} y, 
w_{K} = d_{K} + b_{K} z; \notag \\
&\qquad \qquad \qquad \qquad a_{K}, b_{K}, c_{K}, d_{K} 
\in \mathbb{R};
K \in \mathcal{T}_h\} \\
&\mathcal{P}^{h}:=
\{
p~|~p = \mbox{constant on each tetrahedron}~K \in \mathcal{T}_{h}
\}
\end{align}
\end{subequations}
where $\mathcal{T}_{h}$, in this case,
is a tetrahedralization on $\Omega$.

In addition to H(div) discretizations on
simplicial meshes, we also consider the
corresponding discretizations on non-simplicial
element -- QUAD and HEX.
The velocity spaces for QUAD and HEX elements
are, respectively, $\mathrm{RCTF_1}$ and
$\mathrm{NCF_1}$
\citep{McRae2016,arnold2014periodic}.
The (macro- and micro-) pressures are
element-wise constants, and $\mathrm{DG}_0$
is commonly used to denote element-wise
constant discretization on non-simplicial
elements. 
See Figure \ref{Fig:elements}
and Table \ref{tab:ELE_level_discretization}
for a description of these discretizations. 
The finite dimensional subspaces for the
$\mathrm{RCTF_1}$ and $\mathrm{NCF}_1$
discretizations can be written precisely
using the language of finite element
exterior calculus. But such a description
needs introduction of additional jargon and
notation, which is beyond the scope of
this paper. We, therefore,
refer the reader to \citep{arnold2006finite,
  arnold2010finite,arnold2014periodic}.
However, to guide the reader, the
degrees-of-freedom for these discretizations
are shown in Figure~\ref{Fig:elements} and
Table \ref{tab:ELE_level_discretization}.

\begin{table}[h]
  \caption{The element-level discretization
    for different mesh types and the chosen
    three formulations. $\bigoplus$ denotes
  the direct sum operator between two finite
  element spaces. The notation used in this
  table is based on the \emph{Periodic Table
    of the Finite Elements} \citep{arnold2014periodic}.}
    \begin{tabular}{ccc}
      \Xhline{2\arrayrulewidth}
      \multirow{2}{*}{
	\begin{tabular}[c]{@{}c@{}} Mesh\\ type
      \end{tabular}} & \multicolumn{2}{c}{Finite element formulation} \\ \cline{2-3} 
      & H(div) & CG-VMS/DG-VMS \\ \hline
      TRI & $[\mbox{RTF}_{1} \bigoplus \mbox{DP}_{0}]^{2}$ 
      & $[\mbox{P}_{1} \bigoplus \mbox{P}_{1}]^{2}$ \\
      QUAD & $[\mbox{RTCF}_{1} \bigoplus \mbox{DQ}_{0}]^{2}$ 
      & $[\mbox{Q}_{1} \bigoplus \mbox{Q}_{1}]^{2}$ \\
      TET & $[\mbox{N1F}_{1} \bigoplus \mbox{DP}_{0}]^{2}$ 
      & $[\mbox{P}_{1} \bigoplus \mbox{P}_{1}]^{2}$ \\
      HEX & $[\mbox{NCF}_{1} \bigoplus \mbox{DQ}_{0}]^{2}$  
      & $[\mbox{Q}_{1} \bigoplus \mbox{Q}_{1}]^{2}$ \\
      \Xhline{2\arrayrulewidth}
    \end{tabular}
    \label{tab:ELE_level_discretization}
\end{table}

\subsection{Stabilized mixed continuous
  Galerkin formulation (CG-VMS)}
The weak form of the CG-VMS formulation
can be written as follows:~Find
$\left(\mathbf{u}_1(\mathbf{x}),
\mathbf{u}_2(\mathbf{x}) \right) 
\in \mathcal{U}_1 \times \mathcal{U}_2$
and $\left(p_1(\mathbf{x}),p_2(\mathbf{x})\right)
\in \mathcal{Q}$ such that we have
\begin{align}
  \mathcal{B}^{\mathrm{CG}}_{\mathrm{stab}}(\mathbf{w}_1,\mathbf{w}_2,q_1,q_2;
  \mathbf{u}_1,\mathbf{u}_2,p_1,p_2)
  = \mathcal{L}^{\mathrm{CG}}_{\mathrm{stab}}(\mathbf{w}_1,\mathbf{w}_2,q_1,q_2)
  \notag \\
  \quad \forall
  \left(\mathbf{w}_1(\mathbf{x}), \mathbf{w}_2(\mathbf{x})\right) ~
  \in \mathcal{W}_1 \times \mathcal{W}_2,
  \left(q_1(\mathbf{x}),q_2(\mathbf{x})\right)
  \in \mathcal{Q} 
  \label{Eqn:VMS_Galerkin_Weak_Form} 
\end{align}
where the bilinear form and the linear
functional are defined, respectively,
as follows:
\begin{subequations}
  \begin{align}
    \label{Eqn:CG_VMS_Weak_form}
    \mathcal{B}^{\mathrm{CG}}_{\mathrm{stab}}
    &:= \mathcal{B}_{\mathrm{Gal}}(\mathbf{w}_1,\mathbf{w}_2,q_1,q_2;
    \mathbf{u}_1,\mathbf{u}_2,p_1,p_2) \nonumber \\
    &-\frac{1}{2} \left(\mu k_1^{-1}
    \mathbf{w}_1 - \mathrm{grad}[q_1];\frac{1}{\mu}
    k_1 (\mu k_1^{-1}
    \mathbf{u}_1 + \mathrm{grad}[p_1])\right)
    \nonumber \\
    &-\frac{1}{2} \left(\mu k_2^{-1} \mathbf{w}_2
    - \mathrm{grad}[q_2]; \frac{1}{\mu} k_2
    (\mu k_2^{-1}
    \mathbf{u}_2 + \mathrm{grad}[p_2])\right) \\ 
    \mathcal{L}_{\mathrm{stab}}^{\mathrm{CG}}
    &:= \mathcal{L}_{\mathrm{Gal}}(\mathbf{w}_1,\mathbf{w}_2,q_1,q_2)
    -\frac{1}{2} \left(\mu k_1^{-1}
    \mathbf{w}_1 - \mathrm{grad}[q_1];
    \frac{1}{\mu} k_1 \gamma \mathbf{b}\right) \nonumber \\
    &\qquad -\frac{1}{2} \left(\mu k_2^{-1} \mathbf{w}_2
    - \mathrm{grad}[q_2]; \frac{1}{\mu} k_2
    \gamma \mathbf{b}\right)
  \end{align}
\end{subequations}

An attractive feature of the CG-VMS formulation
is that nodal-based equal-order interpolation
for all the field variables (micro- and macro-
velocities and pressures) is stable, which is
not the case with the classical mixed formulation.
The stability is achieved by the addition of
stabilization terms, which circumvent the LBB
condition.

\subsection{Stabilized mixed discontinuous
  Galerkin formulation (DG-VMS)}
Formulations under the discontinuous Galerkin
(DG) method inherit attractive features of
both finite element and finite volume methods
by allowing discontinuous basis functions
(e.g., in the form of piecewise polynomials)
\citep{Warburton2007nodal}.
The DG method supports non-matching grids
and hanging nodes, and hence ideal for
$hp$ adaptivity \citep{Cockburn_DG}.
Moreover, the method can naturally
handle jumps in the profiles of the
solution variables
\citep{hughes2006stabilized,joshaghani2018stabilized}.
Since the DG method offers several attractive
features, we choose the DG formulation that
is recently proposed by \citep{joshaghani2018stabilized}
for the DPP model, which will be referred
to as the DG-VMS formulation, as one of
the three representative formulations to
illustrate the performance of the proposed
composable block solvers.

We now document the weak form under the DG-VMS
formulation. To this end, we decompose the domain
into $Nele$ open subdomains such that
\begin{align}
  \overline{\Omega} = \bigcup_{i=1}^{Nele} \overline{\omega}^{i} 
\end{align}
where $\omega^{i}$ denotes the $i$-th subdomain.
The union of all open subdomains will be denoted
by
\begin{align}
  \label{Eqn:DG_Omega_tilde}
  \widetilde{\Omega} =
  \bigcup_{i=1}^{Nele} \omega^{i} 
\end{align}
We denote the two adjacent subdomains 
sharing a given interior edge by $\omega^{+}$ and $\omega^{-}$. 
The unit normal vectors on the shared interface $\Gamma^{\pm}$
pointing outwards to $\omega^{+}$ and $\omega^{-}$ are, 
respectively, denoted by $\widehat{\mathbf{n}}^{+}$ and 
$\widehat{\mathbf{n}}^{-}$. 
The jump and average operators on an interior facet
for a scalar field $\varphi(\mathbf{x})$ are, 
respectively, defined as follows: 
\begin{align}
	\llbracket \varphi \rrbracket 
	:= \varphi^{+} \widehat{\mathbf{n}}^{+} 
	+ \varphi^{-} \widehat{\mathbf{n}}^{-} 
	\quad \mathrm{and} \quad 
	\{\varphi\} := \frac{\varphi^{+} + \varphi^{-}}{2} 
\end{align}
where 
\begin{align}
	\label{Eqn:DG_wplus_wminus}
	\varphi^{+} = \varphi|_{\partial \omega^{+}} 
	\quad \mathrm{and} \quad
	\varphi^{-} = \varphi|_{\partial \omega^{-}} 
\end{align}
For a vector field 
$\boldsymbol{\tau}(\mathbf{x})$ these operators are defined as follows: 
\begin{align}
	\llbracket \boldsymbol{\tau} \rrbracket :=
	\boldsymbol{\tau}^{+} \cdot \widehat{\mathbf{n}}^{+}
	+ \boldsymbol{\tau}^{-} \cdot \widehat{\mathbf{n}}^{-} 
	\quad \mathrm{and} \quad 
	\{ \boldsymbol{\tau} \}  := 
	\frac{\boldsymbol{\tau}^{+} + \boldsymbol{\tau}^{-} }{2} 
	\quad \mathrm{on} \; \Gamma^{\mathrm{int}} 
\end{align}
where $\boldsymbol{\tau}^{+}$ and $\boldsymbol{\tau}^{-}$ 
are defined similar to equation \eqref{Eqn:DG_wplus_wminus}. 

We denote the set of all square-integrable functions
on $\omega^{i}$ by $L_2(\omega^{i})$. We denote the
set of all functions that belong to $L_2(\omega^{i})$
and are continuously differentiable by $H^{1}(\omega^{i})$.
We then introduce the following broken Sobolev
spaces (which are piece-wise discontinuous
spaces): 
\begin{subequations}
	\begin{align}
	\mathcal{U}^{\mathrm{dg}} &:= \left\{\mathbf{u}(\mathbf{x}) \; \big| \;
	\mathbf{u}(\mathbf{x})\big|_{\omega^i} \in \left(L_{2}(\omega^i)
	\right)^{nd}; \; \mathrm{div}[\mathbf{u}] \in L_{2}(\omega^i); \;
	i = 1, \cdots, Nele \right\} \\
        \mathcal{Q}^{\mathrm{dg}} &:=
        \left\{(p_1(\mathbf{x}),p_2(\mathbf{x})) \; \big| \;
        p_{1}(\mathbf{x})\big|_{\omega^i} \in H^{1}(\omega^i), \;
        p_{2}(\mathbf{x})\big|_{\omega^i} \in H^{1}(\omega^i), \;
        \right. \notag \\
        &\qquad \qquad \qquad \qquad \qquad \qquad \qquad \left.
        \left(\int_{\widetilde{\Omega}} p_{1}(\mathbf{x}) \mathrm{d} \Omega\right)
        \left(\int_{\widetilde{\Omega}} p_{2}(\mathbf{x}) \mathrm{d} \Omega\right)
        = 0 \right\} 
        \end{align}
\end{subequations}

The weak form under the DG-VMS formulation
reads as follows:~Find
$\left(\mathbf{u}_1(\mathbf{x}),\mathbf{u}_2(\mathbf{x})\right)
\in \mathcal{U}^{\mathrm{dg}} \times \mathcal{U}^{\mathrm{dg}}
$, $\left(p_1(\mathbf{x}),
p_2(\mathbf{x})\right)
\in \mathcal{Q}^{\mathrm{dg}}
$ such that we have
\begin{align}
	\mathcal{B}_{\mathrm{stab}}^{\mathrm{DG}}(\mathbf{w}_1,\mathbf{w}_2,q_1,q_2;
	\mathbf{u}_1,\mathbf{u}_2,p_1,p_2)
	= \mathcal{L}_{\mathrm{stab}}^{\mathrm{DG}}(\mathbf{w}_1,\mathbf{w}_2,q_1,q_2) \notag \\ 
	\quad \forall \left(\mathbf{w}_1(\mathbf{x}),
	\mathbf{w}_2(\mathbf{x})\right) \in
	\mathcal{U}^{\mathrm{dg}} \times \mathcal{U}^{\mathrm{dg}},~
	\left(q_1(\mathbf{x}),q_2(\mathbf{x})\right) \in \mathcal{Q}^{\mathrm{dg}}
	\label{Eqn:VMS_DG_Weak_Form} 
\end{align}
where the bilinear form and the linear
functional are defined, respectively,
as follows:
\begin{subequations}
  \begin{align}
    \mathcal{B}^{\mathrm{DG}}_{\mathrm{stab}}
    &~:= \left( \mathbf{w}_1 ; \mu k_{1}^{-1} \mathbf{u}_{1} \right)
    - \left(\mathrm{div}[\mathbf{w}_{1}] ; p_{1} \right)
    +\left( \mathbf{w}_2 ; \mu k_{2}^{-1} \mathbf{u}_{2} \right)
    - \left(\mathrm{div}[\mathbf{w}_{2}] ; p_{2} \right) \notag \\ 
    &~+\left(\llbracket \mathbf{w}_{1} \rrbracket; \{\!\!\{ p_{1} \}\!\!\}
    \right)_{\Gamma^{\mathrm{int}}}
    - \left(\{\!\!\{q_{1} \}\!\!\} ; \llbracket \mathbf{u}_{1} \rrbracket
    \right)_{\Gamma^{\mathrm{int}}}  
    + \left(q_1;\mathrm{div}[\mathbf{u}_{1}]\right)
    + \left(q_2;\mathrm{div}[\mathbf{u}_{2}]\right) \notag \\ 
    &~+ \left(\llbracket \mathbf{w}_{2} \rrbracket; \{\!\!\{p_{2} \}\!\!\}
    \right)_{\Gamma^{\mathrm{int}}}
    - \left(\{\!\!\{q_{2} \}\!\!\} ; \llbracket \mathbf{u}_{2} \rrbracket
    \right)_{\Gamma^{\mathrm{int}}} 
    + \left(q_1 - q_2; \frac{\beta}{\mu}(p_{1} - p_{2}) \right) \notag \\
    &~+ \left(\mathbf{w}_{1} \cdot \widehat{\mathbf{n}}; p_1
    \right)_{\Gamma_{1}^{u}}
    + \left(\mathbf{w}_{2} \cdot \widehat{\mathbf{n}}; p_2
    \right)_{\Gamma_{2}^{u}} 
    -\left(q_{1};\mathbf{u}_{1} \cdot
    \widehat{\mathbf{n}}\right)_{\Gamma^{u}_{1}}
    -\left(q_{2};\mathbf{u}_{2} \cdot
    \widehat{\mathbf{n}}\right)_{\Gamma^{u}_{2}} \notag \\
    &-\frac{1}{2} \left(\mu k_1^{-1}
    \mathbf{w}_1 - \mathrm{grad}[q_1];
    \mu^{-1} k_1 (\mu k^{-1}_1
    \mathbf{u}_1 + \mathrm{grad}[p_1])\right)
    \nonumber \\
    &-\frac{1}{2} \left(\mu k_2^{-1}
    \mathbf{w}_2 - \mathrm{grad}[q_2];
    \mu^{-1} k_2(\mu k^{-1}_2 \mathbf{u}_2 + \mathrm{grad}[p_2])\right) 
    \nonumber\\
    &+ \eta_{u} h \left( \{\!\!\{\mu k_1^{-1}\}\!\!\}
    \llbracket \mathbf{w}_{1} \rrbracket ; \llbracket
    \mathbf{u}_1 \rrbracket
    \right)_{\Gamma^{\mathrm{int}}}
    + \eta_{u} h \left( \{\!\!\{\mu k_2^{-1}\}\!\!\}
    \llbracket \mathbf{w}_{2} \rrbracket ; \llbracket
    \mathbf{u}_2 \rrbracket\right)_{\Gamma^{\mathrm{int}}} \nonumber \\
    &+ \frac{\eta_{p}}{h} \left( \{\!\!\{\mu^{-1} k_1\}\!\!\}
    \llbracket q_{1} \rrbracket ; \llbracket p_1 \rrbracket
    \right)_{\Gamma^{\mathrm{int}}}
    + \frac{\eta_{p}}{h} \left( \{\!\!\{\mu^{-1} k_2\}\!\!\}
    \llbracket q_{2} \rrbracket ; \llbracket p_2
    \rrbracket\right)_{\Gamma^{\mathrm{int}}}  \\
    \mathcal{L}_{\mathrm{stab}}^{\mathrm{DG}} 
    &:= \left( \mathbf{w}_{1} ; \gamma \mathbf{b}_1 \right)
    + \left( \mathbf{w}_{2} ; \gamma \mathbf{b}_2 \right)
    -(\mathbf{w}_1 \cdot \widehat{\mathbf{n}}; p_{01})_{\Gamma_{1}^{p}} 
    -(\mathbf{w}_2 \cdot \widehat{\mathbf{n}}; p_{02})_{\Gamma_{2}^{p}} 
    - \left(q_{1};u_{n1} \right)_{\Gamma^{u}_{1}}
    - \left(q_{2};u_{n2}\right)_{\Gamma^{u}_{2}} \notag \\
    &-\frac{1}{2} \left(\mu k_1^{-1} \mathbf{w}_1
    - \mathrm{grad}[q_1]; \mu^{-1} k_1 \gamma
    \mathbf{b}_{1}\right)
    -\frac{1}{2} \left(\mu k_2^{-1} \mathbf{w}_2 - \mathrm{grad}[q_2];
    \mu^{-1} k_2 \gamma \mathbf{b}_{2}\right) 
  \end{align}
\end{subequations}
where $\eta_{u}$ and $\eta_{p}$ are
non-negative, non-dimensional numbers.

The DG-VMS formulation also circumvents
the LBB condition and is stable under
arbitrary combinations of interpolation
functions for the field variables.

\section{PROPOSED FOUR-FIELD SOLVERS}
\label{Sec:S4_Comparison_Solver}
The fully discrete formulations for the DPP model can be
assembled into the following linear
problem:
\begin{align}
  \label{Eqn:S4_linear_problem}
  \boldsymbol{Ku}=\boldsymbol{f}
\end{align}
where $\boldsymbol{K}$ is the stiffness matrix, $\boldsymbol{u}$
is the vector of unknown velocities and pressure, and $\boldsymbol{f}$
is the corresponding forcing or RHS vector. Solving the system of
equations \eqref{Eqn:S4_linear_problem} in a fast and scalable way
requires careful composition and manipulation of the four different 
physical fields. In this section, we demonstrate how this can be done through 
PETSc \citep{petsc-user-ref,petsc-web-page,Dalcin2011} and its 
composable solver capabilities \citep{brown2012composable}. 
The individual block components of the stiffness matrix 
$\boldsymbol{K}$ for the mixed Galerkin formulation using H(div) elements can be 
categorized into the following:
\begin{subequations}
  \label{Eqn:S4_rt0_stiffness_matrix}
  \begin{align}
    \boldsymbol{K}^{1}_{uu} &\leftarrow \left(\mathbf{w}_1;\mu {k}_{1}^{-1}\mathbf{u}_1\right) \\
    \boldsymbol{K}^{1}_{up} &\leftarrow - \left(\mathrm{div}[\mathbf{w}_1];p_1\right)\\
    \boldsymbol{K}^{1}_{pu} &\leftarrow \left(q_1;\;\mathrm{div}[\mathbf{v}_1]\right)\\
    \boldsymbol{K}^{1}_{pp} &\leftarrow \left(q_1;\;\frac{\beta}{\mu}p_1\right)\\
    \boldsymbol{K}^{2}_{uu} &\leftarrow \left( \mathbf{w}_2;\;\mu{k}^{-1}_2\mathbf{u}_2\right) \\
    \boldsymbol{K}^{2}_{up} &\leftarrow - \left( \mathrm{div}[\mathbf{w}_2];\; p_2\right) \\
    \boldsymbol{K}^{2}_{pu} &\leftarrow \left( q_2;\;\mathrm{div}[\mathbf{v}_2]\right) \\
    \boldsymbol{K}^{2}_{pp} &\leftarrow \left(q_2;\;\frac{\beta}{\mu}p_2\right) \\
    \boldsymbol{K}^{12}_{pp} &\leftarrow -\left(q_1;\;\frac{\beta}{\mu}p_2\right) \\
    \boldsymbol{K}^{21}_{pp} &\leftarrow -\left(q_2;\;\frac{\beta}{\mu}p_1\right)
\end{align}
\end{subequations}
For the CG-VMS formulation, the individual block components of 
the stiffness matrix can be categorized into the following:
\begin{subequations}
  \label{Eqn:S4_cg_stiffness_matrix}
  \begin{align}
    \boldsymbol{K}^{1}_{uu} &\leftarrow \frac{1}{2}\left(\mathbf{w}_1;\mu {k}_{1}^{-1}\mathbf{u}_1\right) \\
    \boldsymbol{K}^{1}_{up} &\leftarrow - \left(\mathrm{div}[\mathbf{w}_1];p_1\right) 
    - \frac{1}{2}\left(\mathbf{w}_1;\;\mathrm{grad}[p_1]\right) \\
    \boldsymbol{K}^{1}_{pu} &\leftarrow \left(q_1;\;\mathrm{div}[\mathbf{v}_1]\right) 
    + \frac{1}{2}\left(\mathrm{grad}[q_1];\;\mathbf{u}_1\right) \\
    \boldsymbol{K}^{1}_{pp} &\leftarrow \frac{1}{2}\left(\mathrm{grad}[q_1];\;\frac{1}{\mu}{k}_1\mathrm{grad}[p_1]\right) 
    + \left(q_1;\;\frac{\beta}{\mu}p_1\right)\\
    \boldsymbol{K}^{2}_{uu} &\leftarrow \frac{1}{2}\left( \mathbf{w}_2;\;\mu{k}^{-1}_2\mathbf{u}_2\right) \\
    \boldsymbol{K}^{2}_{up} &\leftarrow - \left( \mathrm{div}[\mathbf{w}_2];\; p_2\right) 
    - \frac{1}{2}\left( \mathbf{w}_2;\;\mathrm{grad}[p_2]\right) \\
    \boldsymbol{K}^{2}_{pu} &\leftarrow \left( q_2;\;\mathrm{div}[\mathbf{v}_2]\right) 
    + \frac{1}{2}\left(\mathrm{grad}[q_2];\;\mathbf{u}_2\right) \\
    \boldsymbol{K}^{2}_{pp} &\leftarrow \frac{1}{2}\left(\mathrm{grad}[q_2];\;\frac{1}{\mu}{k}_2\mathrm{grad}[p_2]\right) 
    + \left(q_2;\;\frac{\beta}{\mu}p_2\right) \\
    \boldsymbol{K}^{12}_{pp} &\leftarrow -\left(q_1;\;\frac{\beta}{\mu}p_2\right) \\
    \boldsymbol{K}^{21}_{pp} &\leftarrow -\left(q_2;\;\frac{\beta}{\mu}p_1\right)
\end{align}
\end{subequations}
Likewise, the block components of the stiffness matrix for the DG-VMS 
formulation read:
\begin{subequations}
  \label{Eqn:S4_dg_stiffness_matrix}
  \begin{align}
    \boldsymbol{K}^{1}_{uu} &\leftarrow \frac{1}{2}\left( \mathbf{w}_1;\;\mu{k}^{-1}_1\mathbf{u}_1\right) 
    + \eta_{u} h \left( \{\!\!\{\mu {k}_1^{-1}\}\!\!\}\llbracket \mathbf{w}_{1} \rrbracket;\;\llbracket\mathbf{u}_1 \rrbracket\right)_{\Gamma^{\mathrm{int}}}\\
    \boldsymbol{K}^{1}_{up} &\leftarrow - \left( \mathrm{div}[\mathbf{w}_1];\; p_1\right) 
    - \frac{1}{2}\left( \mathbf{w}_1;\;\mathrm{grad}[p_1]\right) 
    + \left(\llbracket \mathbf{w}_{1} \rrbracket;\;\{\!\!\{ p_{1} \}\!\!\}\right)_{\Gamma^{\mathrm{int}}} 
    + \left(\mathbf{w}_{1} \cdot \widehat{\mathbf{n}};\;p_1\right)_{\Gamma_{1}^{u}}\\
    \boldsymbol{K}^{1}_{pu} &\leftarrow \left( q_1;\;\mathrm{div}[\mathbf{v}_1]\right) 
    + \frac{1}{2}\left(\mathrm{grad}[q_1];\;\mathbf{u}_1\right) 
    - \left(\{\!\!\{ q_{1} \}\!\!\};\;\llbracket \mathbf{v}_{1} \rrbracket\right)_{\Gamma^{\mathrm{int}}} 
    - \left(q_1;\;\mathbf{v}_{1} \cdot \widehat{\mathbf{n}}\right)_{\Gamma_{1}^{u}}\\
    \boldsymbol{K}^{1}_{pp} &\leftarrow \frac{1}{2}\left(\mathrm{grad}[q_1];\;\frac{1}{\mu}{k}_1\mathrm{grad}[p_1]\right) 
    + \left(q_1;\;\frac{\beta}{\mu}p_1\right)
    + \frac{\eta_{p}}{h} \left( \{\!\!\{\mu^{-1} {k}_1\}\!\!\}\llbracket q_{1} \rrbracket ; \llbracket p_1 \rrbracket\right)_{\Gamma^{\mathrm{int}}}\\
    \boldsymbol{K}^{2}_{uu} &\leftarrow \frac{1}{2}\left( \mathbf{w}_2;\;\mu{k}^{-1}_2\mathbf{u}_2\right)
    + \eta_{u} h \left( \{\!\!\{\mu {k}_2^{-1}\}\!\!\}\llbracket \mathbf{w}_{2} \rrbracket;\;\llbracket\mathbf{u}_2 \rrbracket\right)_{\Gamma^{\mathrm{int}}}\\
    \boldsymbol{K}^{2}_{up} &\leftarrow - \left( \mathrm{div}[\mathbf{w}_2];\; p_2\right) 
    - \frac{1}{2}\left( \mathbf{w}_2;\;\mathrm{grad}[p_2]\right) 
    + \left(\llbracket \mathbf{w}_{2} \rrbracket;\;\{\!\!\{ p_{2} \}\!\!\}\right)_{\Gamma^{\mathrm{int}}} 
    + \left(\mathbf{w}_{2} \cdot \widehat{\mathbf{n}};\;p_2\right)_{\Gamma_{2}^{u}}\\
    \boldsymbol{K}^{2}_{pu} &\leftarrow \left( q_2;\;\mathrm{div}[\mathbf{v}_2]\right) 
    + \frac{1}{2}\left(\mathrm{grad}[q_2];\;\mathbf{u}_2\right)
    - \left(\{\!\!\{ q_{2} \}\!\!\};\;\llbracket \mathbf{v}_{2} \rrbracket\right)_{\Gamma^{\mathrm{int}}} 
    - \left(q_2;\;\mathbf{v}_{2} \cdot \widehat{\mathbf{n}}\right)_{\Gamma_{2}^{u}}\\
    \boldsymbol{K}^{2}_{pp} &\leftarrow \frac{1}{2}\left(\mathrm{grad}[q_2];\;\frac{1}{\mu}{k}_2\mathrm{grad}[p_2]\right) 
    + \left(q_2;\;\frac{\beta}{\mu}p_2\right) 
    + \frac{\eta_{p}}{h} \left( \{\!\!\{\mu^{-1} {k}_2\}\!\!\}\llbracket q_{2} \rrbracket ; \llbracket p_2 \rrbracket\right)_{\Gamma^{\mathrm{int}}}\\
    \boldsymbol{K}^{12}_{pp} &\leftarrow -\left(q_1;\;\frac{\beta}{\mu}p_2\right) \\
    \boldsymbol{K}^{21}_{pp} &\leftarrow -\left(q_2;\;\frac{\beta}{\mu}p_1\right)
\end{align}
\end{subequations}
The components of the corresponding RHS vector $\boldsymbol{f}$ for equations
\eqref{Eqn:S4_rt0_stiffness_matrix}, \eqref{Eqn:S4_cg_stiffness_matrix} and 
\eqref{Eqn:S4_dg_stiffness_matrix} are
\begin{subequations}
  \label{Eqn:S4_rt0_rhs}
  \begin{align}
    \boldsymbol{f}^1_{u} &\leftarrow \left(\mathbf{w}_{1};\;\gamma\mathbf{b}\right)
    - \left(\mathbf{w}_1\cdot\widehat{\mathbf{n}};\;p_{01}\right)_{\Gamma^{\mathrm{p}}_{1}}\\
    \boldsymbol{f}^1_{p} &\leftarrow \boldsymbol{0} \\ 
    \boldsymbol{f}^2_{u} &\leftarrow \left(\mathbf{w}_{2};\;\gamma\mathbf{b}\right)
    - \left(\mathbf{w}_2\cdot\widehat{\mathbf{n}};\;p_{02}\right)_{\Gamma^{\mathrm{p}}_{2}}\\
    \boldsymbol{f}^2_{p} &\leftarrow \boldsymbol{0}
\end{align}
\end{subequations}
and
\begin{subequations}
  \label{Eqn:S4_cg_rhs}
  \begin{align}
    \boldsymbol{f}^1_{u} &\leftarrow \frac{1}{2}\left(\mathbf{w}_{1};\;\gamma\mathbf{b}\right)
    - \left(\mathbf{w}_1\cdot\widehat{\mathbf{n}};\;p_{01}\right)_{\Gamma^{\mathrm{p}}_{1}}\\
    \boldsymbol{f}^1_{p} &\leftarrow \frac{1}{2}\left(\mathrm{grad}[q_1];\;\frac{1}{\mu}{k}_1\gamma\mathbf{b}\right) \\ 
    \boldsymbol{f}^2_{u} &\leftarrow \frac{1}{2}\left(\mathbf{w}_{2};\;\gamma\mathbf{b}\right)
    - \left(\mathbf{w}_2\cdot\widehat{\mathbf{n}};\;p_{02}\right)_{\Gamma^{\mathrm{p}}_{2}}\\
    \boldsymbol{f}^2_{p} &\leftarrow \frac{1}{2}\left(\mathrm{grad}[q_2];\;\frac{1}{\mu}{k}_2\gamma\mathbf{b}\right) 
\end{align}
\end{subequations}
and
\begin{subequations}
  \label{Eqn:S4_dg_rhs}
  \begin{align}
    \boldsymbol{f}^1_{u} &\leftarrow \frac{1}{2}\left(\mathbf{w}_{1};\;\gamma\mathbf{b}\right)
    - \left(\mathbf{w}_1\cdot\widehat{\mathbf{n}};\;p_{01}\right)_{\Gamma^{\mathrm{p}}_{1}}\\
    \boldsymbol{f}^1_{p} &\leftarrow \frac{1}{2}\left(\mathrm{grad}[q_1];\;\frac{1}{\mu}{k}_1\gamma\mathbf{b}\right) - \left(q_1;\;u_{n1}\right)_{\Gamma^{u}_{1}}\\ 
    \boldsymbol{f}^2_{u} &\leftarrow \frac{1}{2}\left(\mathbf{w}_{2};\;\gamma\mathbf{b}\right)
    - \left(\mathbf{w}_2\cdot\widehat{\mathbf{n}};\;p_{02}\right)_{\Gamma^{\mathrm{p}}_{2}}\\
    \boldsymbol{f}^2_{p} &\leftarrow \frac{1}{2}\left(\mathrm{grad}[q_2];\;\frac{1}{\mu}{k}_2\gamma\mathbf{b}\right) - \left(q_2;\;u_{n2}\right)_{\Gamma^{u}_{2}}
\end{align}
\end{subequations}
respectively. Specifically, we employ PETSc's block solver 
capabilities, in the PCFIEDLSPLIT class, taking two fields at a time. 
However, the global DPP model is a four field problem so we subdivide 
our problem recursively such that we end up with 2$\times$2 blocks. Conceptually, 
PETSc can employ a wide variety of block solver methodologies on a 2$\times$2 
matrix:
\begin{align}
\label{Eqn:S4_2by2_block_matrix}
\boldsymbol{K} = \begin{bmatrix}
\boldsymbol{A} & \boldsymbol{B}\\
\boldsymbol{C} & \boldsymbol{D}
\end{bmatrix},
\end{align}
where $\boldsymbol{A}$, $\boldsymbol{B}$, $\boldsymbol{C}$, and $\boldsymbol{D}$
are individual block matrices which also consist of 2$\times$2 blocks. Although
equation \eqref{Eqn:S4_2by2_block_matrix} is conceptually a 4$\times$4 block 
matrix, PETSc's field-splitting capabilities enables us to break the
system down dynamically at runtime into two levels of 2$\times$2 blocks.

We now propose two different
ways one can compose scalable and efficient solvers and preconditioners for blocks 
$\boldsymbol{A}$, $\boldsymbol{B}$, $\boldsymbol{C}$, and $\boldsymbol{D}$ with
the individual components shown in equations \eqref{Eqn:S4_rt0_stiffness_matrix}, 
\eqref{Eqn:S4_cg_stiffness_matrix}, and \eqref{Eqn:S4_dg_stiffness_matrix}.
%
\subsection{Method 1: splitting by scales}
One option is to split the global problem by scales. That is, each 
macro- or micro- scale 2$\times$2 block will contain its corresponding 
velocity and pressure fields. Under this solver strategy, equation 
\eqref{Eqn:S4_linear_problem} is then rewritten as:
\begin{align}
  \begin{bmatrix}
  \boldsymbol{K}^{1}_{uu} & \boldsymbol{K}^{1}_{up} & \boldsymbol{0} & \boldsymbol{0}\\ 
  \boldsymbol{K}^{1}_{pu} & \boldsymbol{K}^{1}_{pp} & \boldsymbol{0} & \boldsymbol{K}^{12}_{pp} \\
  \boldsymbol{0} & \boldsymbol{0} & \boldsymbol{K}^2_{uu} & \boldsymbol{K}^{2}_{up}\\ 
  \boldsymbol{0} & \boldsymbol{K}^{21}_{pp} & \boldsymbol{K}^{2}_{pu} & \boldsymbol{K}^2_{pp} 
  \end{bmatrix}
  \begin{pmatrix}
  \boldsymbol{u}_1 \\
  \boldsymbol{p}_1 \\
  \boldsymbol{u}_2 \\
  \boldsymbol{p}_2 
  \end{pmatrix}
  =
  \begin{pmatrix}
  \boldsymbol{f}_u^1 \\
  \boldsymbol{f}_p^1 \\
  \boldsymbol{f}_u^2 \\
  \boldsymbol{f}_p^2 
  \end{pmatrix}
\end{align}
where $\boldsymbol{0}$ is a zero matrix, $\boldsymbol{u}_1$ and $\boldsymbol{p}_1$
are the respective macro-scale velocity and pressure vectors, $\boldsymbol{v}_2$ 
and $\boldsymbol{p}_2$ are the respective micro-scale velocity and pressure 
vectors. The individual 2$\times$2 blocks from equation 
\eqref{Eqn:S4_2by2_block_matrix} would be
\begin{align}
&\boldsymbol{A} := 
  \begin{bmatrix}
    \boldsymbol{K}^1_{uu} & \boldsymbol{K}^1_{up} \\
    \boldsymbol{K}^1_{pu} & \boldsymbol{K}^1_{pp}
  \end{bmatrix},\qquad
\boldsymbol{B} := 
  \begin{bmatrix}
    \boldsymbol{0} & \boldsymbol{0} \\
    \boldsymbol{0} & \boldsymbol{K}^{12}_{pp}
  \end{bmatrix},\nonumber\\
&\boldsymbol{C} := 
  \begin{bmatrix}
    \boldsymbol{0} & \boldsymbol{0} \\
    \boldsymbol{0} & \boldsymbol{K}^{21}_{pp}
  \end{bmatrix},\qquad
\boldsymbol{D} := 
  \begin{bmatrix}
    \boldsymbol{K}^2_{uu} & \boldsymbol{K}^2_{up} \\
    \boldsymbol{K}^2_{pu} & \boldsymbol{K}^2_{pp}
  \end{bmatrix}
\end{align}
Although the off diagonal blocks $\boldsymbol{B}$ and $\boldsymbol{C}$ 
contain the inter-scale pressure coupling terms, they are very sparse 
so we will ignore these blocks for now. The composition of the 
$\boldsymbol{A}$ and $\boldsymbol{D}$ blocks
are similar to the classical mixed Poisson problem so the Schur complement
approach outlined in \citep{Chang_CMAME_2017,Mapakshi_JCP_2018} and the 
references within can be applied.

The task is to individually precondition the decoupled 
$\boldsymbol{A}$ and $\boldsymbol{D}$ blocks. We note that they admit factorizations of
\begin{align}
\boldsymbol{A} = \begin{bmatrix}
\boldsymbol{I} & \boldsymbol{0}\\
\boldsymbol{K}^1_{pu}\left(\boldsymbol{K}_{uu}^1\right)^{-1} & \boldsymbol{I}
\end{bmatrix}
\begin{bmatrix}
\boldsymbol{K}_{uu}^1 & \boldsymbol{0}\\
\boldsymbol{0} & \boldsymbol{S}^{1}
\end{bmatrix}
\begin{bmatrix}
\boldsymbol{I} & \left(\boldsymbol{K}^1_{uu}\right)^{-1}\boldsymbol{K}^1_{up}\\
\boldsymbol{0} & \boldsymbol{I}
\end{bmatrix}\\
\boldsymbol{D} = \begin{bmatrix}
\boldsymbol{I} & \boldsymbol{0}\\
\boldsymbol{K}^2_{pu}\left(\boldsymbol{K}_{uu}^2\right)^{-1} & \boldsymbol{I}
\end{bmatrix}
\begin{bmatrix}
\boldsymbol{K}_{uu}^2 & \boldsymbol{0}\\
\boldsymbol{0} & \boldsymbol{S}^{2}
\end{bmatrix}
\begin{bmatrix}
\boldsymbol{I} & \left(\boldsymbol{K}^2_{uu}\right)^{-1}\boldsymbol{K}^2_{up}\\
\boldsymbol{0} & \boldsymbol{I}
\end{bmatrix}
\end{align}
where $\boldsymbol{I}$ is the identity matrix and
\begin{align}
\boldsymbol{S}^1=\boldsymbol{K}^1_{pp}-\boldsymbol{K}^1_{pu}\left(\boldsymbol{K}^1_{uu}\right)^{-1}\boldsymbol{K}^1_{up}\\
\boldsymbol{S}^2=\boldsymbol{K}^2_{pp}-\boldsymbol{K}^2_{pu}\left(\boldsymbol{K}^2_{uu}\right)^{-1}\boldsymbol{K}^2_{up}
\end{align}
are the Schur complements for the $\boldsymbol{A}$ and $\boldsymbol{D}$ blocks, 
respectively. The inverses can therefore be written as
\begin{align}
\boldsymbol{A}^{-1} = \begin{bmatrix}
\boldsymbol{I} & -\left(\boldsymbol{K}^1_{uu}\right)^{-1}\boldsymbol{K}^1_{up}\\
\boldsymbol{0} & \boldsymbol{I}
\end{bmatrix}
\begin{bmatrix}
\left(\boldsymbol{K}^1_{uu}\right)^{-1} & \boldsymbol{0}\\
\boldsymbol{0} & \left(\boldsymbol{S}^1\right)^{-1}
\end{bmatrix}
\begin{bmatrix}
\boldsymbol{I} & \boldsymbol{0}\\
-\boldsymbol{K}_{pu}^1\left(\boldsymbol{K}^1_{uu}\right)^{-1} & \boldsymbol{I}
\end{bmatrix} \\
\boldsymbol{D}^{-1} = \begin{bmatrix}
\boldsymbol{I} & -\left(\boldsymbol{K}^2_{uu}\right)^{-1}\boldsymbol{K}^2_{up}\\
\boldsymbol{0} & \boldsymbol{I}
\end{bmatrix}
\begin{bmatrix}
\left(\boldsymbol{K}^2_{uu}\right)^{-1} & \boldsymbol{0}\\
\boldsymbol{0} & \left(\boldsymbol{S}^2\right)^{-1}
\end{bmatrix}
\begin{bmatrix}
\boldsymbol{I} & \boldsymbol{0}\\
-\boldsymbol{K}_{pu}^2\left(\boldsymbol{K}^2_{uu}\right)^{-1} & \boldsymbol{I}
\end{bmatrix} 
\end{align}
The task at hand is to approximate the inverses of the $\boldsymbol{K}^{1}_{vv}$,
$\boldsymbol{K}^2_{uu}$, $\boldsymbol{S}^1$, and $\boldsymbol{S}^2$ blocks. 
The first two blocks are simply mass matrices so we can invert them using the
ILU$(0)$ (incomplete lower upper) solver. For the Schur complement blocks, 
we employ a diagonal mass-lumping of $\boldsymbol{K}^1_{uu}$ and $\boldsymbol{K}^2_{uu}$ to 
estimate $\left(\boldsymbol{K}^1_{uu}\right)^{-1}$ and 
$\left(\boldsymbol{K}^2_{uu}\right)^{-1}$ because they are spectrally equivalent
to the identity. That is,
\begin{align}
\label{Eqn:S4_sp1}
\boldsymbol{S}^1_p = \boldsymbol{K}^1_{pp}-\boldsymbol{K}^1_{pu}\mathrm{diag}\left(
\boldsymbol{K}^1_{uu}\right)^{-1}\boldsymbol{K}^{1}_{up}\\
\label{Eqn:S4_sp2}
\boldsymbol{S}^2_p = \boldsymbol{K}^2_{pp}-\boldsymbol{K}^2_{pu}\mathrm{diag}\left(
\boldsymbol{K}^2_{uu}\right)^{-1}\boldsymbol{K}^{2}_{up}
\end{align}
to precondition the inner solvers responsible for inverting $\boldsymbol{S}^1$ 
and $\boldsymbol{S}^2$. For these blocks we employ the multigrid V-cycle 
on $\boldsymbol{S}^1_p$ and $\boldsymbol{S}^2_p$ from the HYPRE BoomerAMG 
package \citep{falgout2002hypre}. We expect these to work because the 
$\boldsymbol{S}$ blocks are spectrally 
equivalent to the Laplacian, modulo the penalty terms. In 
\citep{Mapakshi_JCP_2018} it turns out the presence 
of the VMS stabilization terms in the $\boldsymbol{K}^1_{pp}$ and 
$\boldsymbol{K}^2_{pp}$ blocks do not drastically affect the 
performance or scalability of this solver strategy. 

Instead of completely solving for the $\boldsymbol{K}_{uu}^{-1}$ and $\boldsymbol{S}_p$ 
of both scales, we apply only a single sweep of ILU($0$)/block Jacobi and V-cycle, 
respectively, and rely on GMRES \citep{saad1986gmres} to solve the entire 
4$\times$4 block system. Thus this outer GMRES is able to pick up the inter-scale 
pressure coupling blocks $\boldsymbol{B}$ and $\boldsymbol{C}$. The PETSc command-line
options for this solver methodology is given in listing \ref{Code:ex_fields}.
\begin{lstlisting}[language=Python,caption=PETSc command-line options for splitting by fields , label=Code:ex_fields,frame=single]
-ksp_type gmres
-pc_type fieldsplit
-pc_fieldsplit_0_fields 0,1
-pc_fieldsplit_1_fields 2,3
-pc_fieldsplit_type additive
-fieldsplit_0_ksp_type preonly
-fieldsplit_0_pc_type fieldsplit
-fieldsplit_0_pc_fieldsplit_type schur
-fieldsplit_0_pc_fieldsplit_schur_fact_type full
-fieldsplit_0_pc_fieldsplit_schur_precondition selfp
-fieldsplit_0_fieldsplit_0_ksp_type preonly
-fieldsplit_0_fieldsplit_0_pc_type bjacobi
-fieldsplit_0_fieldsplit_1_ksp_type preonly
-fieldsplit_0_fieldsplit_1_pc_type hypre
-fieldsplit_1_ksp_type preonly
-fieldsplit_1_pc_type fieldsplit
-fieldsplit_1_pc_fieldsplit_type schur
-fieldsplit_1_pc_fieldsplit_schur_fact_type full
-fieldsplit_1_pc_fieldsplit_schur_precondition selfp
-fieldsplit_1_fieldsplit_0_ksp_type preonly
-fieldsplit_1_fieldsplit_0_pc_type bjacobi
-fieldsplit_1_fieldsplit_1_ksp_type preonly
-fieldsplit_1_fieldsplit_1_pc_type hypre
\end{lstlisting}
where we assume that the global ordering of the mixed function space 
is macro-scale velocity (0), macro-scale pressure (1), micro-scale velocity
(2), and micro-scale pressure (3).
\subsection{Method 2: splitting by fields}
Another option is to group the velocities and pressures of both scales into
two different blocks. If this approach is taken, equation \eqref{Eqn:S4_linear_problem} 
is then rewritten as:
\begin{align}
  \begin{bmatrix}
  \boldsymbol{K}^{1}_{uu} & \boldsymbol{0} & \boldsymbol{K}^1_{up} & \boldsymbol{0}\\ 
  \boldsymbol{0} & \boldsymbol{K}^{2}_{uu} & \boldsymbol{0} & \boldsymbol{K}^2_{up}\\
  \boldsymbol{K}^{1}_{pu} & \boldsymbol{0} & \boldsymbol{K}^1_{pp} & \boldsymbol{K}^{12}_{pp}\\ 
  \boldsymbol{0} & \boldsymbol{K}^{2}_{pu} & \boldsymbol{K}^{21}_{pp} & \boldsymbol{K}^2_{pp} 
  \end{bmatrix}
  \begin{pmatrix}
  \boldsymbol{u}_1 \\
  \boldsymbol{u}_2 \\
  \boldsymbol{p}_1 \\
  \boldsymbol{p}_2 
  \end{pmatrix}
  =
  \begin{pmatrix}
  \boldsymbol{f}_u^1 \\
  \boldsymbol{f}_u^2 \\
  \boldsymbol{f}_p^1 \\
  \boldsymbol{f}_p^2 
  \end{pmatrix}
\end{align}
and the individual blocks in equation \eqref{Eqn:S4_2by2_block_matrix}
would now look like
\begin{align}
&\boldsymbol{A} := 
  \begin{bmatrix}
    \boldsymbol{K}^1_{uu} & \boldsymbol{0} \\
    \boldsymbol{0} & \boldsymbol{K}^2_{uu}
  \end{bmatrix}, \qquad
\boldsymbol{B} := 
  \begin{bmatrix}
    \boldsymbol{K}^1_{up} & \boldsymbol{0} \\
    \boldsymbol{0} & \boldsymbol{K}^2_{up}
  \end{bmatrix},\nonumber\\
&\boldsymbol{C} := 
  \begin{bmatrix}
    \boldsymbol{K}^1_{pu} & \boldsymbol{0} \\
    \boldsymbol{0} & \boldsymbol{K}^2_{pu}
  \end{bmatrix}, \qquad
\boldsymbol{D} := 
  \begin{bmatrix}
    \boldsymbol{K}^1_{pp} & \boldsymbol{K}^{12}_{pp} \\
    \boldsymbol{K}^{21}_{pp} & \boldsymbol{K}^2_{pp}
  \end{bmatrix}
\end{align}
Unlike the previous methodology, we can work directly with the above stiffness
matrix, which admits a factorization of
\begin{align}
\boldsymbol{K} &= \begin{bmatrix}
\boldsymbol{I} & \boldsymbol{0}\\
\boldsymbol{C}\boldsymbol{A}^{-1} & \boldsymbol{I}
\end{bmatrix}
\begin{bmatrix}
\boldsymbol{A} & \boldsymbol{0}\\
\boldsymbol{0} & \boldsymbol{S}
\end{bmatrix}
\begin{bmatrix}
\boldsymbol{I} & \boldsymbol{A}^{-1}\boldsymbol{B}\\
\boldsymbol{0} & \boldsymbol{I}
\end{bmatrix},
\end{align}
where the Schur complement $\boldsymbol{S}$ is
\begin{align}
\label{Eqn:S4_schurfields}
\boldsymbol{S}=\boldsymbol{D}-\boldsymbol{C}\boldsymbol{A}^{-1}\boldsymbol{B}.
\end{align}
The inverse can therefore be written as
\begin{align}
\boldsymbol{K}^{-1} = \begin{bmatrix}
\boldsymbol{I} & -\boldsymbol{A}^{-1}\boldsymbol{B}\\
\boldsymbol{0} & \boldsymbol{I}
\end{bmatrix}
\begin{bmatrix}
\boldsymbol{A}^{-1} & \boldsymbol{0}\\
\boldsymbol{0} & \boldsymbol{S}^{-1}
\end{bmatrix}
\begin{bmatrix}
\boldsymbol{I} & \boldsymbol{0}\\
-\boldsymbol{C}\boldsymbol{A}^{-1} & \boldsymbol{I}
\end{bmatrix}.
\end{align}
Although $\boldsymbol{A}$ is a 2$\times$2 block containing velocities spanning across two
different scales, we can still approximate $\boldsymbol{A}^{-1}$ by inverting the entire 
$\boldsymbol{A}$ block using ILU$(0)$ because the off-diagonal blocks are zero and the 
diagonal blocks consist of only mass matrices. Approximating 
$\boldsymbol{S}^{-1}$ is a little trickier because equation \eqref{Eqn:S4_schurfields} 
is a dense 2$\times$2 block with off-diagonal terms. However,
we can still employ a diagonal mass-lumping of $\boldsymbol{A}$ to 
estimate $\boldsymbol{A}^{-1}$ because it is again spectrally equivalent
to the identity. The preconditioner needed for $\boldsymbol{S}^{-1}$ is:
\begin{align}
\boldsymbol{S}_p &= \boldsymbol{D}-\boldsymbol{C}\mathrm{diag}\left(
\boldsymbol{A}\right)^{-1}\boldsymbol{B}\nonumber\\
  &= \begin{bmatrix}
    \boldsymbol{K}^1_{pp} & \boldsymbol{K}^{12}_{pp} \\
    \boldsymbol{K}^{21}_{pp} & \boldsymbol{K}^2_{pp}
  \end{bmatrix} - 
  \begin{bmatrix}
    \boldsymbol{K}^1_{pu} & \boldsymbol{0} \\
    \boldsymbol{0} & \boldsymbol{K}^2_{pu}
  \end{bmatrix}
  \mathrm{diag}\left(
  \begin{bmatrix}
    \boldsymbol{K}^1_{uu} & \boldsymbol{0} \\
    \boldsymbol{0} & \boldsymbol{K}^2_{uu}
  \end{bmatrix}
  \right)^{-1}
  \begin{bmatrix}
    \boldsymbol{K}^1_{up} & \boldsymbol{0} \\
    \boldsymbol{0} & \boldsymbol{K}^2_{up}
  \end{bmatrix}\nonumber\\
  &= \begin{bmatrix}
    \boldsymbol{K}^1_{pp} - \boldsymbol{K}^1_{pu}\mathrm{diag}\left(\boldsymbol{K}^1_{uu}\right)\boldsymbol{K}^1_{up} & \boldsymbol{K}^{12}_{pp} \\
    \boldsymbol{K}^{21}_{pp} & \boldsymbol{K}^2_{pp} - \boldsymbol{K}^2_{pu}\mathrm{diag}\left(\boldsymbol{K}^2_{uu}\right)\boldsymbol{K}^2_{up}
  \end{bmatrix}
\end{align}
The off-diagonal blocks only consist of mass-matrix terms but the decoupled 
diagonal blocks are identical to equations \eqref{Eqn:S4_sp1} and \eqref{Eqn:S4_sp2}.
Thus, we individually employ multigrid V-cycle on each of the diagonal blocks.
As in the previous solver methodology, only a single sweep of ILU($0$) and the two
multigrid V-cycles are needed for the $\boldsymbol{A}^{-1}$ matrix and 
the two diagonal terms within the $\boldsymbol{S}_p$ matrix, 
respectively, and the GMRES method is employed to solve the entire block system.
The PETSc implementation is shown in listing \ref{Code:ex_scales}.
\begin{lstlisting}[language=Python,caption=PETSc command-line options for splitting by scale , label=Code:ex_scales,frame=single]
-ksp_type gmres
-pc_type fieldsplit
-pc_fieldsplit_0_fields 0,2
-pc_fieldsplit_1_fields 1,3
-pc_fieldsplit_type schur
-pc_fieldsplit_schur_fact_type full
-pc_fieldsplit_schur_precondition selfp
-fieldsplit_0_ksp_type preonly
-fieldsplit_0_pc_type bjacobi
-fieldsplit_1_ksp_type preonly
-fieldsplit_1_pc_type fieldsplit
-fieldsplit_1_pc_fieldsplit_type additive
-fieldsplit_1_fieldsplit_0_ksp_type preonly
-fieldsplit_1_fieldsplit_0_pc_type hypre
-fieldsplit_1_fieldsplit_1_ksp_type preonly
-fieldsplit_1_fieldsplit_1_pc_type hypre
\end{lstlisting}
where we again assume that the global ordering of the mixed function space 
is macro-scale velocity (0), macro-scale pressure (1), micro-scale velocity
(2), and micro-scale pressure (3).
\subsection{Computer implementation}
The finite element capabilities are provided by the Firedrake Project package
\citep{rathgeber2016firedrake,Luporini2015,Homolya2016,McRae2016,
Bercea2016,Luporini2016,Homolya2017,Homolya2017a} with GNU compilers. This
sophisticate finite element simulation package and its software dependencies 
can be found at \citep{zenodo/firedrake:2018,zenodo/PyOP2:2018,
zenodo/tsfc:2018,zenodo/COFFEE:2017,zenodo/ufl:2018,zenodo/FInAT:2018,
zenodo/fiat:2018,zenodo/petsc:2018,zenodo/petsc4py:2018}. The computational meshes 
are built on top of the DMPlex unstructured grid format \citep{KnepleyKarpeev09,
LangeKnepleyGorman2015,LangeMitchellKnepleyGorman2015} and partitioned through 
the Chaco package \citep{Chaco95}. Pictorial descriptions
of the specific elements represented by this mesh format and utilized in this paper 
are illustrated in Figure \ref{Fig:elements}. The DMPlex data structure interfaces very 
nicely with PETSc's suite of parallel solvers and provide excellent scalability
across thousands of MPI processes \citep{Chang_JOMP_2017,chang2018performance}. 
Sample Firedrake codes for some of these benchmark problems 
can be found in Appendix \ref{App:code}. 

\textit{In our PETSc implementation, the same global 
	matrix will be assembled for both solvers. The preconditioners differ by the subblocks which are extracted. 
	The different sparsity pattern of the subblocks contributes to the performance differences seen in the solvers, but the overall
	assembly time remains unchanged for either solvers.}

\section{PERFORMANCE SPECTRUM MODELING}
\label{Sec:S5_Comparison_Performance}
To understand the parallel performance and algorithmic scalability of the proposed 
DPP composable block solver methodologies for the three finite element formulations, 
a performance model is needed. The performance model based on the 
Time-Accuracy-Size (TAS) spectrum analysis outlined in \citep{chang2018comparative} 
shall be used as the basis for understanding the quality of these finite element 
formulations with the proposed block solvers.
We now briefly highlight the performance metrics used in this section and why they are
each important in each of their own ways.
\subsection{Mesh convergence}
This criterion uses the convergence notion to account for numerical accuracy of a 
solution in the performance spectrum. In this paper, we are adopting $L_2$ norm 
of the error defined as:
\begin{align}
  L_2^\mathrm{norm}= \Vert u_h-u\Vert_{L_2}
\end{align}
where $u$ is the exact solution, $u_h$ is the the finite element solution, 
and $h$ is measure of element size. Based on theory, most finite element 
discretizations will have an upper-bound for $L_2$ error norm as follows:
\begin{align}
  L_2^\mathrm{norm}\leq Ch^{\alpha}
\end{align} 
where $\alpha$ is known as \textit{convergence rate} and $C$ is some constant. 
When reporting and comparing how much accuracy is attained for each discretization, 
we use the notation of \textit{Digits of Accuracy} (DoA) defined as:
\begin{align}
    \mathrm{DoA} := -\mathrm{log}_{10} ( L_2^{\mathrm{norm}})
\end{align}
and plot DoA against \textit{Digits of Size} (DoS), which is defined as:
\begin{align}
    \mathrm{DoS} := -\mathrm{log}_{10} (\mathrm{DoF})
\end{align}
Noting that for most formulations $DoF=Dh^{-d}$, where $d$ is the spatial dimension 
and $D$ is some constant, the slope of DoA vs DoS plot is in the order of 
$\frac{\alpha}{d}$. Any tailing off from the line plot is an indicator 
of incorrect implementation or solver convergence tolerances being too relaxed.
Furthermore, the ratio DoA/DoS can be a good indicator of how much accuracy is achieved
per DoF.
\subsection{Strong-scaling}
In this basic parallel scaling while the size of problem remains unchanged, the 
number of processes increases. In general, this metric comments on the marginal 
efficiency of each additional processes assigned to a problem. It is conventional 
to plot number of processes against the parallel efficiency defined as:
\begin{align}
\mbox{Parallel eff. $(\%)$} = \frac{T_1}{T_p\times proc}\times100\%
\end{align}
where $proc$ is the number of MPI processes, $T_1$ is the total wall-clock time needed
on a single MPI processes, and $T_p$ is the total wall-clock time needed with
$proc$ MPI processes. However, this metric must be interpreted carefully for the following
reasons:
\begin{enumerate}
\item \textsf{Solver iteration counts:} The number of solver iterations may 
fluctuate as the number of MPI processes changes. This can happen for a number 
of reasons, whether it is algorithmic implementation or relaxed convergence 
criterion. It is necessary to also report the number of KSP iterations required 
as the number of processes changes.
\item \textsf{Problems too small:} If the DoF count is too small for a particular MPI
concurrency, communication time will swamp the computation time, thus reducing the
parallel efficiency. This issue may arise when making comparative studies between different
finite element discretizations, as different formulations have different DoF counts for
a given $h$-size. Furthermore, for Python-based simulation packages like Firedrake, 
overheads from just-in-time compilation and instantiation of objects can also affect the
strong-scaling .
\item \textsf{Problems too large:} If the DoF count is too large for a particular MPI 
concurrency, the problems not only drop out of the various levels of cache in the 
memory hierarchy but also invoke several expensive cache misses which can slow down 
the overall performance. This may result in superlinear speedups, like the BLMVM 
bound-constrained optimization solver in \citep{Chang_JOMP_2017}.
\end{enumerate}
Lastly, the global problem size for this scaling analysis is fixed, so we need an 
additional parallel scaling metric which explains whether the performance 
of our solvers might degrade due to increased KSP iteration counts or memory contention as 
the problem size increases.
\subsection{Static-scaling}
As described in \citep{chang2018performance},
static-scaling is a scaling analysis where
the MPI concurrency is fixed but the problem
size is increased. The essential metric
for this analysis is the computation rate (DoF over Time). In this paper, we run a series of 
problem sizes at a fixed parallelism and plot the computation rate against the  
wall-clock time. Note that the time need not be the total time to solution, instead one 
could look at various phases like the finite element assembly or solver computation rates.

Static-scaling returns information on performance and scalability of software and 
solvers across different hardware architectures. 
This scaling analysis also captures both strong-scaling and weak-scaling effects.  
Assuming that the block solvers are of $\mathcal{O}(N)$ scalability, where $N = DoF$, 
optimal scaling is indicated by a horizontal curve. Any tail offs at small problem 
sizes suggests strong-scaling effects whereas tail offs at large problem sizes 
indicate suboptimal algorithmic or memory effects. The exact reasoning for the tail offs
towards the right can be verified through arithmetic intensity, which is the measure of
the total work over the total bytes transferred (see \citep{chang2018performance} and
the references within).
\subsection{Digits-of-Efficacy ($\mathrm{DoE}$)}
The final metric needed for our performance spectrum study is the 
\textit{Digits of Efficacy} (DoE). This metric measures the accuracy production
by a particular scheme in a given amount of time. The DoE could be defined as:
\begin{align}
\mathrm{DoE}:=-\mathrm{log}_{10}(L_2^{\mathrm{norm}}\times \mathrm{Time})
\end{align}
Assuming that straight lines are captured in both the mesh convergence and static-scaling
diagrams, the DoE has a linear dependence on problem size and returns a slope of 
$d-\alpha$ (see \citep{chang2018comparative} for details on the exact derivation). 
This efficacy measure is analogous to the \emph{action} of a mechanical system, 
that is the product of energy and time. In the TAS spectrum analysis, the DoE 
represents an analogous action for computation, and we speculate that an
optimal algorithm minimizes this product over its runtime. Since the DoE takes
the negative logarithm of \emph{action}, a higher DoE is desirable.

\section{REPRESENTATIVE NUMERICAL RESULTS }
\label{Sec:S6_Comparison_NR}
In this section, after clarifying the terminology and framework adopted 
for the performance spectrum model, we solve the four-field DPP model in two- and three-dimensional 
settings in order to demonstrate the implementation of the proposed composable block solvers and gauge their performances.
The two-dimensional problem will be conducted in serial (one MPI process) on a dual
socket Intel Xeon E5-2609v3 server node. The three-dimensional problems will be conducted
on a dual socket Intel Xeon E5-2698v3 server node and will utilize up to
16 MPI processes (8 MPI processes per socket).
On different performance metrics, H(div), CG-VMS, and DG VMS 
formulations are compared for both simplicial (TRI, TET) and non-simplicial 
(QUAD, HEX) meshes.
 Both two-dimensional and three-dimensional problems were adopted by
\citep{joodat2018modeling} for the convergence analysis of continuous
stabilized mixed formulation (CG-VMS) and by \citep{joshaghani2018stabilized}
for the convergence analysis of discontinuous stabilized mixed
formulation (DG-VMS) for the DPP model. We are generating three
series of outputs for first-order CG-VMS, DG-VMS with
$\eta_p=\eta_u=10$, and H(div) formulations.
\subsection{Two-dimensional study}
For this first problem, let us consider a two-dimensional DPP boundary value problem 
with governing equations stated in equations \eqref{Eqn:DG_GE_Darcy_BLM}--\eqref{Eqn:Composable_mass_transfer} and \eqref{Eqn:DG_GE_vBC}--\eqref{Eqn:DG_GE_Darcy_pBC}. The homogeneous (i.e., constant macro and micro-permeabilities) 
bi-unit square computational domain and boundary conditions for this study 
are shown in Figure \ref{Fig1:2D_BVP}, and 
the corresponding parameters are described in Table \ref{tab:2_D_dataset}. 
%
\begin{table}[]
\centering
\caption{ Parameters for two-dimensional problem.}
\label{tab:2_D_dataset}
\begin{tabular}{ll}
\Xhline{2\arrayrulewidth}
 Parameter & Value \\ \hline
 $L$ & $1.0$ \\
 $\gamma \mathbf{b}$ & $\{0.0,0.0\}$  \\
 $\mu$ & $1.0$  \\
 $\beta$ & $1.0$  \\
 $k_1$ & $1.0$  \\
 $k_2$ & $0.1$  \\
 $\eta$ & $\sqrt{11}$ \\
 $\eta_p $ & 10 \\
 $\eta_u$ & 10\\
 \Xhline{2\arrayrulewidth}
\end{tabular}
\end{table}

The analytical solution for the pressure and velocity fields takes the following form:
\begin{subequations}
  \begin{align}
    &\mathbf{u}_{1}(x,y) = -k_1
    \begin{pmatrix}
      e^{\pi x}\sin(\pi y)       \\
      e^{\pi x}\cos(\pi y)-\frac{\eta}{\beta k_1} e^{\eta y}       
    \end{pmatrix} \\
    &p_1(x,y)=\frac{\mu}{\pi} e^{\pi x}\sin(\pi y)
    -\frac{\mu}{\beta k_1}e^{\eta y} \\
    &\mathbf{u}_{2}(x,y) = -k_2
    \begin{pmatrix}
      e^{\pi x}\sin(\pi y)       \\
      e^{\pi x}\cos(\pi y)+\frac{\eta}{\beta k_2} e^{\eta y}       
    \end{pmatrix} \\
    &p_2(x,y)=\frac{\mu}{\pi} e^{\pi x}\sin(\pi y)
    +\frac{\mu}{\beta k_2}e^{\eta y}
  \end{align}
\end{subequations}
where $\eta$ is defined as:
\begin{align}
\eta:=\sqrt{\beta \frac{k_1+k_2}{k_1 k_2}}
\end{align}

For two-dimensional performance spectrum analysis, all three finite element formulations will start 
off with the same $h$-sizes and will be refined up to $6$ times. The initial TRI and QUAD 
coarse meshes are shown in Figures \ref{Fig1:Mesh_T3} and \ref{Fig1:Mesh_Q4} and the corresponding
DoF counts for each formulation is shown in Table \ref{tab:2D_h-size_vs_DoF}. 
%
{\scriptsize
\begin{table}[]
	\centering
	\caption{This table illustrates degrees-of-freedom for two-dimensional h-size refinement study.}
	\label{tab:2D_h-size_vs_DoF}
\begin{tabular}{c|c|c|c|c|c|c|c|cccc}
	\Xhline{2\arrayrulewidth}
	\multicolumn{4}{c|}{CG-VMS}                                                                                     & \multicolumn{4}{c|}{DG-VMS}                                                                                     & \multicolumn{4}{c}{H(div)}                                                                                           \\ \hline
	\multicolumn{2}{c|}{TRI}                               & \multicolumn{2}{c|}{QUAD}                               & \multicolumn{2}{c|}{TRI}                               & \multicolumn{2}{c|}{QUAD}                               & \multicolumn{2}{c|}{TRI}                                  & \multicolumn{2}{c}{QUAD}                               \\ \hline
	\multicolumn{1}{l|}{h-size} & \multicolumn{1}{l|}{DoF} & \multicolumn{1}{l|}{h-size} & \multicolumn{1}{l|}{DoF} & \multicolumn{1}{l|}{h-size} & \multicolumn{1}{l|}{DoF} & \multicolumn{1}{l|}{h-size} & \multicolumn{1}{l|}{DoF} & \multicolumn{1}{l|}{h-size} & \multicolumn{1}{l|}{DoF}    & \multicolumn{1}{l|}{h-size} & \multicolumn{1}{l}{DoF} \\ \hline
	5                           & 216                      & 5                           & 216                      & 5                           & 900                      & 5                           & 600                      & \multicolumn{1}{c|}{5}      & \multicolumn{1}{c|}{270}    & \multicolumn{1}{c|}{5}      & 170                     \\
	10                          & 726                      & 10                          & 726                      & 10                          & 3600                     & 10                          & 2400                     & \multicolumn{1}{c|}{10}     & \multicolumn{1}{c|}{1040}   & \multicolumn{1}{c|}{10}     & 640                     \\
	20                          & 2646                     & 20                          & 2646                     & 20                          & 14400                    & 20                          & 9600                     & \multicolumn{1}{c|}{20}     & \multicolumn{1}{c|}{4080}   & \multicolumn{1}{c|}{20}     & 2480                    \\
	40                          & 10086                    & 40                          & 10086                    & 40                          & 57600                    & 40                          & 38400                    & \multicolumn{1}{c|}{40}     & \multicolumn{1}{c|}{16160}  & \multicolumn{1}{c|}{40}     & 9760                    \\
	80                          & 39366                    & 80                          & 39366                    & 80                          & 230400                   & 80                          & 153600                   & \multicolumn{1}{c|}{80}     & \multicolumn{1}{c|}{64320}  & \multicolumn{1}{c|}{80}     & 38720                   \\
	160                         & 155526                   & 160                         & 155526                   & 160                         & 921600                   & 160                         & 614400                   & \multicolumn{1}{c|}{160}    & \multicolumn{1}{c|}{256640} & \multicolumn{1}{c|}{160}    & 154240     \\           
	\Xhline{2\arrayrulewidth}
\end{tabular}
\end{table}
}
The mesh convergence results with respect to DoA and DoS are performed under field-splitting
solver and are shown in Figures \ref{Fig:2D_convergence_T3}
and \ref{Fig:2D_convergence_Q4} respectively for TRI and QUAD meshes. 
It should be noted that by applying scale-splitting solver, very same results could be obtained and for brevity, we decided not to plot them in figures.
It can be seen in these
diagrams that the CG-VMS and DG-VMS lines exhibit a slope $\frac{\alpha}{d}=1$, which verifies
that our Firedrake implementation of these discretizations is correct. 
The H(div) lines 
exhibit a slope of 0.5 for TRI meshes but appear to have superlinear convergence for the
QUAD meshes, which has also been observed in other Firedrake endeavors \citep{gibson2018domain}. It can also be seen that if the solver tolerances are not strict enough, the
mesh convergence lines will tail off. Nonetheless, the CG-VMS and DG-VMS have the highest ratios 
of DoA over DoS in most of these diagrams which suggests that each DoF in VMS 
formulations has a greater level of contribution to the overall numerical 
accuracy than their H(div) counterparts.

Static-scaling results for both block solver strategies are shown in Figure 
\ref{Fig:2D_Static_scaling}, and we see that the total wall clock time is almost equally
distributed among the assemble and solve phases.
The field-splitting methodologies are slightly worse than their scale-splitting 
counterparts for the VMS formulations. However, the difference in performance is
almost negligible when we look at the total time. The DoF counts are too small as 
the line curves for both the assembly and solve phases flatten out when all three
formulations have roughly $10$K DoF or more. No matter which mesh is utilized, 
the H(div) formulation processes its DoF count faster than either VMS formulations.

Figures \ref{Fig:2D_DOE_T3} and \ref{Fig:2D_DOE_Q4} contain DoE diagrams for TRI and QUAD meshes, 
respectively.
Although H(div) appears to have the highest computation rates, it has a lower DoA than
its VMS counterparts which results in a much smaller DoE. The QUAD mesh on the other hand has a very 
high DoA and it beats out its VMS counterparts for all the fields. 
%
\subsection{Three-dimensional study}
In this section, we are solving a three-dimensional problem which is constructed by the Method of Manufactured Solutions (MMS) \citep{oberkampf2010verification}.
The homogeneous computational domain and boundary conditions for this problem are
illustrated in Figure \ref{Fig9:2D_schematic}, and related parameters are listed in Table \ref{tab:3_D_dataset}. Also, a representative TET and HEX coarse meshes are shown in Figures \ref{Fig1:Mesh_TET} and \ref{Fig1:Mesh_HEX}, respectively.
%
%
\begin{table}[]
	\centering
	\caption{ Parameters for three-dimensional problem.}
	\label{tab:3_D_dataset}
	\begin{tabular}{ll}
		\Xhline{2\arrayrulewidth}
		Parameter & Value \\ \hline
		$L$ & $1.0$ \\
		$\gamma \mathbf{b}$ & $\{0.0,0.0,0.0\}$  \\
		$\mu$ & $1.0$  \\
		$\beta$ & $1.0$  \\
		$k_1$ & $1.0$  \\
		$k_2$ & $0.1$  \\
		$\eta$ & $\sqrt{11}$ \\
		\Xhline{2\arrayrulewidth}
	\end{tabular}
\end{table}
The analytical solution for the pressure and velocity fields in the two pore-networks 
takes the following form:
\begin{subequations}
  \begin{align}
    &\mathbf{u}_{1}(x,y,z) = -k_1
    \begin{pmatrix}
      e^{\pi x}( \sin(\pi y)+\sin(\pi z)  )       \\
      e^{\pi x}\cos(\pi y)-\frac{\eta}{\beta k_1} e^{\eta y}\\      
      e^{\pi x}\cos(\pi z)-\frac{\eta}{\beta k_1} e^{\eta z} 
    \end{pmatrix} \\
    &p_1(x,y,z)=\frac{\mu}{\pi} e^{\pi x}(\sin(\pi y)+\sin(\pi z))
    -\frac{\mu}{\beta k_1}(e^{\eta y}+e^{\eta z}) \\
    &\mathbf{u}_{2}(x,y,z) = -k_2
    \begin{pmatrix}
      e^{\pi x}( \sin(\pi y)+\sin(\pi z)  )       \\
      e^{\pi x}\cos(\pi y)+\frac{\eta}{\beta k_2} e^{\eta y}\\      
      e^{\pi x}\cos(\pi z)+\frac{\eta}{\beta k_2} e^{\eta z} 
    \end{pmatrix} \\
    &p_2(x,y,z)=\frac{\mu}{\pi} e^{\pi x}(\sin(\pi y)+\sin(\pi z))
    +\frac{\mu}{\beta k_1}(e^{\eta y}+e^{\eta z})
  \end{align}
\end{subequations}

\subsubsection{Test 1: Strong-scaling results}
First we investigate the strong-scaling performance of the proposed block solvers 
when applied to different finite element formulations. Two case studies are shown:
first we fix the $h$-size for all finite element formulations, and second we
modify each formulation's $h$-size such that they all have roughly matching
DoF counts.  Table \ref{Tab:3D_strong_scaling} contains the corresponding $h$-sizes
and DoF counts needed for both case studies.

First, we consider when all discretizations have an $h$-size = 1/16. Strong-scaling 
results for both field-splitting and scale-splitting block 
solver methodologies for H(div), CG-VMS, and DG-VMS can be found in Tables
\ref{tab:Strong_RT$0$_hsize}, \ref{tab:Strong_CGVMS_hsize}, and \ref{tab:Strong_DGVMS_hsize}, respectively.
All three Tables indicate that the KSP iteration counts between 
field-splitting and scale-splitting are identical whereas the wall-clock time
for scale-splitting is slightly smaller. The KSP counts for H(div) and CG-VMS 
do not change much when the number of MPI processes increases, whereas 
DG-VMS's KSP counts increase drastically. This increase in KSP iteration counts
will affect the parallel efficiency so one has to be careful when interpreting
these results. Nonetheless, we see that the DG-VMS parallel efficiency is the highest,
even with its proliferated KSP counts. This is attributed to the fact that 
the DoF count for DG-VMS is larger than CG-VMS and H(div). All three tables indicate
that higher DoF counts bring in more efficiency in the parallel sense.

Second, we consider the case when all discretizations contain 
approximately 200K degrees-of-freedom. Strong-scaling results for 
both field-splitting and scale-splitting block solver methodologies 
for H(div), CG-VMS, and DG-VMS can be found in Tables
\ref{tab:Strong_RT$0$_DoF}, \ref{tab:Strong_CGVMS_DoF}, and \ref{tab:Strong_DGVMS_DoF}, respectively.
Like with the same $h$-size case, the scale-splitting method appears to be 
more efficient in terms of wall-clock time needed despite having the same 
KSP counts as the field-splitting method. It can also be seen that the
H(div) and CG-VMS KSP counts do not fluctuate much with MPI processes and that
DG-VMS KSP counts still increase dramatically. However, tuning the mesh sizes
such that all finite element discretizations have the same DoF count enables
us to have better understanding of the parallel performance, especially for
three-dimensional problems. 
It can be seen that H(div) requires the least amount of wall-clock 
time resulting in the lowest parallel efficiency, but that does not mean 
this is a bad formulation. In order to understand the quality of the H(div)
discretizations, we need to take into consideration the numerical accuracy
and perform a TAS spectrum analysis.

{\scriptsize
\begin{table}[]
	\centering
	\caption{This table shows h-size and corresponding degrees-of-freedom for three-dimensional strong-scaling studies.}
	\label{Tab:3D_strong_scaling}
	\begin{tabular}{ccccccccccccc}
		\Xhline{2\arrayrulewidth}
		\multicolumn{1}{c|}{\multirow{3}{*}{\begin{tabular}[c]{@{}c@{}}Strong\\ scaling\\ tests\end{tabular}}} & \multicolumn{4}{c|}{CG-VMS}                                                                                           & \multicolumn{4}{c|}{DG-VMS}                                                                                           & \multicolumn{4}{c}{H(div)}                                                                                              \\ \cline{2-13} 
		\multicolumn{1}{c|}{}                                                                                  & \multicolumn{2}{c|}{TET}                                  & \multicolumn{2}{c|}{HEX}                                  & \multicolumn{2}{c|}{TET}                                  & \multicolumn{2}{c|}{HEX}                                  & \multicolumn{2}{c|}{TET}                                  & \multicolumn{2}{c}{HEX}                                  \\ \cline{2-13} 
		\multicolumn{1}{c|}{}                                                                                  & \multicolumn{1}{c|}{h-size} & \multicolumn{1}{c|}{DoF}    & \multicolumn{1}{c|}{h-size} & \multicolumn{1}{c|}{DoF}    & \multicolumn{1}{c|}{h-size} & \multicolumn{1}{c|}{DoF}    & \multicolumn{1}{c|}{h-size} & \multicolumn{1}{c|}{DoF}    & \multicolumn{1}{c|}{h-size} & \multicolumn{1}{c|}{DoF}    & \multicolumn{1}{c|}{h-size} & \multicolumn{1}{c}{DoF}    \\ \hline
		\multicolumn{1}{c|}{Test 1}                                                                            & \multicolumn{1}{c|}{1/16}     & \multicolumn{1}{c|}{39305}  & \multicolumn{1}{c|}{1/16}     & \multicolumn{1}{c|}{39305}  & \multicolumn{1}{c|}{1/16}     & \multicolumn{1}{c|}{786432} & \multicolumn{1}{c|}{1/16}     & \multicolumn{1}{c|}{262144} & \multicolumn{1}{c|}{1/16}     & \multicolumn{1}{c|}{150528} & \multicolumn{1}{c|}{1/16}     & \multicolumn{1}{c}{34304}  \\
		\multicolumn{1}{c|}{Test 2}                                                                            & \multicolumn{1}{c|}{1/28}     & \multicolumn{1}{c|}{195112} & \multicolumn{1}{c|}{1/29}     & \multicolumn{1}{c|}{216000} & \multicolumn{1}{c|}{1/10}     & \multicolumn{1}{c|}{192000} & \multicolumn{1}{c|}{1/15}     & \multicolumn{1}{c|}{216000} & \multicolumn{1}{c|}{1/17}     & \multicolumn{1}{c|}{180336} & \multicolumn{1}{c|}{1/29}     & \multicolumn{1}{c}{200158} \\ 	\Xhline{2\arrayrulewidth}                                                                   
	\end{tabular}
\end{table}
}
\medskip
{\scriptsize
\begin{table}[]
	\caption{\textsf{3D problem:} Strong-scaling results for H(div) formulation with same h-size.}
	\label{tab:Strong_RT$0$_hsize}
\begin{tabular}{ccccccccccc}
\Xhline{2\arrayrulewidth}
\multicolumn{11}{l}{Field-splitting} \\ \hline
\multicolumn{1}{c|}{\multirow{3}{*}{\begin{tabular}[c]{@{}c@{}}No.\\ of MPI\\ proc.\end{tabular}}} & \multicolumn{5}{c|}{TET mesh} & \multicolumn{5}{c}{HEX mesh} \\ \cline{2-11} 
\multicolumn{1}{c|}{} & \multicolumn{3}{l}{Time} & \multicolumn{1}{l}{\multirow{2}{*}{KSP}} & \multicolumn{1}{l|}{\multirow{2}{*}{\begin{tabular}[c]{@{}l@{}}Parallel\\ eff. (\%)\end{tabular}}} & \multicolumn{3}{l}{Time} & \multicolumn{1}{l}{\multirow{2}{*}{KSP}} & \multicolumn{1}{l}{\multirow{2}{*}{\begin{tabular}[c]{@{}l@{}}Parallel\\ eff. (\%)\end{tabular}}} \\ \cline{2-4} \cline{7-9}
\multicolumn{1}{c|}{} & Assembly & Solver & Total & \multicolumn{1}{l}{} & \multicolumn{1}{l|}{} & Assembly & Solver & Total & \multicolumn{1}{l}{} & \multicolumn{1}{l}{} \\ \hline
\multicolumn{1}{c|}{1}  & 6.82E-01 & 5.80E-01 & 1.26E+00 & 15 & \multicolumn{1}{c|}{100}     & 5.45E-01 & 2.97E-01 & 8.42E-01 & 17 & 100     \\
\multicolumn{1}{c|}{2}  & 4.77E-01 & 4.16E-01 & 8.93E-01 & 15 & \multicolumn{1}{c|}{70.6363} & 5.47E-01 & 2.66E-01 & 8.13E-01 & 17 & 51.8154 \\
\multicolumn{1}{c|}{4}  & 3.67E-01 & 2.91E-01 & 6.58E-01 & 16 & \multicolumn{1}{c|}{47.8812} & 5.15E-01 & 2.44E-01 & 7.59E-01 & 17 & 27.7229 \\
\multicolumn{1}{c|}{8}  & 3.27E-01 & 2.43E-01 & 5.70E-01 & 16 & \multicolumn{1}{c|}{27.6535} & 5.58E-01 & 2.89E-01 & 8.47E-01 & 17 & 12.4233 \\
\multicolumn{1}{c|}{12} & 3.96E-01 & 2.50E-01 & 6.46E-01 & 16 & \multicolumn{1}{c|}{16.2693} & 6.18E-01 & 2.89E-01 & 9.07E-01 & 18 & 7.73527 \\
\multicolumn{1}{c|}{16} & 3.38E-01 & 2.83E-01 & 6.21E-01 & 16 & \multicolumn{1}{c|}{12.6851} & 6.60E-01 & 3.98E-01 & 1.06E+00 & 17 & 4.97401 \\
\Xhline{2\arrayrulewidth}
\multicolumn{11}{l}{Scale-splitting} \\ \hline
\multicolumn{1}{c|}{\multirow{3}{*}{\begin{tabular}[c]{@{}c@{}}No.\\ of MPI\\ proc.\end{tabular}}} & \multicolumn{5}{c|}{TET mesh} & \multicolumn{5}{c}{HEX mesh} \\ \cline{2-11} 
\multicolumn{1}{c|}{} & \multicolumn{3}{l}{Time} & \multirow{2}{*}{KSP} & \multicolumn{1}{c|}{\multirow{2}{*}{\begin{tabular}[c]{@{}c@{}}Parallel\\ eff. (\%)\end{tabular}}} & \multicolumn{3}{l}{Time} & \multirow{2}{*}{KSP} & \multirow{2}{*}{\begin{tabular}[c]{@{}c@{}}Parallel\\ eff. (\%)\end{tabular}} \\ \cline{2-4} \cline{7-9}
\multicolumn{1}{c|}{} & Assembly & Solver & Total &  & \multicolumn{1}{c|}{} & Assembly & Solver & Total &  &  \\ \hline
\multicolumn{1}{c|}{1}  & 6.81E-01 & 5.34E-01 & 1.22E+00 & 15 & \multicolumn{1}{c|}{100}     & 5.53E-01 & 2.93E-01 & 8.46E-01 & 17 & 100     \\
\multicolumn{1}{c|}{2}  & 4.77E-01 & 3.97E-01 & 8.75E-01 & 15 & \multicolumn{1}{c|}{69.4603} & 5.45E-01 & 2.72E-01 & 8.18E-01 & 17 & 51.7248 \\
\multicolumn{1}{c|}{4}  & 3.63E-01 & 2.95E-01 & 6.58E-01 & 16 & \multicolumn{1}{c|}{46.1907} & 5.15E-01 & 2.63E-01 & 7.78E-01 & 17 & 27.1754 \\
\multicolumn{1}{c|}{8}  & 3.26E-01 & 2.55E-01 & 5.81E-01 & 16 & \multicolumn{1}{c|}{26.1448} & 5.74E-01 & 3.68E-01 & 9.42E-01 & 17 & 11.2209 \\
\multicolumn{1}{c|}{12} & 3.81E-01 & 2.78E-01 & 6.59E-01 & 16 & \multicolumn{1}{c|}{15.3642} & 6.06E-01 & 2.90E-01 & 8.96E-01 & 18 & 7.86639 \\
\multicolumn{1}{c|}{16} & 3.64E-01 & 2.81E-01 & 6.44E-01 & 16 & \multicolumn{1}{c|}{11.7897} & 6.77E-01 & 3.31E-01 & 1.01E+00 & 17 & 5.24368 \\ 
\Xhline{2\arrayrulewidth}
\end{tabular}
\end{table}
}

%
{\scriptsize
\begin{table}[]
		\caption{\textsf{3D problem:} Strong-scaling results for CG-VMS formulation with same h-size.}
		\label{tab:Strong_CGVMS_hsize}
\begin{tabular}{ccccccccccc}
\Xhline{2\arrayrulewidth}
\multicolumn{11}{l}{Field-splitting} \\ \hline
\multicolumn{1}{c|}{\multirow{3}{*}{\begin{tabular}[c]{@{}c@{}}No.\\ of MPI\\ proc.\end{tabular}}} & \multicolumn{5}{c|}{TET mesh} & \multicolumn{5}{c}{HEX mesh} \\ \cline{2-11} 
\multicolumn{1}{c|}{} & \multicolumn{3}{l}{Time} & \multicolumn{1}{l}{\multirow{2}{*}{KSP}} & \multicolumn{1}{l|}{\multirow{2}{*}{\begin{tabular}[c]{@{}l@{}}Parallel\\ eff. (\%)\end{tabular}}} & \multicolumn{3}{l}{Time} & \multicolumn{1}{l}{\multirow{2}{*}{KSP}} & \multicolumn{1}{l}{\multirow{2}{*}{\begin{tabular}[c]{@{}l@{}}Parallel\\ eff. (\%)\end{tabular}}} \\ \cline{2-4} \cline{7-9}
\multicolumn{1}{c|}{} & Assembly & Solver & Total & \multicolumn{1}{l}{} & \multicolumn{1}{l|}{} & Assembly & Solver & Total & \multicolumn{1}{l}{} & \multicolumn{1}{l}{} \\ \hline
\multicolumn{1}{c|}{1}  & 2.37E+00 & 1.02E+00 & 3.39E+00 & 12 & \multicolumn{1}{c|}{100}  &  2.67E+00 & 1.67E+00 & 4.34E+00 & 16 & 100     \\
\multicolumn{1}{c|}{2}  & 2.06E+00 & 7.28E-01 & 2.78E+00 & 13 & \multicolumn{1}{c|}{60.8118}  & 2.16E+00 & 1.20E+00 & 3.36E+00 & 18 & 64.5344 \\
\multicolumn{1}{c|}{4}  & 1.55E+00 & 4.77E-01 & 2.02E+00 & 14 & \multicolumn{1}{c|}{41.8438}  & 1.65E+00 & 7.54E-01 & 2.40E+00 & 19 & 45.1499 \\
\multicolumn{1}{c|}{8}  & 1.06E+00 & 3.75E-01 & 1.44E+00 & 14 & \multicolumn{1}{c|}{29.4128}  & 1.41E+00 & 5.94E-01 & 2.00E+00 & 20 & 27.0719 \\
\multicolumn{1}{c|}{12} & 1.01E+00 & 3.61E-01 & 1.37E+00 & 15 & \multicolumn{1}{c|}{20.6564}  & 1.24E+00 & 6.18E-01 & 1.86E+00 & 21 & 19.4146 \\
\multicolumn{1}{c|}{16} & 8.90E-01 & 3.64E-01 & 1.25E+00 & 15 & \multicolumn{1}{c|}{16.876}  & 1.19E+00 & 5.13E-01 & 1.70E+00 & 20 & 15.9673 \\ 
\Xhline{2\arrayrulewidth}
\multicolumn{11}{l}{Scale-splitting} \\ \hline
\multicolumn{1}{c|}{\multirow{3}{*}{\begin{tabular}[c]{@{}c@{}}No.\\ of MPI\\ proc.\end{tabular}}} & \multicolumn{5}{c|}{TET mesh} & \multicolumn{5}{c}{HEX mesh} \\ \cline{2-11} 
\multicolumn{1}{c|}{} & \multicolumn{3}{l}{Time} & \multirow{2}{*}{KSP} & \multicolumn{1}{c|}{\multirow{2}{*}{\begin{tabular}[c]{@{}c@{}}Parallel\\ eff. (\%)\end{tabular}}} & \multicolumn{3}{l}{Time} & \multirow{2}{*}{KSP} & \multirow{2}{*}{\begin{tabular}[c]{@{}c@{}}Parallel\\ eff. (\%)\end{tabular}} \\ \cline{2-4} \cline{7-9}
\multicolumn{1}{c|}{} & Assembly & Solver & Total &  & \multicolumn{1}{c|}{} & Assembly & Solver & Total &  &  \\ \hline
\multicolumn{1}{c|}{1}  & 2.39E+00 & 8.62E-01 & 3.25E+00 & 12 & \multicolumn{1}{c|}{100}     & 2.67E+00 & 1.28E+00 & 3.96E+00 & 16 & 100     \\
\multicolumn{1}{c|}{2}  & 2.03E+00 & 6.35E-01 & 2.67E+00 & 13 & \multicolumn{1}{c|}{61.0049} & 2.17E+00 & 9.79E-01 & 3.15E+00 & 18 & 62.8295 \\
\multicolumn{1}{c|}{4}  & 1.57E+00 & 4.34E-01 & 2.00E+00 & 14 & \multicolumn{1}{c|}{40.6141} & 1.66E+00 & 6.57E-01 & 2.32E+00 & 19 & 42.6401 \\
\multicolumn{1}{c|}{8}  & 1.04E+00 & 3.49E-01 & 1.39E+00 & 14 & \multicolumn{1}{c|}{29.2416} & 1.39E+00 & 5.27E-01 & 1.92E+00 & 20 & 25.8155 \\
\multicolumn{1}{c|}{12} & 9.76E-01 & 3.41E-01 & 1.32E+00 & 15 & \multicolumn{1}{c|}{20.5897} & 1.29E+00 & 5.60E-01 & 1.85E+00 & 21 & 17.8629 \\
\multicolumn{1}{c|}{16} & 9.62E-01 & 3.56E-01 & 1.32E+00 & 15 & \multicolumn{1}{c|}{15.4423} & 1.21E+00 & 5.10E-01 & 1.72E+00 & 20 & 14.3786 \\ 
\Xhline{2\arrayrulewidth}
\end{tabular}
\end{table}
}

%
{\scriptsize
\begin{table}[]
	\caption{\textsf{3D problem:} Strong-scaling results for DG-VMS formulation with same h-size.}
	\label{tab:Strong_DGVMS_hsize}
\begin{tabular}{ccccccccccc}
\Xhline{2\arrayrulewidth}
\multicolumn{11}{l}{Field-splitting} \\ \hline
\multicolumn{1}{c|}{\multirow{3}{*}{\begin{tabular}[c]{@{}c@{}}No.\\ of MPI\\ proc.\end{tabular}}} & \multicolumn{5}{c|}{TET mesh} & \multicolumn{5}{c}{HEX mesh} \\ \cline{2-11} 
\multicolumn{1}{c|}{} & \multicolumn{3}{l}{Time} & \multicolumn{1}{l}{\multirow{2}{*}{KSP}} & \multicolumn{1}{l|}{\multirow{2}{*}{\begin{tabular}[c]{@{}l@{}}Parallel\\ eff. (\%)\end{tabular}}} & \multicolumn{3}{l}{Time} & \multicolumn{1}{l}{\multirow{2}{*}{KSP}} & \multicolumn{1}{l}{\multirow{2}{*}{\begin{tabular}[c]{@{}l@{}}Parallel\\ eff. (\%)\end{tabular}}} \\ \cline{2-4} \cline{7-9}
\multicolumn{1}{c|}{} & Assembly & Solver & Total & \multicolumn{1}{l}{} & \multicolumn{1}{l|}{} & Assembly & Solver & Total & \multicolumn{1}{l}{} & \multicolumn{1}{l}{} \\ \hline
\multicolumn{1}{c|}{1}  & 5.51E+01 & 2.42E+01 & 7.93E+01 & 19  & \multicolumn{1}{c|}{100}     & 5.32E+01 & 2.53E+01 & 7.85E+01 & 22 & 100     \\
\multicolumn{1}{c|}{2}  & 2.64E+01 & 2.37E+01 & 5.01E+01 & 44  & \multicolumn{1}{c|}{79.1343} & 2.62E+01 & 2.08E+01 & 4.70E+01 & 39 & 83.4645 \\
\multicolumn{1}{c|}{4}  & 9.65E+00 & 1.88E+01 & 2.85E+01 & 81  & \multicolumn{1}{c|}{69.6209} & 1.14E+01 & 1.10E+01 & 2.24E+01 & 42 & 87.6787 \\
\multicolumn{1}{c|}{8}  & 5.45E+00 & 1.26E+01 & 1.80E+01 & 100 & \multicolumn{1}{c|}{55.0972} & 7.70E+00 & 6.81E+00 & 1.45E+01 & 45 & 67.6172 \\
\multicolumn{1}{c|}{12} & 5.08E+00 & 1.01E+01 & 1.52E+01 & 108 & \multicolumn{1}{c|}{43.6126} & 7.53E+00 & 5.64E+00 & 1.32E+01 & 52 & 49.7024 \\
\multicolumn{1}{c|}{16} & 3.66E+00 & 8.67E+00 & 1.23E+01 & 110 & \multicolumn{1}{c|}{40.2496} & 6.48E+00 & 4.35E+00 & 1.08E+01 & 46 & 45.3385 \\ 
\Xhline{2\arrayrulewidth}
\multicolumn{11}{l}{Scale-splitting} \\ \hline
\multicolumn{1}{c|}{\multirow{3}{*}{\begin{tabular}[c]{@{}c@{}}No.\\ of MPI\\ proc.\end{tabular}}} & \multicolumn{5}{c|}{TET mesh} & \multicolumn{5}{c}{HEX mesh} \\ \cline{2-11} 
\multicolumn{1}{c|}{} & \multicolumn{3}{l}{Time} & \multirow{2}{*}{KSP} & \multicolumn{1}{c|}{\multirow{2}{*}{\begin{tabular}[c]{@{}c@{}}Parallel\\ eff. (\%)\end{tabular}}} & \multicolumn{3}{l}{Time} & \multirow{2}{*}{KSP} & \multirow{2}{*}{\begin{tabular}[c]{@{}c@{}}Parallel\\ eff. (\%)\end{tabular}} \\ \cline{2-4} \cline{7-9}
\multicolumn{1}{c|}{} & Assembly & Solver & Total &  & \multicolumn{1}{c|}{} & Assembly & Solver & Total &  &  \\ \hline
\multicolumn{1}{c|}{1}  & 5.52E+01 & 1.86E+01 & 7.37E+01 & 19  & \multicolumn{1}{c|}{100}     & 5.35E+01 & 1.82E+01 & 7.17E+01 & 22 & 100     \\
\multicolumn{1}{c|}{2}  & 2.65E+01 & 1.88E+01 & 4.52E+01 & 44  & \multicolumn{1}{c|}{81.4987} & 2.62E+01 & 1.51E+01 & 4.13E+01 & 39 & 86.6804 \\
\multicolumn{1}{c|}{4}  & 1.00E+01 & 1.57E+01 & 2.56E+01 & 81  & \multicolumn{1}{c|}{71.8994} & 1.10E+01 & 8.13E+00 & 1.91E+01 & 42 & 93.881  \\
\multicolumn{1}{c|}{8}  & 5.68E+00 & 1.04E+01 & 1.61E+01 & 100 & \multicolumn{1}{c|}{57.1805} & 7.92E+00 & 5.11E+00 & 1.30E+01 & 45 & 68.7356 \\
\multicolumn{1}{c|}{12} & 5.01E+00 & 8.48E+00 & 1.35E+01 & 108 & \multicolumn{1}{c|}{45.5861} & 7.49E+00 & 4.34E+00 & 1.18E+01 & 52 & 50.472  \\
\multicolumn{1}{c|}{16} & 3.78E+00 & 7.45E+00 & 1.12E+01 & 110 & \multicolumn{1}{c|}{41.0762} & 6.63E+00 & 3.40E+00 & 1.00E+01 & 46 & 44.6028 \\ 
\Xhline{2\arrayrulewidth}
\end{tabular}
\end{table}
}

%
\subsubsection{Test 2: TAS Spectrum Analysis}
For the TAS spectrum analysis, we consider a range of problems, shown in Table
\ref{tab:h-size_vs_DoF}, such that all finite element formulations in each
refinement step have roughly the same DoF count.
The mesh convergence results with respect to DoA and DoS, for both TET and 
HEX meshes, are shown in Figure \ref{Fig:3D_convergence_e-7}. 
CG-VMS and DG-VMS lines indicate a slope of $\frac{2}{3}$, which again corroborates 
that our Firedrake implementation of these formulations are correct. The H(div) 
lines exhibit a slope of $\frac{1}{3}$ for TET mesh. However, similar to the
two-dimensional problem for non-simplicial element QUAD, H(div) exhibits super linear 
convergence for the HEX meshes. We are not observing any tail-offs in these 
results as the solver relative convergence tolerance of $1e-7$ was strict enough. 
The observation that both CG- and DG-VMS have the highest DoA over DoS ratio for almost 
all velocity and pressure fields implies that they have greater levels of 
contribution to the overall numerical accuracy than the H(div) schemes.

Static-scaling results for both block solver strategies are presented in 
Figure \ref{Fig:3D_Static_scaling}. Flat lines appear in all six subfigures, 
indicating that the proposed block-solver methodologies are scalable under the
chosen $h$-sizes and hardware environment. 
It is a common belief among application
scientists that a solver exhibits worse
scaling than an assembly procedure, since
assembly is almost entirely local. However,
the results show that for all the chosen
discretizations\textemdash no matter
what solver methodology is employed\textemdash
time to assemble stiffness matrix is higher 
than the solver time.  This infers that we
have successfully optimized solvers to such
an extent that the assembly procedure is
more dominant.

Analogous to the two-dimensional problem, 
the scale-splitting methodologies are slightly better than their field-splitting 
counterparts for the all formulations. Evidently, this disparity is more 
clear for VMS formulations at the solve time level. However, the difference 
in performance is almost inconsequential when we look at the total time. 
It can be seen that the DoF counts are sufficient to level out the curves 
when all three formulations have roughly $20$K DoFs or more.
Regardless of the mesh type, the H(div) formulation processes its DoF count 
faster than either VMS formulations.

%
Figures \ref{Fig:3D_DoE_TET} and \ref{Fig:3D_DoE_HEX} contain 
DoE diagrams for TET and HEX meshes, respectively. For the case of 
TET mesh type, in spite of H(div) having the fastest computation rates, 
it has a lower DoA than its VMS counterparts which in turns lead to a 
much smaller DoE with steep declining curve. On the contrary, for the 
case of the HEX mesh, H(div) surpasses its VMS counterparts due to its 
high DoA values. These diagrams demonstrate how numerical accuracy can have a
drastic effect on the overall computational performance of these various 
finite element formulations.

{\scriptsize
\begin{table}[]
	\caption{\textsf{3D problem:} Strong-scaling results for H(div) formulation with same DoF count.}
	\label{tab:Strong_RT$0$_DoF}
\begin{tabular}{ccccccccccc}
\Xhline{2\arrayrulewidth}
\multicolumn{11}{l}{Field-splitting} \\ \hline
\multicolumn{1}{c|}{\multirow{3}{*}{\begin{tabular}[c]{@{}c@{}}No.\\ of MPI\\ proc.\end{tabular}}} & \multicolumn{5}{c|}{TET mesh} & \multicolumn{5}{c}{HEX mesh} \\ \cline{2-11} 
\multicolumn{1}{c|}{} & \multicolumn{3}{l}{Time} & \multicolumn{1}{l}{\multirow{2}{*}{KSP}} & \multicolumn{1}{l|}{\multirow{2}{*}{\begin{tabular}[c]{@{}l@{}}Parallel\\ eff. (\%)\end{tabular}}} & \multicolumn{3}{l}{Time} & \multicolumn{1}{l}{\multirow{2}{*}{KSP}} & \multicolumn{1}{l}{\multirow{2}{*}{\begin{tabular}[c]{@{}l@{}}Parallel\\ eff. (\%)\end{tabular}}} \\ \cline{2-4} \cline{7-9}
\multicolumn{1}{c|}{} & Assembly & Solver & Total & \multicolumn{1}{l}{} & \multicolumn{1}{l|}{} & Assembly & Solver & Total & \multicolumn{1}{l}{} & \multicolumn{1}{l}{} \\ \hline
\multicolumn{1}{c|}{1}  & 9.11E-01 & 6.62E-01 & 1.57E+00 & 15 & \multicolumn{1}{c|}{100}     & 1.39E+00 & 1.13E+00 & 2.52E+00 & 20 & 100     \\
\multicolumn{1}{c|}{2}  & 6.25E-01 & 5.20E-01 & 1.15E+00 & 16 & \multicolumn{1}{c|}{68.7336} & 9.78E-01 & 7.72E-01 & 1.75E+00 & 20 & 71.9143 \\
\multicolumn{1}{c|}{4}  & 4.10E-01 & 3.51E-01 & 7.60E-01 & 16 & \multicolumn{1}{c|}{51.7627} & 7.13E-01 & 5.21E-01 & 1.23E+00 & 20 & 50.9927 \\
\multicolumn{1}{c|}{8}  & 3.63E-01 & 2.70E-01 & 6.33E-01 & 16 & \multicolumn{1}{c|}{31.092}  & 6.72E-01 & 4.58E-01 & 1.13E+00 & 21 & 27.8429 \\
\multicolumn{1}{c|}{12} & 3.87E-01 & 3.04E-01 & 6.90E-01 & 16 & \multicolumn{1}{c|}{18.9986} & 7.24E-01 & 4.18E-01 & 1.14E+00 & 21 & 18.3669 \\
\multicolumn{1}{c|}{16} & 3.63E-01 & 3.00E-01 & 6.62E-01 & 16 & \multicolumn{1}{c|}{14.8513} & 7.45E-01 & 4.19E-01 & 1.16E+00 & 21 & 13.5148 \\ 
\Xhline{2\arrayrulewidth}
\multicolumn{11}{l}{Scale-splitting} \\ \hline
\multicolumn{1}{c|}{\multirow{3}{*}{\begin{tabular}[c]{@{}c@{}}No.\\ of MPI\\ proc.\end{tabular}}} & \multicolumn{5}{c|}{TET mesh} & \multicolumn{5}{c}{HEX mesh} \\ \cline{2-11} 
\multicolumn{1}{c|}{} & \multicolumn{3}{l}{Time} & \multirow{2}{*}{KSP} & \multicolumn{1}{c|}{\multirow{2}{*}{\begin{tabular}[c]{@{}c@{}}Parallel\\ eff. (\%)\end{tabular}}} & \multicolumn{3}{l}{Time} & \multirow{2}{*}{KSP} & \multirow{2}{*}{\begin{tabular}[c]{@{}c@{}}Parallel\\ eff. (\%)\end{tabular}} \\ \cline{2-4} \cline{7-9}
\multicolumn{1}{c|}{} & Assembly & Solver & Total &  & \multicolumn{1}{c|}{} & Assembly & Solver & Total &  &  \\ \hline
\multicolumn{1}{c|}{1}  & 9.14E-01 & 6.19E-01 & 1.53E+00 & 15 & \multicolumn{1}{c|}{100}     & 1.39E+00 & 1.02E+00 & 2.41E+00 & 20 & 100     \\
\multicolumn{1}{c|}{2}  & 6.07E-01 & 4.92E-01 & 1.10E+00 & 16 & \multicolumn{1}{c|}{69.8543} & 9.72E-01 & 7.26E-01 & 1.70E+00 & 20 & 71.0542 \\
\multicolumn{1}{c|}{4}  & 4.25E-01 & 3.36E-01 & 7.61E-01 & 16 & \multicolumn{1}{c|}{50.4008} & 7.07E-01 & 5.32E-01 & 1.24E+00 & 20 & 48.6885 \\
\multicolumn{1}{c|}{8}  & 3.81E-01 & 2.83E-01 & 6.63E-01 & 16 & \multicolumn{1}{c|}{28.9085} & 7.48E-01 & 4.57E-01 & 1.21E+00 & 21 & 25.0311 \\
\multicolumn{1}{c|}{12} & 3.36E-01 & 2.68E-01 & 6.04E-01 & 16 & \multicolumn{1}{c|}{21.168}  & 7.13E-01 & 4.28E-01 & 1.14E+00 & 21 & 17.6389 \\
\multicolumn{1}{c|}{16} & 3.82E-01 & 2.73E-01 & 6.55E-01 & 16 & \multicolumn{1}{c|}{14.6352} & 7.95E-01 & 4.50E-01 & 1.25E+00 & 21 & 12.1135 \\ 
\Xhline{2\arrayrulewidth}
\end{tabular}
\end{table}
}

%
{\scriptsize
\begin{table}[]
	\caption{\textsf{3D problem:} Strong-scaling results for CG-VMS formulation with same DoF count.}
	\label{tab:Strong_CGVMS_DoF}
\begin{tabular}{ccccccccccc}
\Xhline{2\arrayrulewidth}
\multicolumn{11}{l}{Field-splitting} \\ \hline
\multicolumn{1}{c|}{\multirow{3}{*}{\begin{tabular}[c]{@{}c@{}}No.\\ of MPI\\ proc.\end{tabular}}} & \multicolumn{5}{c|}{TET mesh} & \multicolumn{5}{c}{HEX mesh} \\ \cline{2-11} 
\multicolumn{1}{c|}{} & \multicolumn{3}{l}{Time} & \multicolumn{1}{l}{\multirow{2}{*}{KSP}} & \multicolumn{1}{l|}{\multirow{2}{*}{\begin{tabular}[c]{@{}l@{}}Parallel\\ eff. (\%)\end{tabular}}} & \multicolumn{3}{l}{Time} & \multicolumn{1}{l}{\multirow{2}{*}{KSP}} & \multicolumn{1}{l}{\multirow{2}{*}{\begin{tabular}[c]{@{}l@{}}Parallel\\ eff. (\%)\end{tabular}}} \\ \cline{2-4} \cline{7-9}
\multicolumn{1}{c|}{} & Assembly & Solver & Total & \multicolumn{1}{l}{} & \multicolumn{1}{l|}{} & Assembly & Solver & Total & \multicolumn{1}{l}{} & \multicolumn{1}{l}{} \\ \hline
\multicolumn{1}{c|}{1}  & 1.23E+01 & 5.01E+00 & 1.73E+01 & 13 & \multicolumn{1}{c|}{100}     & 1.53E+01 & 9.33E+00 & 2.46E+01 & 17 & 100     \\
\multicolumn{1}{c|}{2}  & 7.95E+00 & 3.39E+00 & 1.13E+01 & 15 & \multicolumn{1}{c|}{76.455}  & 9.20E+00 & 6.22E+00 & 1.54E+01 & 19 & 79.9481 \\
\multicolumn{1}{c|}{4}  & 4.81E+00 & 1.99E+00 & 6.80E+00 & 15 & \multicolumn{1}{c|}{63.7406} & 5.08E+00 & 3.37E+00 & 8.45E+00 & 20 & 72.8908 \\
\multicolumn{1}{c|}{8}  & 3.16E+00 & 1.26E+00 & 4.42E+00 & 16 & \multicolumn{1}{c|}{49.0052} & 3.53E+00 & 2.34E+00 & 5.87E+00 & 21 & 52.4791 \\
\multicolumn{1}{c|}{12} & 2.99E+00 & 1.06E+00 & 4.05E+00 & 16 & \multicolumn{1}{c|}{35.6702} & 3.11E+00 & 2.01E+00 & 5.11E+00 & 21 & 40.1669 \\
\multicolumn{1}{c|}{16} & 2.42E+00 & 9.54E-01 & 3.38E+00 & 16 & \multicolumn{1}{c|}{32.1111} & 2.78E+00 & 1.66E+00 & 4.44E+00 & 21 & 34.6691 \\ 
\Xhline{2\arrayrulewidth}
\multicolumn{11}{l}{Scale-splitting} \\ \hline
\multicolumn{1}{c|}{\multirow{3}{*}{\begin{tabular}[c]{@{}c@{}}No.\\ of MPI\\ proc.\end{tabular}}} & \multicolumn{5}{c|}{TET mesh} & \multicolumn{5}{c}{HEX mesh} \\ \cline{2-11} 
\multicolumn{1}{c|}{} & \multicolumn{3}{l}{Time} & \multirow{2}{*}{KSP} & \multicolumn{1}{c|}{\multirow{2}{*}{\begin{tabular}[c]{@{}c@{}}Parallel\\ eff. (\%)\end{tabular}}} & \multicolumn{3}{l}{Time} & \multirow{2}{*}{KSP} & \multirow{2}{*}{\begin{tabular}[c]{@{}c@{}}Parallel\\ eff. (\%)\end{tabular}} \\ \cline{2-4} \cline{7-9}
\multicolumn{1}{c|}{} & Assembly & Solver & Total &  & \multicolumn{1}{c|}{} & Assembly & Solver & Total &  &  \\ \hline
\multicolumn{1}{c|}{1}  & 1.25E+01 & 4.14E+00 & 1.66E+01 & 13 & \multicolumn{1}{c|}{100}     & 1.53E+01 & 6.93E+00 & 2.22E+01 & 17 & 100     \\
\multicolumn{1}{c|}{2}  & 7.84E+00 & 2.80E+00 & 1.06E+01 & 15 & \multicolumn{1}{c|}{77.9605} & 9.16E+00 & 4.64E+00 & 1.38E+01 & 19 & 80.5435 \\
\multicolumn{1}{c|}{4}  & 4.84E+00 & 1.64E+00 & 6.48E+00 & 15 & \multicolumn{1}{c|}{64.0244} & 5.23E+00 & 2.68E+00 & 7.91E+00 & 20 & 70.2592 \\
\multicolumn{1}{c|}{8}  & 3.32E+00 & 1.11E+00 & 4.43E+00 & 16 & \multicolumn{1}{c|}{46.8009} & 3.62E+00 & 1.93E+00 & 5.55E+00 & 21 & 50.0676 \\
\multicolumn{1}{c|}{12} & 2.90E+00 & 9.59E-01 & 3.86E+00 & 16 & \multicolumn{1}{c|}{35.8439} & 3.13E+00 & 1.64E+00 & 4.77E+00 & 21 & 38.8365 \\
\multicolumn{1}{c|}{16} & 2.44E+00 & 8.38E-01 & 3.28E+00 & 16 & \multicolumn{1}{c|}{31.6313} & 2.75E+00 & 1.37E+00 & 4.12E+00 & 21 & 33.7555 \\ 
\Xhline{2\arrayrulewidth}
\end{tabular}
\end{table}
}

{\scriptsize
\begin{table}[]
	\caption{\textsf{3D problem:} Strong-scaling results for DG-VMS formulation with same DoF count.}
	\label{tab:Strong_DGVMS_DoF}
\begin{tabular}{ccccccccccc}
\Xhline{2\arrayrulewidth}
\multicolumn{11}{l}{Field-splitting} \\ \hline
\multicolumn{1}{c|}{\multirow{3}{*}{\begin{tabular}[c]{@{}c@{}}No.\\ of MPI\\ proc.\end{tabular}}} & \multicolumn{5}{c|}{TET mesh} & \multicolumn{5}{c}{HEX mesh} \\ \cline{2-11} 
\multicolumn{1}{c|}{} & \multicolumn{3}{l}{Time} & \multicolumn{1}{l}{\multirow{2}{*}{KSP}} & \multicolumn{1}{l|}{\multirow{2}{*}{\begin{tabular}[c]{@{}l@{}}Parallel\\ eff. (\%)\end{tabular}}} & \multicolumn{3}{l}{Time} & \multicolumn{1}{l}{\multirow{2}{*}{KSP}} & \multicolumn{1}{l}{\multirow{2}{*}{\begin{tabular}[c]{@{}l@{}}Parallel\\ eff. (\%)\end{tabular}}} \\ \cline{2-4} \cline{7-9}
\multicolumn{1}{c|}{} & Assembly & Solver & Total & \multicolumn{1}{l}{} & \multicolumn{1}{l|}{} & Assembly & Solver & Total & \multicolumn{1}{l}{} & \multicolumn{1}{l}{} \\ \hline
\multicolumn{1}{c|}{1}  & 8.08E+00 & 5.40E+00 & 1.35E+01 & 19  & \multicolumn{1}{c|}{100}     & 3.40E+01 & 2.03E+01 & 5.42E+01 & 22 & 100     \\
\multicolumn{1}{c|}{2}  & 4.76E+00 & 8.27E+00 & 1.30E+01 & 79  & \multicolumn{1}{c|}{51.7268} & 1.82E+01 & 1.71E+01 & 3.54E+01 & 40 & 76.6544 \\
\multicolumn{1}{c|}{4}  & 2.55E+00 & 5.45E+00 & 7.99E+00 & 102 & \multicolumn{1}{c|}{42.1566} & 1.02E+01 & 9.67E+00 & 1.99E+01 & 45 & 68.2402 \\
\multicolumn{1}{c|}{8}  & 1.91E+00 & 3.37E+00 & 5.27E+00 & 109 & \multicolumn{1}{c|}{31.9492} & 7.32E+00 & 5.92E+00 & 1.32E+01 & 46 & 51.1801 \\
\multicolumn{1}{c|}{12} & 1.98E+00 & 2.69E+00 & 4.67E+00 & 111 & \multicolumn{1}{c|}{24.0646} & 6.41E+00 & 4.42E+00 & 1.08E+01 & 50 & 41.7514 \\
\multicolumn{1}{c|}{16} & 1.61E+00 & 2.52E+00 & 4.13E+00 & 119 & \multicolumn{1}{c|}{20.4045} & 6.04E+00 & 3.98E+00 & 1.00E+01 & 52 & 33.8136 \\ 
\Xhline{2\arrayrulewidth}
\multicolumn{11}{l}{Scale-splitting} \\ \hline
\multicolumn{1}{c|}{\multirow{3}{*}{\begin{tabular}[c]{@{}c@{}}No.\\ of MPI\\ proc.\end{tabular}}} & \multicolumn{5}{c|}{TET mesh} & \multicolumn{5}{c}{HEX mesh} \\ \cline{2-11} 
\multicolumn{1}{c|}{} & \multicolumn{3}{l}{Time} & \multirow{2}{*}{KSP} & \multicolumn{1}{c|}{\multirow{2}{*}{\begin{tabular}[c]{@{}c@{}}Parallel\\ eff. (\%)\end{tabular}}} & \multicolumn{3}{l}{Time} & \multirow{2}{*}{KSP} & \multirow{2}{*}{\begin{tabular}[c]{@{}c@{}}Parallel\\ eff. (\%)\end{tabular}} \\ \cline{2-4} \cline{7-9}
\multicolumn{1}{c|}{} & Assembly & Solver & Total &  & \multicolumn{1}{c|}{} & Assembly & Solver & Total &  &  \\ \hline
\multicolumn{1}{c|}{1}  & 7.81E+00 & 4.11E+00 & 1.19E+01 & 19  & \multicolumn{1}{c|}{100}     & 3.25E+01 & 1.45E+01 & 4.70E+01 & 22 & 100     \\
\multicolumn{1}{c|}{2}  & 4.57E+00 & 6.62E+00 & 1.12E+01 & 79  & \multicolumn{1}{c|}{53.2618} & 1.82E+01 & 1.24E+01 & 3.06E+01 & 40 & 76.8715 \\
\multicolumn{1}{c|}{4}  & 2.54E+00 & 4.43E+00 & 6.97E+00 & 102 & \multicolumn{1}{c|}{42.7301} & 1.02E+01 & 6.95E+00 & 1.71E+01 & 45 & 68.6369 \\
\multicolumn{1}{c|}{8}  & 1.85E+00 & 2.88E+00 & 4.73E+00 & 109 & \multicolumn{1}{c|}{31.5077} & 7.41E+00 & 4.19E+00 & 1.16E+01 & 46 & 50.6789 \\
\multicolumn{1}{c|}{12} & 1.88E+00 & 2.36E+00 & 4.24E+00 & 111 & \multicolumn{1}{c|}{23.4498} & 6.19E+00 & 3.64E+00 & 9.83E+00 & 50 & 39.8694 \\
\multicolumn{1}{c|}{16} & 1.61E+00 & 2.26E+00 & 3.87E+00 & 119 & \multicolumn{1}{c|}{19.2556} & 6.00E+00 & 3.40E+00 & 9.40E+00 & 52 & 31.2566 \\ 
\Xhline{2\arrayrulewidth}
\end{tabular}
\end{table}
}

{\scriptsize
\begin{table}[]
	\centering
	\caption{This table illustrates three-dimensional h-size refinement for each discretization such that at each step DoF approximately doubles.}
	\label{tab:h-size_vs_DoF}
\begin{tabular}{c|c|c|c|c|c|c|c|cccc}
	\Xhline{2\arrayrulewidth}
	\multicolumn{4}{c|}{CG-VMS}                                                                                     & \multicolumn{4}{c|}{DG-VMS}                                                                                     & \multicolumn{4}{c}{H(div)}                                                                                           \\ \hline
	\multicolumn{2}{c|}{TET}                               & \multicolumn{2}{c|}{HEX}                               & \multicolumn{2}{c|}{TET}                               & \multicolumn{2}{c|}{HEX}                               & \multicolumn{2}{c|}{TET}                                  & \multicolumn{2}{c}{HEX}                               \\ \hline
	\multicolumn{1}{l|}{h-size} & \multicolumn{1}{l|}{DoF} & \multicolumn{1}{l|}{h-size} & \multicolumn{1}{l|}{DoF} & \multicolumn{1}{l|}{h-size} & \multicolumn{1}{l|}{DoF} & \multicolumn{1}{l|}{h-size} & \multicolumn{1}{l|}{DoF} & \multicolumn{1}{l|}{h-size} & \multicolumn{1}{l|}{DoF}    & \multicolumn{1}{l|}{h-size} & \multicolumn{1}{l}{DoF} \\ \hline
		1/13       & 21952         & 1/13       & 21952         & 1/5        & 24000         & 1/7        & 21952         & \multicolumn{1}{c|}{1/8}  & \multicolumn{1}{c|}{19200}   & \multicolumn{1}{c|}{1/13} & 18590   \\
		1/17       & 46656         & 1/17       & 46656         & 1/6        & 41472         & 1/9        & 46686         & \multicolumn{1}{c|}{1/10} & \multicolumn{1}{c|}{37200}   & \multicolumn{1}{c|}{1/17} & 41038   \\
		1/21       & 85184         & 1/21       & 85184         & 1/8        & 98304         & 1/11       & 85184         & \multicolumn{1}{c|}{1/13} & \multicolumn{1}{c|}{81120}   & \multicolumn{1}{c|}{1/22} & 88088   \\
		1/28       & 195112        & 1/28       & 195112        & 1/10       & 192000        & 1/14       & 175616        & \multicolumn{1}{c|}{1/17} & \multicolumn{1}{c|}{180336}  & \multicolumn{1}{c|}{1/28} & 180320  \\
		1/36       & 405224        & 1/36       & 405224        & 1/13       & 421824        & 1/19       & 438976        & \multicolumn{1}{c|}{1/22} & \multicolumn{1}{c|}{389136}  & \multicolumn{1}{c|}{1/37} & 413438  \\
		1/45       & 778688        & 1/45       & 778688        & 1/16       & 786432        & 1/23       & 778688        & \multicolumn{1}{c|}{1/28} & \multicolumn{1}{c|}{799686}  & \multicolumn{1}{c|}{1/46} & 791384  \\
		1/57       & 1560896       & 1/57       & 1560896       & 1/20       & 1536000       & 1/29       & 1560896       & \multicolumn{1}{c|}{1/35} & \multicolumn{1}{c|}{1558200} & \multicolumn{1}{c|}{1/58} & 1581080 \\ 
		\Xhline{2\arrayrulewidth}
	\end{tabular}
\end{table}
}

\section{CLOSURE}
\label{Sec:S7_Comparison_Conclusion}
We have developed two block solver methodologies
which are capable of solving large-scale problems
under the four-field DPP mathematical model. We
have also presented a systematic performance
analysis of various finite element discretizations
for the DPP model using the recently proposed
Time-Accuracy-Size (TAS) spectrum model, which takes
into consideration important metrics such as
mesh convergence, static-scaling, and Digits
of Efficacy (DoE). We have also identified 
strong-scaling issues one needs to be cognizant 
of when the block solvers are applied to various finite
elements. In our numerical studies, two- and
three-dimensional problems had analogous
performance trends, despite their marked
discrepancy in time to solution.

Some of the salient features of the
proposed composable block solver
methodologies are as follows: 
\begin{enumerate}
\item Both composable solvers are
  compatible with different kinds of
  mixed finite element formulations: 
  H(div) and non-H(div) elements,
  simplicial and non-simplicial
  elements, node- and edge-based
  discretizations, and continuous and
  discontinuous approximations. 
\item Both composable solvers
  are scalable in both parallel
  and algorithmic senses.
\item The solvers can be implemented
  seamlessly using the existing PETSc's
  composable solver options. Hence, one
  can leverage on the existing parallel
  computing tools to implement these
  composable solvers into existing
  simulators. 
\end{enumerate}

Some of the main conclusions from the
performance analysis based on the TAS
spectrum model are as follows: 
\begin{enumerate}
\item \textbf{Scale-split vs. field-split.}
  For a fixed problem size, the scale-splitting
  methodology tends to be slightly more efficient
  in terms of wall-clock time needed despite having
  the same KSP counts as the field-splitting method.
  However, selecting either solver methodologies
  will be left to the programmer's convenience
  and limitations as switching from one strategy
  to another exerts negligible overall effects
  on performance metrics.
\item \textbf{H(div) vs. VMS formulations.}
  (a) No matter what mesh type is chosen, DoFs
  are processed the fastest under the H(div)
  formulation compared to the CG-VMS or
  DG-VMS formulations.
  (b) The VMS formulations yield much
    higher overall numerical accuracy
    for all velocity and pressure fields
    than their H(div) counterparts. The
    exception is for non-simplicial meshes,
    where the H(div) formulation exhibits
    super linear convergence.
    %
\end{enumerate}


\appendix
\section{Computer codes}

\label{App:code}
In the following, we have provided Firedrake-based computer codes for 
CG-VMS formulation (listing \ref{Code:ex1}), DG-VMS formulation (listing \ref{Code:ex2}), and H(div) formulation (listing \ref{Code:ex3}), which was earlier discussed in 
Section \ref{Sec:S3_Comparison_Weak}.

\begin{lstlisting}[language=Python,caption=Firedrake code
for 3D problem with TET mesh using H(div) formulation, label=Code:ex3,
frame=single]
from firedrake import *
import numpy as np

#== Create mesh ==#
mesh = BoxMesh(5,5,5,1,1,1)

#== Function spaces ==#
vSpace = FunctionSpace(mesh,"RT",1)
pSpace = FunctionSpace(mesh,"DG",0)
wSpace = MixedFunctionSpace([vSpace,pSpace,vSpace,pSpace])

#== Define trial and test functions ==#
(v1,p1,v2,p2) = TrialFunctions(wSpace)
(w1,q1,w2,q2) = TestFunctions(wSpace)

#== Parameters and material properties ==#
rhob1, rhob2 = Constant((0.0,0.0,0.0)), Constant((0.0,0.0,0.0))
mu = Constant(1.0)
beta = Constant(1.0) 
fact = 1.0                     
k1, k2 = Constant(1.0), Constant(0.1)
alpha1, alpha2 = Constant(mu/k1), Constant(mu/k2)
eta = np.sqrt(1.0 *(1.0+0.1)/(1.0 * 0.1))

#== Boundary conditions  ==#
p1_left = interpolate(Expression("(1/pi)*(sin(pi*x[1])+sin(pi*x[2])) -\
                      (exp(eta*x[1])+exp(eta*x[2]))",eta=eta) , pSpace)
p1_right = interpolate(Expression("(1/pi)*exp(pi)*(sin(pi*x[1])+sin(pi*x[2])) -\
                      (exp(eta*x[1])+exp(eta*x[2]))",eta=eta) , pSpace)
p1_bottom = interpolate(Expression("(1/pi)*exp(pi*x[0])*sin(pi*x[2]) -\
                              (1.0 + exp(eta*x[2]))",eta=eta) , pSpace)
p1_top = interpolate(Expression("(1/pi)*exp(pi*x[0])*sin(pi*x[2]) -\
                         (exp(eta) + exp(eta*x[2]))",eta=eta) , pSpace)
p1_back = interpolate(Expression("(1/pi)*exp(pi*x[0])*sin(pi*x[1]) -\
                              (exp(eta*x[1]) + 1.0)",eta=eta) , pSpace)
p1_front = interpolate(Expression("(1/pi)*exp(pi*x[0])*sin(pi*x[1]) -\
                         (exp(eta) + exp(eta*x[1]))",eta=eta) , pSpace)
p2_left = interpolate(Expression("(1/pi)*(sin(pi*x[1])+sin(pi*x[2])) +\
               10.0 * (exp(eta*x[1])+exp(eta*x[2]))",eta=eta) , pSpace)
p2_right = interpolate(Expression("(1/pi)*exp(pi)*(sin(pi*x[1])+sin(pi*x[2])) +\
               10.0 * (exp(eta*x[1])+exp(eta*x[2]))",eta=eta) , pSpace)
p2_bottom = interpolate(Expression("(1/pi)*exp(pi*x[0])*sin(pi*x[2]) +\
                       10.0 * (1.0 + exp(eta*x[2]))",eta=eta) , pSpace)
p2_top = interpolate(Expression("(1/pi)*exp(pi*x[0])*sin(pi*x[2]) +\
                  10.0 * (exp(eta) + exp(eta*x[2]))",eta=eta) , pSpace)
p2_back = interpolate(Expression("(1/pi)*exp(pi*x[0])*sin(pi*x[1]) +\
                      10.0 *  (exp(eta*x[1]) + 1.0)",eta=eta) , pSpace)
p2_front = interpolate(Expression("(1/pi)*exp(pi*x[0])*sin(pi*x[1]) +\
                  10.0 * (exp(eta) + exp(eta*x[1]))",eta=eta) , pSpace)
bcs = []

#== Normal vectors ==#
n = FacetNormal(mesh)

#== Define variational forms ==#
a = dot(w1, alpha1*v1)*dx + dot(w2, alpha2*v2)*dx \
    - div(w1) * p1 *dx - div(w2) * p2 * dx  + q1 * div(v1) * dx + q2 * div(v2) * dx +\
    q1 * fact * (p1 - p2) * dx - q2 * fact * (p1 - p2) * dx

L = dot(w1,rhob1)*dx + dot(w2,rhob2)*dx -\
    dot(w1,n) * p1_left * ds(1) - dot(w2,n) * p2_left * ds(1) -\
    dot(w1,n) * p1_right * ds(2) - dot(w2,n) * p2_right * ds(2) -\
    dot(w1,n) * p1_bottom * ds(3) - dot(w2,n) * p2_bottom * ds(3) -\
    dot(w1,n) * p1_top * ds(4) - dot(w2,n) * p2_top * ds(4) -\
    dot(w1,n) * p1_back * ds(5) - dot(w2,n) * p2_back * ds(5) -\
    dot(w1,n) * p1_front * ds(6) - dot(w2,n) * p2_front * ds(6)

#== Solver options ==#
parameters_twofields = {
  "ksp_type": "gmres",
  "pc_type": "fieldsplit",
  "pc_fieldsplit_0_fields": "0,2",
  "pc_fieldsplit_1_fields": "1,3",
  "pc_fieldsplit_type": "schur",
  "pc_fieldsplit_schur_fact_type": "full",
  "pc_fieldsplit_schur_precondition": "selfp",
  "fieldsplit_0_ksp_type": "preonly",
  "fieldsplit_0_pc_type": "bjacobi",
  "fieldsplit_1_ksp_type": "preonly",
  "fieldsplit_1_pc_type": "fieldsplit",
  "fieldsplit_1_pc_fieldsplit_type": "additive",
  "fieldsplit_1_fieldsplit_0_ksp_type": "preonly",
  "fieldsplit_1_fieldsplit_0_pc_type": "hypre",
  "fieldsplit_1_fieldsplit_0_pc_hypre_boomeramg_strong_threshold": 0.75,
  "fieldsplit_1_fieldsplit_0_pc_hypre_boomeramg_agg_nl": 2,
  "fieldsplit_1_fieldsplit_1_ksp_type": "preonly",
  "fieldsplit_1_fieldsplit_1_pc_type": "hypre",
  "fieldsplit_1_fieldsplit_1_pc_hypre_boomeramg_strong_threshold": 0.75,
  "fieldsplit_1_fieldsplit_1_pc_hypre_boomeramg_agg_nl": 2,
  "ksp_rtol": 1e-5
}

#== Solve problem ==#
solution = Function(wSpace)
A = assemble(a, bcs=bcs, mat_type='aij')
b = assemble(L)
solver = LinearSolver(A,P=None,options_prefix="twofields_",\
                          solver_parameters=parameters_twofields)
solver.solve(solution,b)
v1sol,p1sol,v2sol,p2sol = solution.split()

#== Define exact solutions ==#
p1_ex = Function(pSpace)
p2_ex = Function(pSpace)
v1_ex = Function(vSpace)
v2_ex = Function(vSpace)
p1_exact = Expression("(1/pi)*exp(pi*x[0])*(sin(pi*x[1]) +\
           sin(pi*x[2])) - (1/(1.0*1.0))*(exp(3.316625*x[1]) +\
           exp(3.316625*x[2]))", degree = 5)
p2_exact = Expression("(1/pi)*exp(pi*x[0])*(sin(pi*x[1]) +\
           sin(pi*x[2])) + (1/(1*0.1))*(exp(3.316625*x[1]) +\
           exp(3.316625*x[2]))", degree = 5)
v1_exact = Expression(("-1*exp(pi*x[0])*(sin(pi*x[1]) +\
           sin(pi*x[2]))","-1*exp(pi*x[0])*cos(pi*x[1]) +\
           (3.316625/1.0)*exp(3.316625*x[1])","-1*exp(pi*x[0])*cos(pi*x[2]) +\
           (3.316625/1.0)*exp(3.316625*x[2])"), degree = 5)
v2_exact = Expression(("-0.1*exp(pi*x[0])*(sin(pi*x[1]) +\
           sin(pi*x[2]))","-0.1*exp(pi*x[0])*cos(pi*x[1]) -\
           (3.316625/1.0)*exp(3.316625*x[1])", "-0.1*exp(pi*x[0])*cos(pi*x[2]) -\
           (3.316625/1.0)*exp(3.316625*x[2])"), degree = 5)
p1_ex = project(p1_exact, pSpace)
v1_ex = project(v1_exact, vSpace)
p2_ex = project(p2_exact, pSpace)
v2_ex = project(v2_exact, vSpace)
#== L2 error norms ==#
L2_p1 = errornorm(p1_ex,p1sol,norm_type='L2',degree_rise= 3)
L2_v1 = errornorm(v1_ex,v1sol,norm_type='L2',degree_rise= 3)
L2_p2 = errornorm(p2_ex,p2sol,norm_type='L2',degree_rise= 3)
L2_v2 = errornorm(v2_ex,v2sol,norm_type='L2',degree_rise= 3)
\end{lstlisting}
\begin{lstlisting}[language=Python,caption=Firedrake code
for 3D problem with TET mesh using CG-VMS formulation , label=Code:ex1,frame=single]
from firedrake import *
import numpy as np

#== Create mesh ==#
mesh = BoxMesh(5,5,5,1,1,1)

#== Function spaces ==#
vSpace = VectorFunctionSpace(mesh,"CG",1)
pSpace = FunctionSpace(mesh,"CG",1)
wSpace = MixedFunctionSpace([vSpace,pSpace,vSpace,pSpace])

#== Define trial and test functions ==#
(v1,p1,v2,p2) = TrialFunctions(wSpace)
(w1,q1,w2,q2) = TestFunctions(wSpace)

#== Parameters and material properties ==#
rhob1, rhob2 = Constant((0.0,0.0,0.0)), Constant((0.0,0.0,0.0))
mu = Constant(1.0)
beta = Constant(1.0)                     
fact = 1.0  
k1, k2 = Constant(1.0), Constant(0.1)
alpha1, alpha2 = Constant(mu/k1), Constant(mu/k2)
eta = np.sqrt(1.0 *(1.0+0.1)/(1.0 * 0.1))
invalpha1 = 1.0 / alpha1
invalpha2 = 1.0 / alpha2

#== Boundary conditions ==#
p1_left = interpolate(Expression("(1/pi)*(sin(pi*x[1])+sin(pi*x[2])) -\
                      (exp(eta*x[1])+exp(eta*x[2]))",eta=eta) , pSpace)
p1_right = interpolate(Expression("(1/pi)*exp(pi)*(sin(pi*x[1])+sin(pi*x[2])) -\
                      (exp(eta*x[1])+exp(eta*x[2]))",eta=eta) , pSpace)
p1_bottom = interpolate(Expression("(1/pi)*exp(pi*x[0])*sin(pi*x[2]) -\
                              (1.0 + exp(eta*x[2]))",eta=eta) , pSpace)
p1_top = interpolate(Expression("(1/pi)*exp(pi*x[0])*sin(pi*x[2]) -\
                         (exp(eta) + exp(eta*x[2]))",eta=eta) , pSpace)
p1_back = interpolate(Expression("(1/pi)*exp(pi*x[0])*sin(pi*x[1]) -\
                              (exp(eta*x[1]) + 1.0)",eta=eta) , pSpace)
p1_front = interpolate(Expression("(1/pi)*exp(pi*x[0])*sin(pi*x[1]) -\
                         (exp(eta) + exp(eta*x[1]))",eta=eta) , pSpace)
p2_left = interpolate(Expression("(1/pi)*(sin(pi*x[1])+sin(pi*x[2])) +\
               10.0 * (exp(eta*x[1])+exp(eta*x[2]))",eta=eta) , pSpace)
p2_right = interpolate(Expression("(1/pi)*exp(pi)*(sin(pi*x[1])+sin(pi*x[2])) +\
               10.0 * (exp(eta*x[1])+exp(eta*x[2]))",eta=eta) , pSpace)
p2_bottom = interpolate(Expression("(1/pi)*exp(pi*x[0])*sin(pi*x[2]) +\
                       10.0 * (1.0 + exp(eta*x[2]))",eta=eta) , pSpace)
p2_top = interpolate(Expression("(1/pi)*exp(pi*x[0])*sin(pi*x[2]) +\
                  10.0 * (exp(eta) + exp(eta*x[2]))",eta=eta) , pSpace)
p2_back = interpolate(Expression("(1/pi)*exp(pi*x[0])*sin(pi*x[1]) +\
                      10.0 *  (exp(eta*x[1]) + 1.0)",eta=eta) , pSpace)
p2_front = interpolate(Expression("(1/pi)*exp(pi*x[0])*sin(pi*x[1]) +\
                  10.0 * (exp(eta) + exp(eta*x[1]))",eta=eta) , pSpace)
bcs = []

#== Normal vectors ==#
n = FacetNormal(mesh)

#== Define variational forms ==#
a = dot(w1, alpha1*v1)*dx + dot(w2, alpha2*v2)*dx \
    - div(w1) * p1 *dx - div(w2) * p2 * dx \
    + q1 * div(v1) * dx + q2 * div(v2) * dx +\
    q1 * fact * (p1 - p2) * dx - q2 * fact * (p1 - p2) * dx -\
    0.5 * dot( alpha1 * w1 - grad(q1), invalpha1 * (alpha1 * v1 + grad(p1)) ) * dx -\
    0.5 * dot( alpha2 * w2 - grad(q2), invalpha2 * (alpha2 * v2 + grad(p2)) ) * dx 

L = dot(w1,rhob1)*dx + dot(w2,rhob2)*dx -\
    0.5 * dot( alpha1 * w1 - grad(q1), invalpha1 * rhob1 ) * dx -\
    0.5 * dot( alpha2 * w2 - grad(q2), invalpha2 * rhob2 ) * dx -\
    dot(w1,n) * p1_left * ds(1) - dot(w2,n) * p2_left * ds(1) -\
    dot(w1,n) * p1_right * ds(2) - dot(w2,n) * p2_right * ds(2) -\
    dot(w1,n) * p1_bottom * ds(3) - dot(w2,n) * p2_bottom * ds(3) -\
    dot(w1,n) * p1_top * ds(4) - dot(w2,n) * p2_top * ds(4) -\
    dot(w1,n) * p1_back * ds(5) - dot(w2,n) * p2_back * ds(5) -\
    dot(w1,n) * p1_front * ds(6) - dot(w2,n) * p2_front * ds(6)

#== Solver options ==#
parameters_twofields = {
  "ksp_type": "gmres",
  "pc_type": "fieldsplit",
  "pc_fieldsplit_0_fields": "0,2",
  "pc_fieldsplit_1_fields": "1,3",
  "pc_fieldsplit_type": "schur",
  "pc_fieldsplit_schur_fact_type": "full",
  "pc_fieldsplit_schur_precondition": "selfp",
  "fieldsplit_0_ksp_type": "preonly",
  "fieldsplit_0_pc_type": "bjacobi",
  "fieldsplit_1_ksp_type": "preonly",
  "fieldsplit_1_pc_type": "fieldsplit",
  "fieldsplit_1_pc_fieldsplit_type": "additive",
  "fieldsplit_1_fieldsplit_0_ksp_type": "preonly",
  "fieldsplit_1_fieldsplit_0_pc_type": "hypre",
  "fieldsplit_1_fieldsplit_0_pc_hypre_boomeramg_strong_threshold": 0.75,
  "fieldsplit_1_fieldsplit_0_pc_hypre_boomeramg_agg_nl": 2,
  "fieldsplit_1_fieldsplit_1_ksp_type": "preonly",
  "fieldsplit_1_fieldsplit_1_pc_type": "hypre",
  "fieldsplit_1_fieldsplit_1_pc_hypre_boomeramg_strong_threshold": 0.75,
  "fieldsplit_1_fieldsplit_1_pc_hypre_boomeramg_agg_nl": 2,
  "ksp_rtol": 1e-5
}

#== Solve problem ==#
solution = Function(wSpace)
A = assemble(a, bcs=bcs, mat_type='aij')
b = assemble(L)
solver = LinearSolver(A,P=None,options_prefix="twofields_",\
                          solver_parameters=parameters_twofields)
solver.solve(solution,b)
v1sol,p1sol,v2sol,p2sol = solution.split()

#== Define exact solutions ==#
p1_ex = Function(pSpace)
p2_ex = Function(pSpace)
v1_ex = Function(vSpace)
v2_ex = Function(vSpace)
p1_exact = Expression("(1/pi)*exp(pi*x[0])*(sin(pi*x[1]) +\
           sin(pi*x[2])) - (1/(1.0*1.0))*(exp(3.316625*x[1]) +\
           exp(3.316625*x[2]))", degree = 5)
p2_exact = Expression("(1/pi)*exp(pi*x[0])*(sin(pi*x[1]) +\
           sin(pi*x[2])) + (1/(1*0.1))*(exp(3.316625*x[1]) +\
           exp(3.316625*x[2]))", degree = 5)
v1_exact = Expression(("-1*exp(pi*x[0])*(sin(pi*x[1]) +\
           sin(pi*x[2]))","-1*exp(pi*x[0])*cos(pi*x[1]) +\
           (3.316625/1.0)*exp(3.316625*x[1])","-1*exp(pi*x[0])*cos(pi*x[2]) +\
           (3.316625/1.0)*exp(3.316625*x[2])"), degree = 5)
v2_exact = Expression(("-0.1*exp(pi*x[0])*(sin(pi*x[1]) +\
           sin(pi*x[2]))","-0.1*exp(pi*x[0])*cos(pi*x[1]) -\
           (3.316625/1.0)*exp(3.316625*x[1])", "-0.1*exp(pi*x[0])*cos(pi*x[2]) -\
           (3.316625/1.0)*exp(3.316625*x[2])"), degree = 5)
p1_ex = interpolate(p1_exact, pSpace)
v1_ex = interpolate(v1_exact, vSpace)
p2_ex = interpolate(p2_exact, pSpace)
v2_ex = interpolate(v2_exact, vSpace)

#== L2 error norms ==#
L2_p1 = errornorm(p1_ex,p1sol,norm_type='L2',degree_rise= 3)
L2_v1 = errornorm(v1_ex,v1sol,norm_type='L2',degree_rise= 3)
L2_p2 = errornorm(p2_ex,p2sol,norm_type='L2',degree_rise= 3)
L2_v2 = errornorm(v2_ex,v2sol,norm_type='L2',degree_rise= 3)
\end{lstlisting}
\begin{lstlisting}[language=Python,caption=Firedrake code
for 3D problem with TET mesh using DG-VMS formulation, label=Code:ex2,
frame=single]
from firedrake import *
import numpy as np

#== Create mesh ==#
mesh = BoxMesh(5,5,5,1,1,1)

#== Function spaces ==#
vSpace = VectorFunctionSpace(mesh,"DG",1)
pSpace = FunctionSpace(mesh,"DG",1)
wSpace = MixedFunctionSpace([vSpace,pSpace,vSpace,pSpace])

#== Define trial and test functions ==#
(v1,p1,v2,p2) = TrialFunctions(wSpace)
(w1,q1,w2,q2) = TestFunctions(wSpace)

#== Parameters and material properties ==#
rhob1, rhob2 = Constant((0.0,0.0,0.0)), Constant((0.0,0.0,0.0))
mu = Constant(1.0)
beta = Constant(1.0) 
fact = 1.0                 
k1, k2 = Constant(1.0), Constant(0.1)
alpha1, alpha2 = Constant(mu/k1), Constant(mu/k2)
eta = np.sqrt(1.0 *(1.0+0.1)/(1.0 * 0.1))
invalpha1 = 1.0 / alpha1
invalpha2 = 1.0 / alpha2

#== Boundary conditions  ==#
p1_left = interpolate(Expression("(1/pi)*(sin(pi*x[1])+sin(pi*x[2])) -\
                      (exp(eta*x[1])+exp(eta*x[2]))",eta=eta) , pSpace)
p1_right = interpolate(Expression("(1/pi)*exp(pi)*(sin(pi*x[1])+sin(pi*x[2])) -\
                      (exp(eta*x[1])+exp(eta*x[2]))",eta=eta) , pSpace)
p1_bottom = interpolate(Expression("(1/pi)*exp(pi*x[0])*sin(pi*x[2]) -\
                              (1.0 + exp(eta*x[2]))",eta=eta) , pSpace)
p1_top = interpolate(Expression("(1/pi)*exp(pi*x[0])*sin(pi*x[2]) -\
                         (exp(eta) + exp(eta*x[2]))",eta=eta) , pSpace)
p1_back = interpolate(Expression("(1/pi)*exp(pi*x[0])*sin(pi*x[1]) -\
                              (exp(eta*x[1]) + 1.0)",eta=eta) , pSpace)
p1_front = interpolate(Expression("(1/pi)*exp(pi*x[0])*sin(pi*x[1]) -\
                         (exp(eta) + exp(eta*x[1]))",eta=eta) , pSpace)
p2_left = interpolate(Expression("(1/pi)*(sin(pi*x[1])+sin(pi*x[2])) +\
               10.0 * (exp(eta*x[1])+exp(eta*x[2]))",eta=eta) , pSpace)
p2_right = interpolate(Expression("(1/pi)*exp(pi)*(sin(pi*x[1])+sin(pi*x[2])) +\
               10.0 * (exp(eta*x[1])+exp(eta*x[2]))",eta=eta) , pSpace)
p2_bottom = interpolate(Expression("(1/pi)*exp(pi*x[0])*sin(pi*x[2]) +\
                       10.0 * (1.0 + exp(eta*x[2]))",eta=eta) , pSpace)
p2_top = interpolate(Expression("(1/pi)*exp(pi*x[0])*sin(pi*x[2]) +\
                  10.0 * (exp(eta) + exp(eta*x[2]))",eta=eta) , pSpace)
p2_back = interpolate(Expression("(1/pi)*exp(pi*x[0])*sin(pi*x[1]) +\
                      10.0 *  (exp(eta*x[1]) + 1.0)",eta=eta) , pSpace)
p2_front = interpolate(Expression("(1/pi)*exp(pi*x[0])*sin(pi*x[1]) +\
                  10.0 * (exp(eta) + exp(eta*x[1]))",eta=eta) , pSpace)
bcs = []

#== Define normal vector, h_avg, and penalty parameters ==#
n = FacetNormal(mesh)
h = CellSize(mesh)
h_avg = (h('+') + h('-'))/2
eta_u, eta_p = Constant(10.), Constant(10.)

#== Define variational forms ==#
a = dot(w1, alpha1*v1)*dx + dot(w2, alpha2*v2)*dx \
    - div(w1) * p1 *dx - div(w2) * p2 * dx \
    + q1 * div(v1) * dx + q2 * div(v2) * dx +\
    q1 * fact * (p1 - p2) * dx - q2 * fact * (p1 - p2) * dx -\
    0.5 * dot( alpha1 * w1 - grad(q1),invalpha1 * (alpha1 * v1 + grad(p1)) ) * dx -\
    0.5 * dot( alpha2 * w2 - grad(q2),invalpha2 * (alpha2 * v2 + grad(p2)) ) * dx +\
    jump(w1,n) * avg(p1) * dS + jump(w2,n) * avg(p2) * dS -\
    avg(q1) * jump(v1,n) * dS - avg(q2) * jump(v2,n) * dS +\
    eta_u * h_avg * avg(alpha1) * (jump(w1,n) * jump(v1,n)) * dS +\
    eta_u * h_avg * avg(alpha2) * (jump(w2,n) * jump(v2,n)) * dS +\
    (eta_p/h_avg) * avg(1/alpha1) * dot(jump(q1,n),jump(p1,n)) * dS +\
    (eta_p / h_avg) * avg(1 / alpha2) * dot(jump(q2,n),jump(p2,n)) * dS
	
L = dot(w1,rhob1)*dx + dot(w2,rhob2)*dx -\
     0.5 * dot( alpha1 * w1 - grad(q1),invalpha1 * rhob1 ) * dx -\
     0.5 * dot( alpha2 * w2 - grad(q2),invalpha2 * rhob2 ) * dx -\
     dot(w1,n) * p1_left * ds(1) - dot(w2,n) * p2_left * ds(1) -\
     dot(w1,n) * p1_right * ds(2) - dot(w2,n) * p2_right * ds(2) -\
     dot(w1,n) * p1_bottom * ds(3) - dot(w2,n) * p2_bottom * ds(3) -\
     dot(w1,n) * p1_top * ds(4) - dot(w2,n) * p2_top * ds(4) -\
     dot(w1,n) * p1_back * ds(5) - dot(w2,n) * p2_back * ds(5) -\
     dot(w1,n) * p1_front * ds(6) - dot(w2,n) * p2_front * ds(6)

#== Solver options ==#
parameters_twofields = {
  "ksp_type": "gmres",
  "pc_type": "fieldsplit",
  "pc_fieldsplit_0_fields": "0,2",
  "pc_fieldsplit_1_fields": "1,3",
  "pc_fieldsplit_type": "schur",
  "pc_fieldsplit_schur_fact_type": "full",
  "pc_fieldsplit_schur_precondition": "selfp",
  "fieldsplit_0_ksp_type": "preonly",
  "fieldsplit_0_pc_type": "bjacobi",
  "fieldsplit_1_ksp_type": "preonly",
  "fieldsplit_1_pc_type": "fieldsplit",
  "fieldsplit_1_pc_fieldsplit_type": "additive",
  "fieldsplit_1_fieldsplit_0_ksp_type": "preonly",
  "fieldsplit_1_fieldsplit_0_pc_type": "hypre",
  "fieldsplit_1_fieldsplit_0_pc_hypre_boomeramg_strong_threshold": 0.75,
  "fieldsplit_1_fieldsplit_0_pc_hypre_boomeramg_agg_nl": 2,
  "fieldsplit_1_fieldsplit_1_ksp_type": "preonly",
  "fieldsplit_1_fieldsplit_1_pc_type": "hypre",
  "fieldsplit_1_fieldsplit_1_pc_hypre_boomeramg_strong_threshold": 0.75,
  "fieldsplit_1_fieldsplit_1_pc_hypre_boomeramg_agg_nl": 2,
  "ksp_rtol": 1e-5
}

#== Solve problem ==#
solution = Function(wSpace)
A = assemble(a, bcs=bcs, mat_type='aij')
b = assemble(L)
solver = LinearSolver(A,P=None,options_prefix="twofields_",\
                          solver_parameters=parameters_twofields)
solver.solve(solution,b)
v1sol,p1sol,v2sol,p2sol = solution.split()

#== Define exact solutions ==#
p1_ex = Function(pSpace)
p2_ex = Function(pSpace)
v1_ex = Function(vSpace)
v2_ex = Function(vSpace)
p1_exact = Expression("(1/pi)*exp(pi*x[0])*(sin(pi*x[1]) +\
           sin(pi*x[2])) - (1/(1.0*1.0))*(exp(3.316625*x[1]) +\
           exp(3.316625*x[2]))", degree = 5)
p2_exact = Expression("(1/pi)*exp(pi*x[0])*(sin(pi*x[1]) +\
           sin(pi*x[2])) + (1/(1*0.1))*(exp(3.316625*x[1]) +\
           exp(3.316625*x[2]))", degree = 5)
v1_exact = Expression(("-1*exp(pi*x[0])*(sin(pi*x[1]) +\
           sin(pi*x[2]))","-1*exp(pi*x[0])*cos(pi*x[1]) +\
           (3.316625/1.0)*exp(3.316625*x[1])","-1*exp(pi*x[0])*cos(pi*x[2]) +\
           (3.316625/1.0)*exp(3.316625*x[2])"), degree = 5)
v2_exact = Expression(("-0.1*exp(pi*x[0])*(sin(pi*x[1]) +\
           sin(pi*x[2]))","-0.1*exp(pi*x[0])*cos(pi*x[1]) -\
           (3.316625/1.0)*exp(3.316625*x[1])", "-0.1*exp(pi*x[0])*cos(pi*x[2]) -\
           (3.316625/1.0)*exp(3.316625*x[2])"), degree = 5)
p1_ex = interpolate(p1_exact, pSpace)
v1_ex = interpolate(v1_exact, vSpace)
p2_ex = interpolate(p2_exact, pSpace)
v2_ex = interpolate(v2_exact, vSpace)

#== L2 error norms ==# 
L2_p1 = errornorm(p1_ex,p1sol,norm_type='L2',degree_rise= 3)
L2_v1 = errornorm(v1_ex,v1sol,norm_type='L2',degree_rise= 3)
L2_p2 = errornorm(p2_ex,p2sol,norm_type='L2',degree_rise= 3)
L2_v2 = errornorm(v2_ex,v2sol,norm_type='L2',degree_rise= 3)
\end{lstlisting}

\section*{ACKNOWLEDGMENTS}
KBN acknowledges the support through the 
\emph{High Priority Area Research Seed Grant} 
from the Division of Research, University of Houston.

\bibliographystyle{plainnat}
\bibliography{Master_References/Books,Master_References/Master_References}
\newpage
\begin{figure}
  \includegraphics[clip,width=0.9\linewidth]{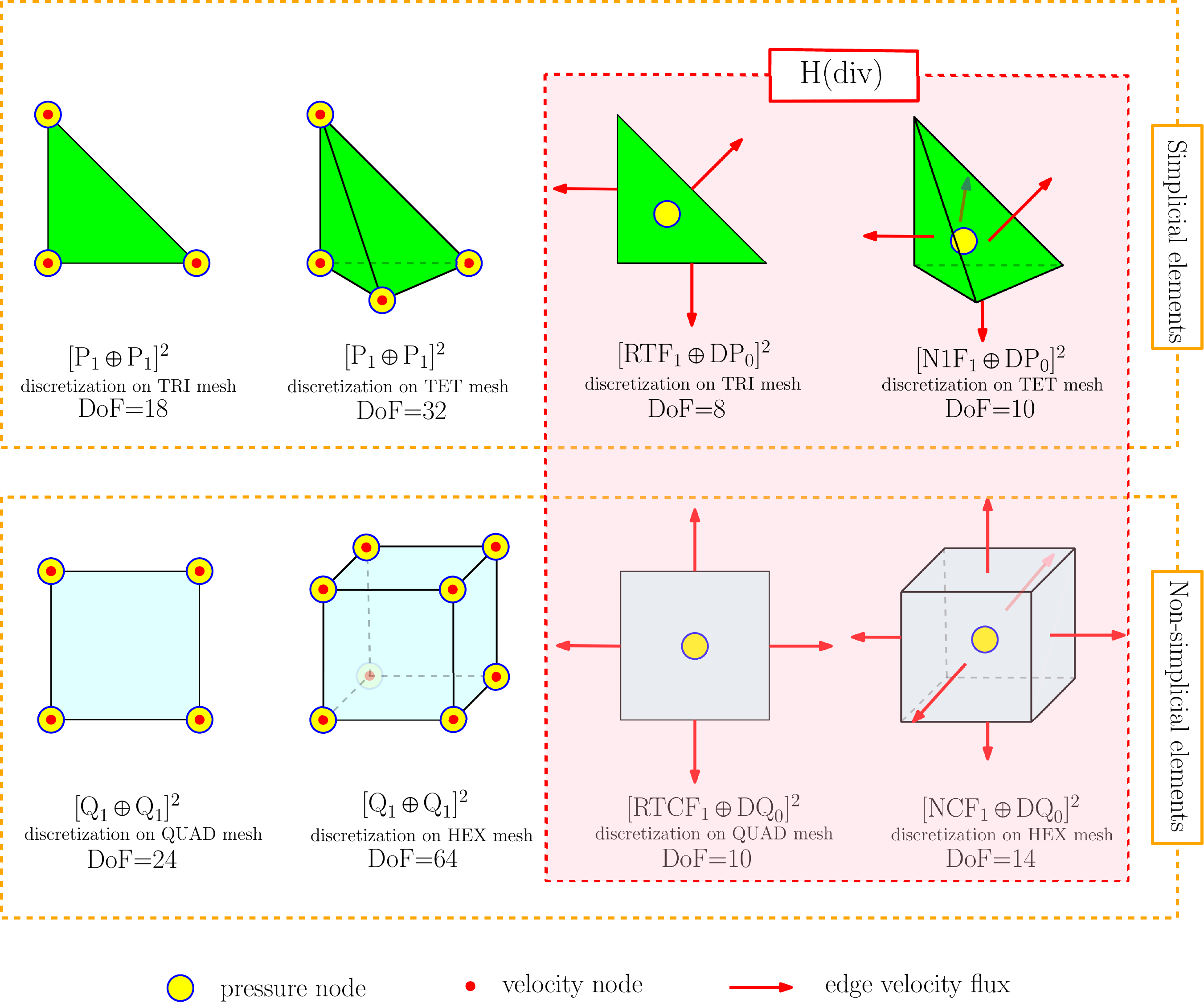}
  \caption{This figure shows the two-dimensional
    and three-dimensional elements that are employed
    in this paper. The degrees-of-freedom (DoF) for
    each element are also indicated. Note that at
    a velocity node, there are two sets of DoFs,
    one for each of the macro- and micro-velocities.
    A similar case exists for each velocity flux
    face and for each pressure node. The notation
    used to denote the discretizations is based
    on the \emph{Periodic Table of the Finite
      Elements} \citep{arnold2014periodic}.
    \label{Fig:elements}}
\end{figure}

\begin{figure}
  \centering
  \sbox{\measurebox}{%
    \begin{minipage}[b]{.4\textwidth}
      \subfigure
	  [Schematic]
	  {\label{Fig1:2D_BVP}\includegraphics[scale=0.45]{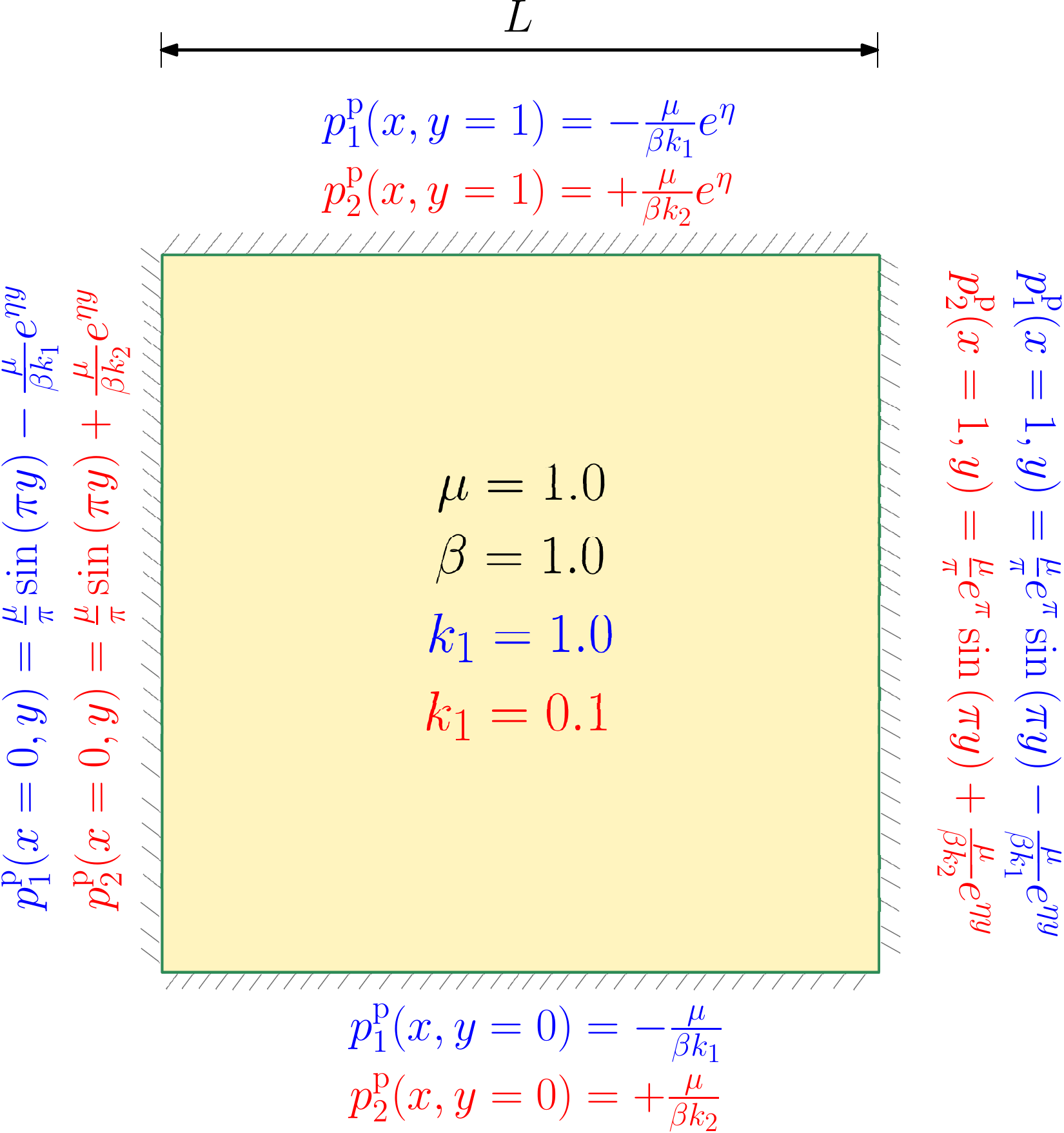}}
  \end{minipage}}
  \usebox{\measurebox}\qquad
  \begin{minipage}[b][\ht\measurebox][s]{.4\textwidth}
    \centering
    \subfigure
	[TRI mesh]
	{\label{Fig1:Mesh_T3}\includegraphics[scale=0.35]{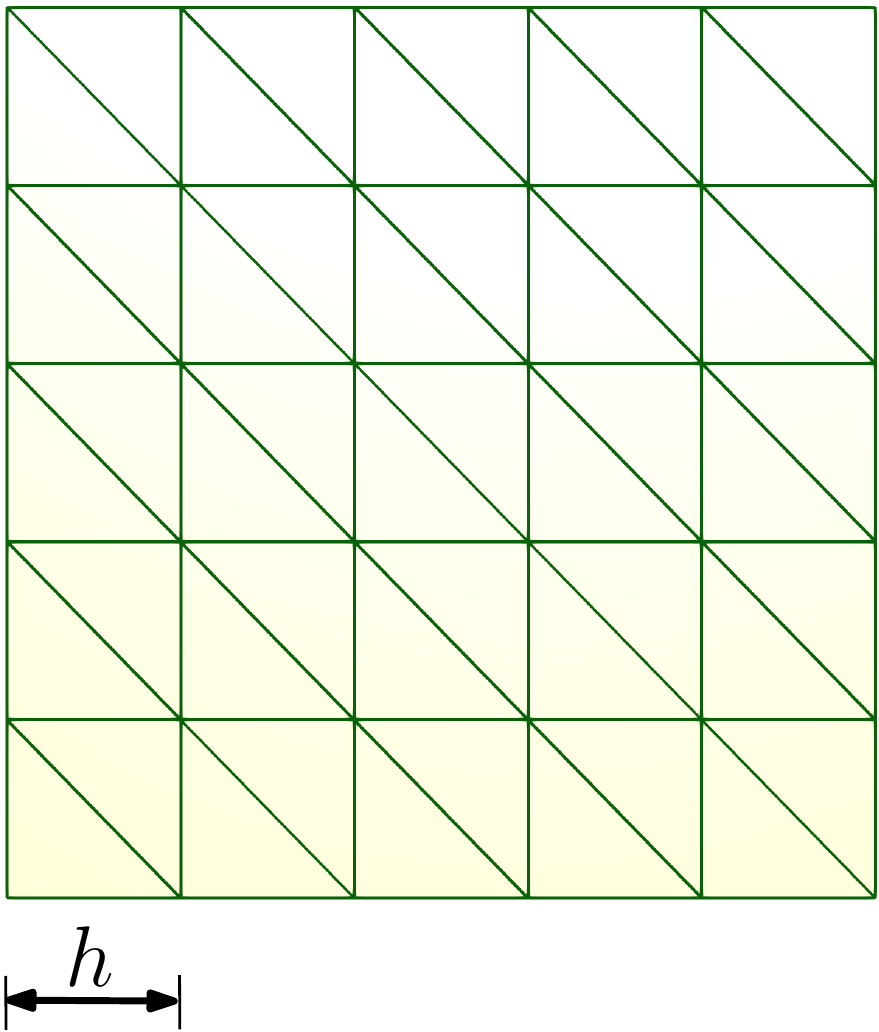}}		
	\vfill		
	\subfigure
	    [QUAD mesh]
	    {\label{Fig1:Mesh_Q4}\includegraphics[scale=0.35]{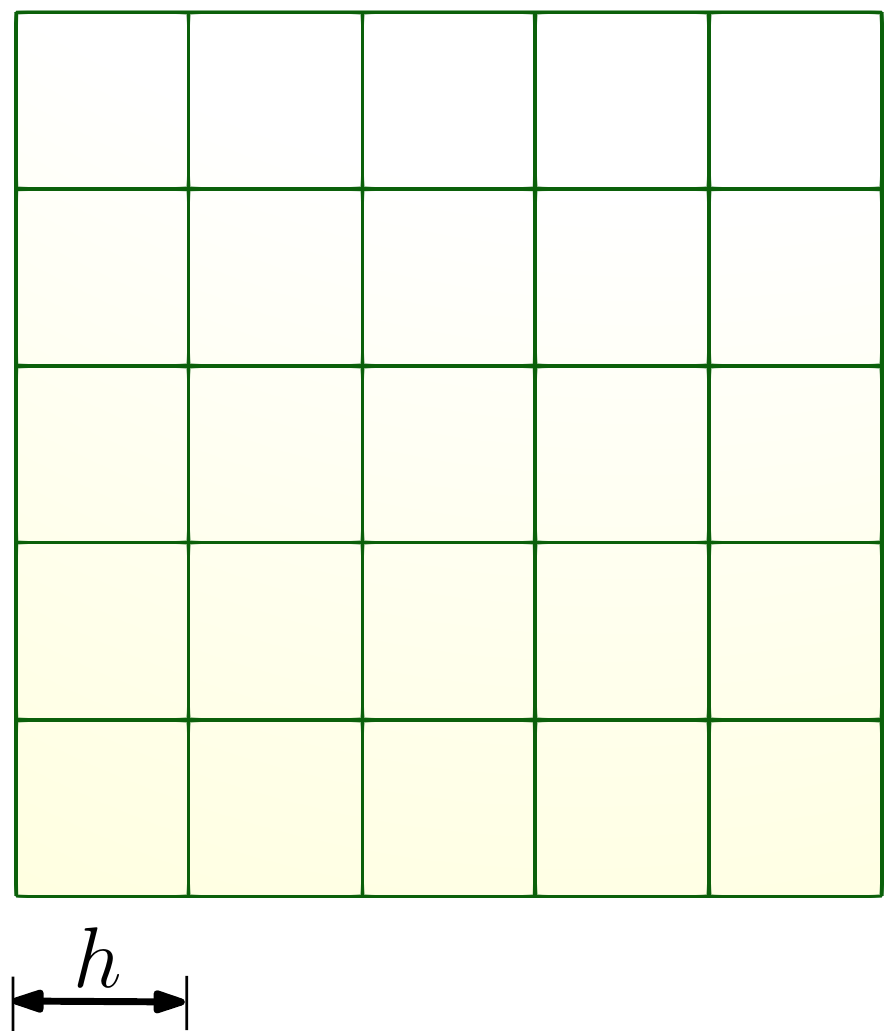}}
  \end{minipage}
  \caption{\textsf{Two-dimensional problem:} 
    This figure provides a pictorial description
    of the boundary value problem and shows the
    typical meshes employed in our numerical
    simulations.\label{Fig:2D_schematic}}
\end{figure}

\begin{figure}
	\subfigure[Macro-pressure  \label{p1_convergence_2D_T3}]{
		\includegraphics[clip,scale=0.37,trim=0 1.5cm 9cm 0]{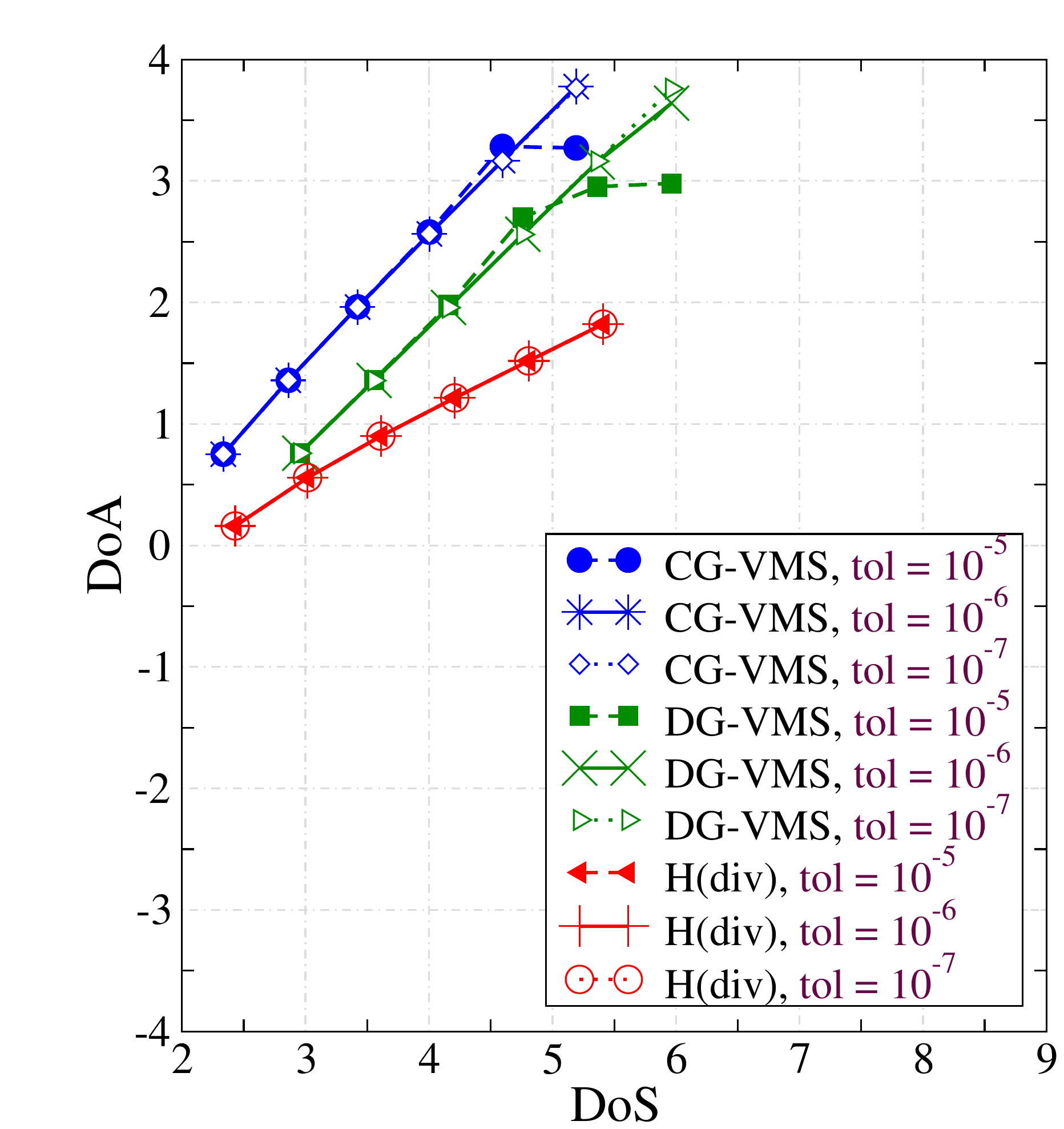}}
	\hspace{1 cm}
	\subfigure[Micro-pressure \label{p2_convergence_2D_T3}]{
		\includegraphics[clip,scale=0.37,trim=0 1.5cm 9cm 0]{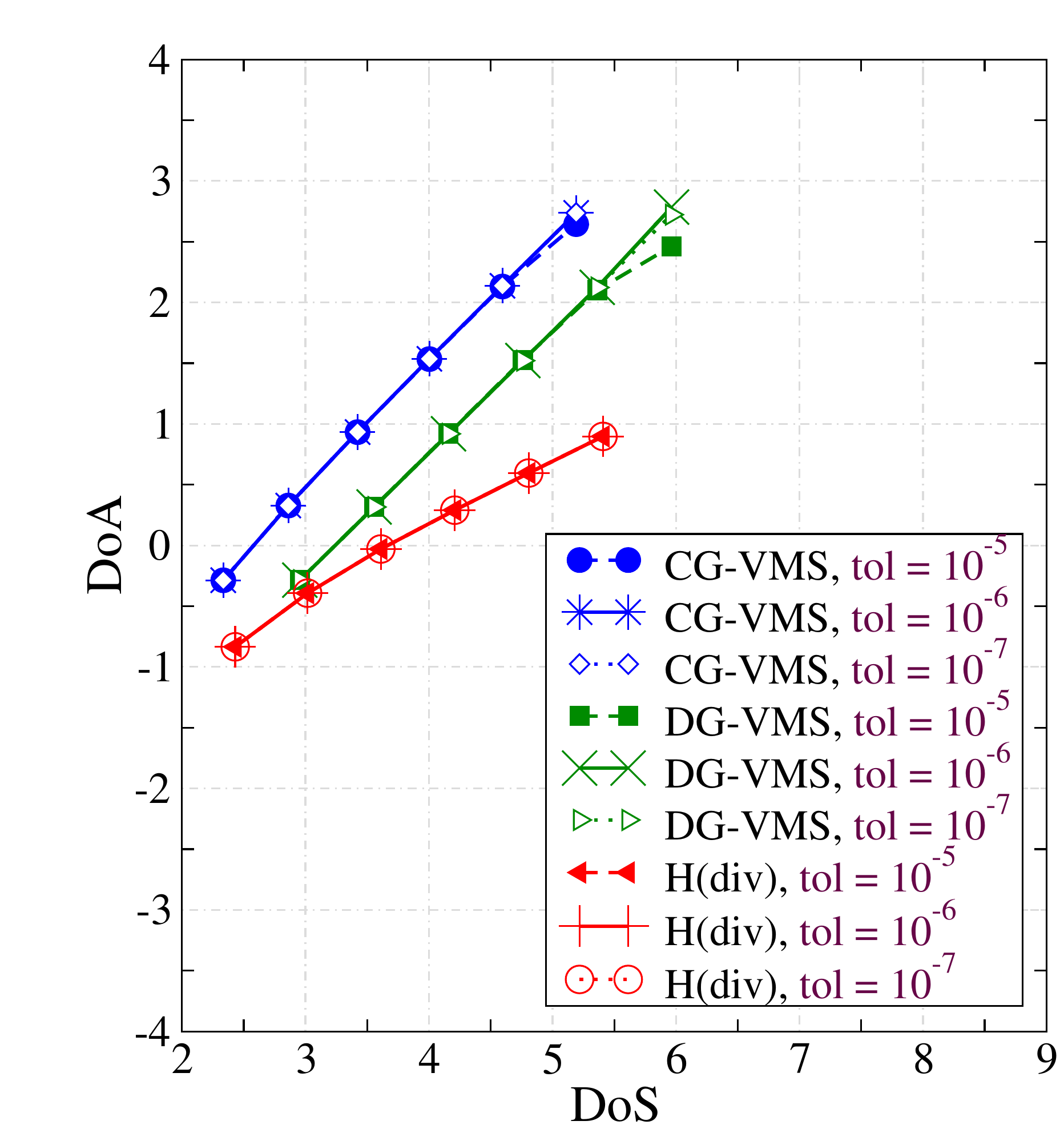}} \\
	\vspace{0.75 cm}
	\subfigure[Macro-velocity \label{v1_convergence_2D_T3}]{
		\includegraphics[clip,scale=0.37,trim=0 1.5cm 9cm 03]{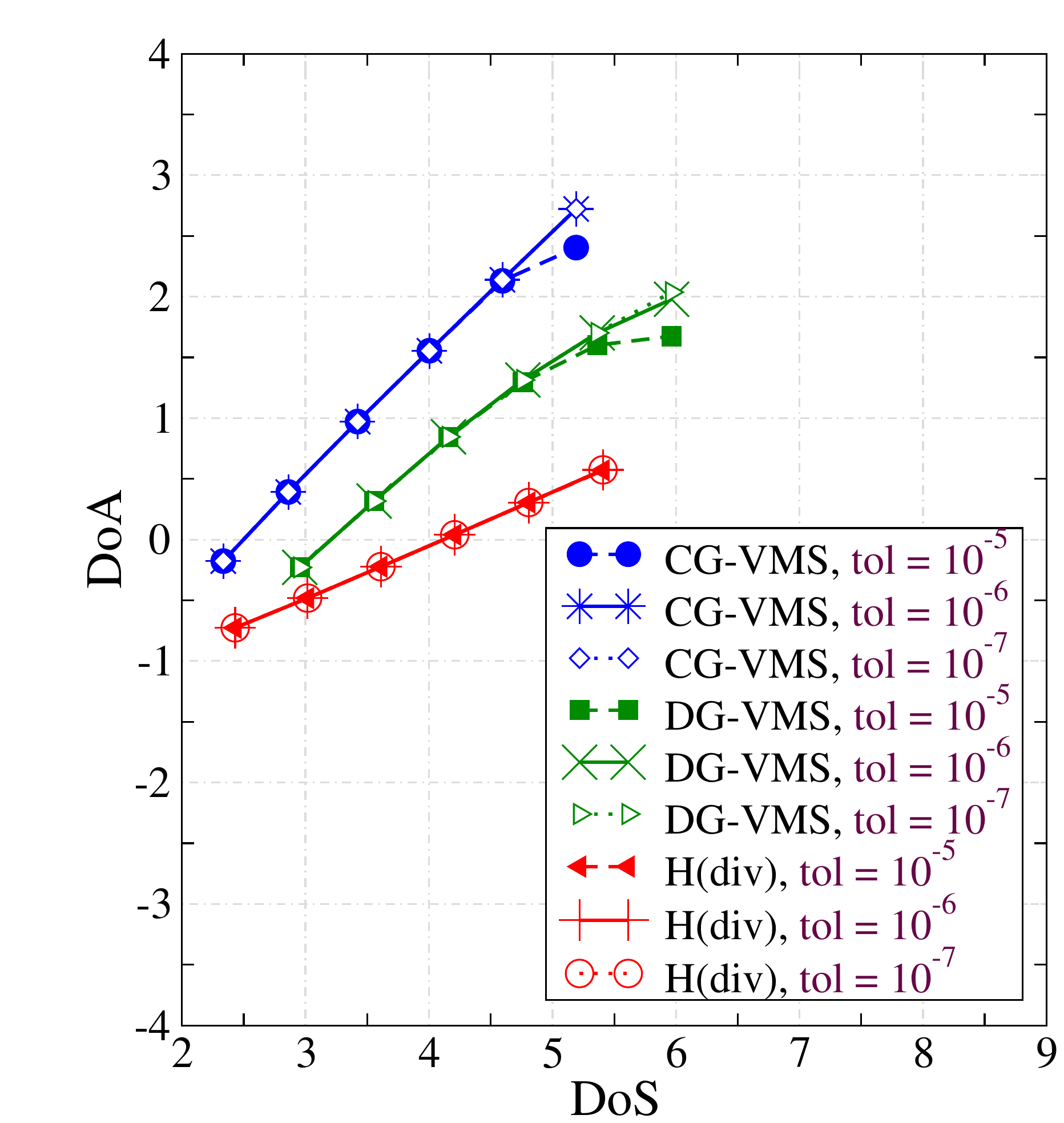}}
	\hspace{1 cm}
	\subfigure[Micro-velocity \label{v2_convergence_2D_T3}]{
		\includegraphics[clip,scale=0.37,trim=0 1.5cm 9cm 0]{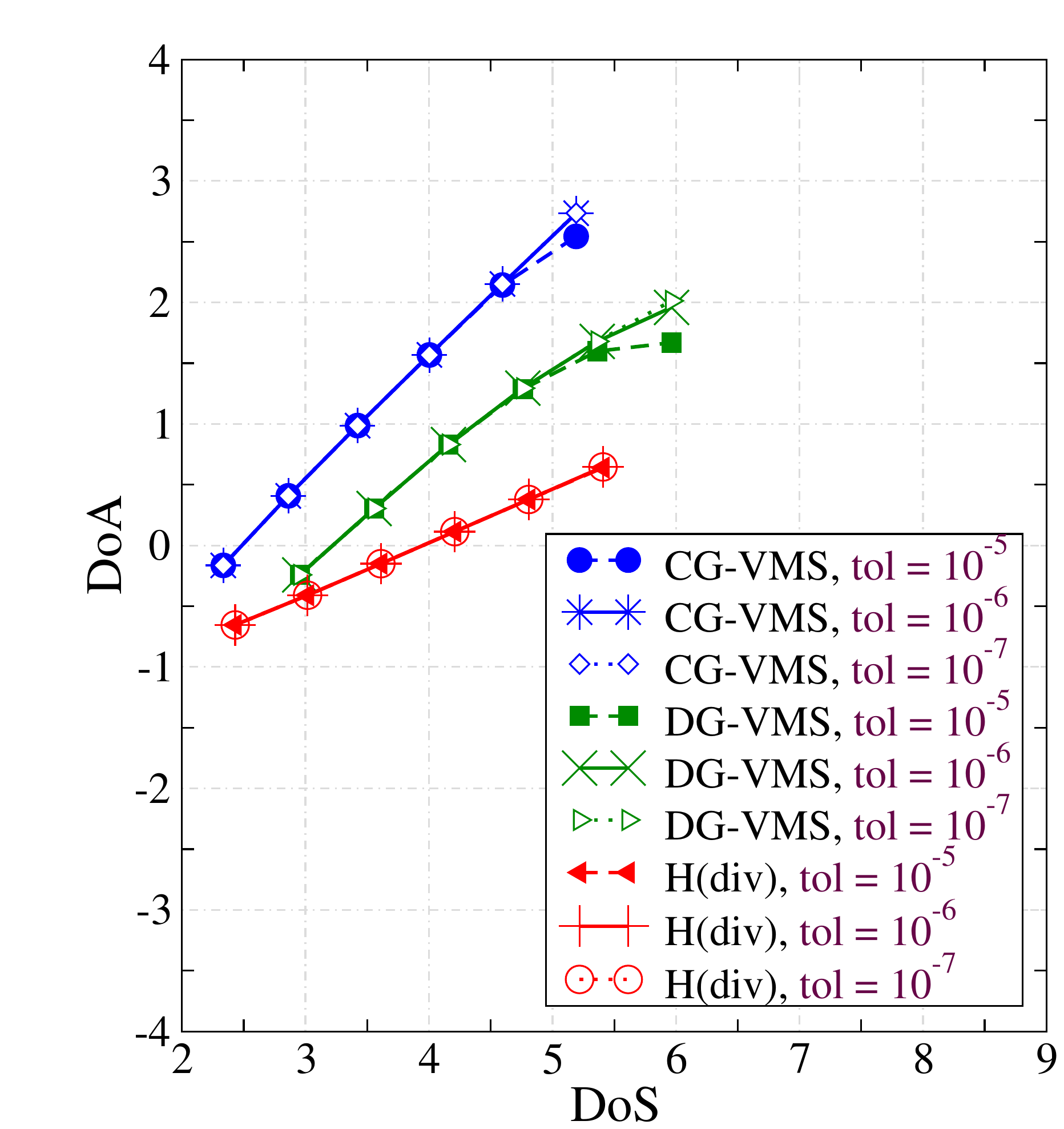}}
	\caption{\textsf{Two-dimensional problem using
            \textbf{TRI} mesh:} 		
	  This figure compares the mesh convergence results
	  for the chosen finite element formulations under
          various solver tolerances. The results are shown
          for field-splitting block solver methodology, but very
          similar results are also obtained under the
          scale-splitting solver (which are not shown for
          brevity). The two main inferences are:
	  (i) The VMS formulations exhibit a slope of
          approximately $1$, while H(div) formulation returns
          a slope of approximately $0.5$ for all fields.
          This indicates that each DoF in the CG- and DG-VMS
          formulations achieves higher overall numerical
          accuracy than the H(div) formulation.
          (ii) Adequate decrease in solver tolerance
          will ensure that the DoA will not flatten as
          the size (which is quantified by DoS) increases.\label{Fig:2D_convergence_T3}}
\end{figure}
\begin{figure}
	\subfigure[Macro-pressure  \label{p1_convergence_2D_Q4}]{
		\includegraphics[clip,scale=0.37,trim=0 1.25cm 9cm 0]{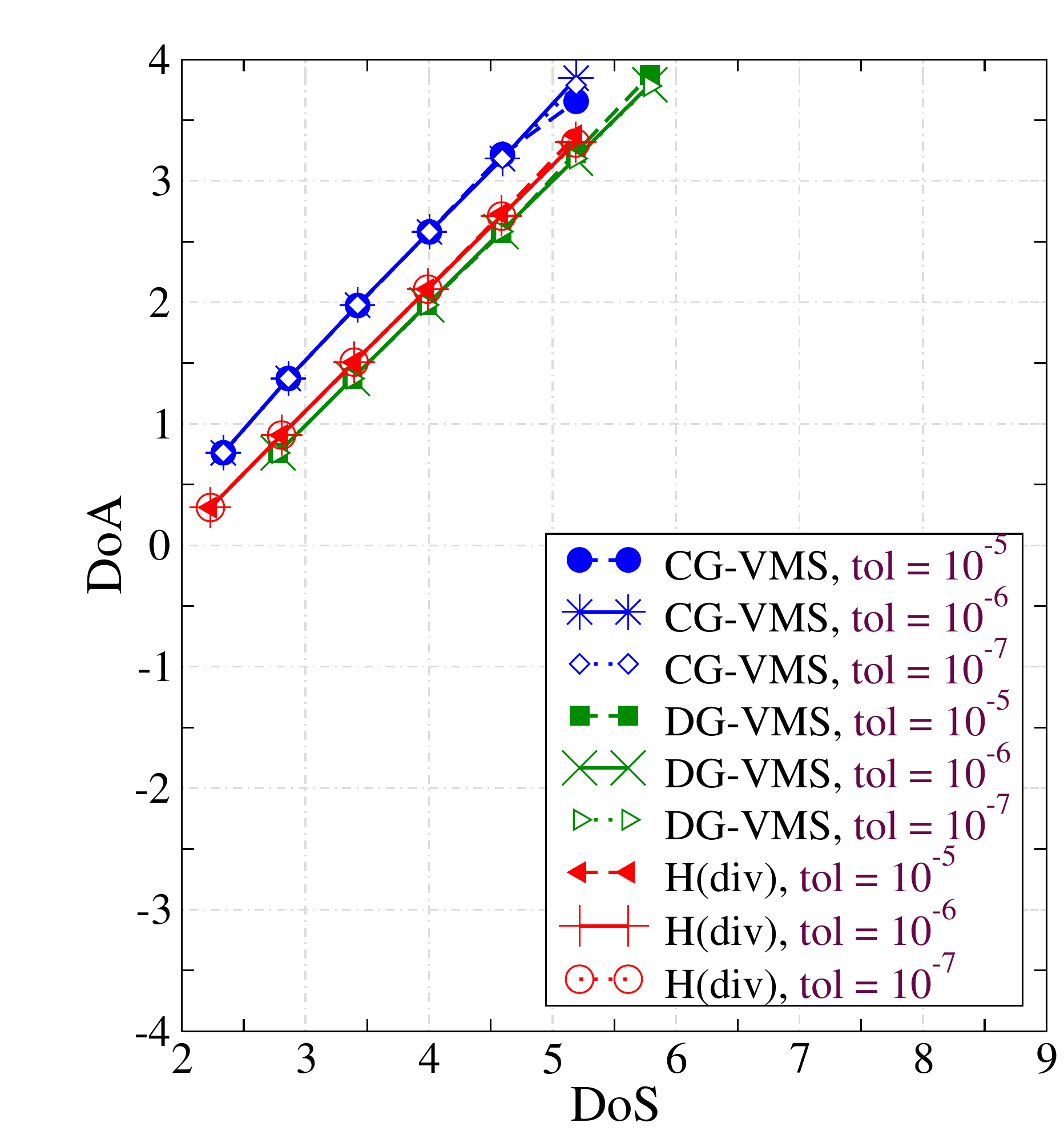}}
	\hspace{1 cm}
	\subfigure[Micro-pressure \label{p2_convergence_2D_Q4}]{
		\includegraphics[clip,scale=0.37,trim=0 1.25cm 9cm 0]{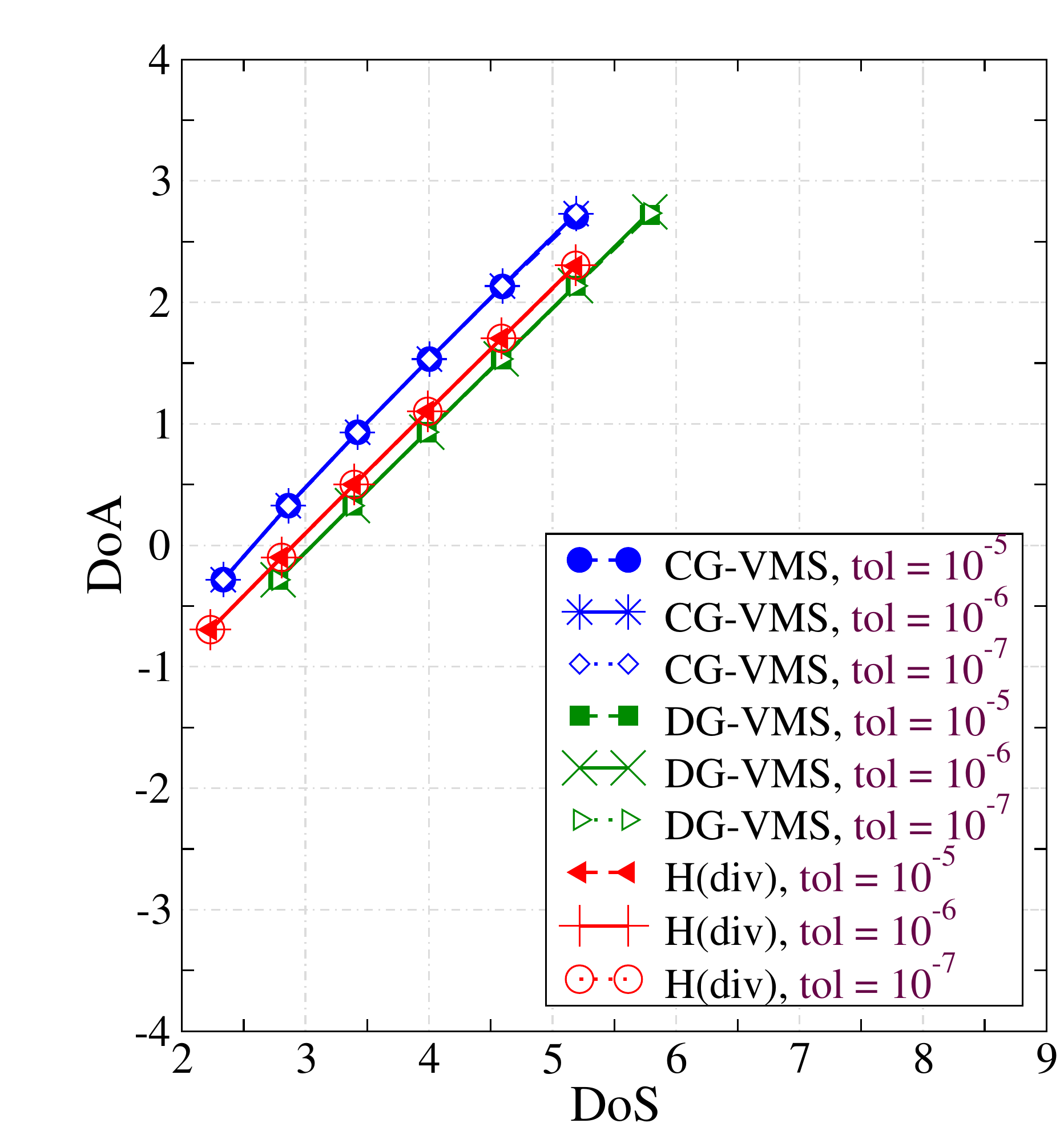}} \\
	\vspace{0.75 cm}
	\subfigure[Macro-velocity \label{v1_convergence_2D_Q4}]{
		
		\includegraphics[clip,scale=0.37,trim=0 1.25cm 9cm 03]{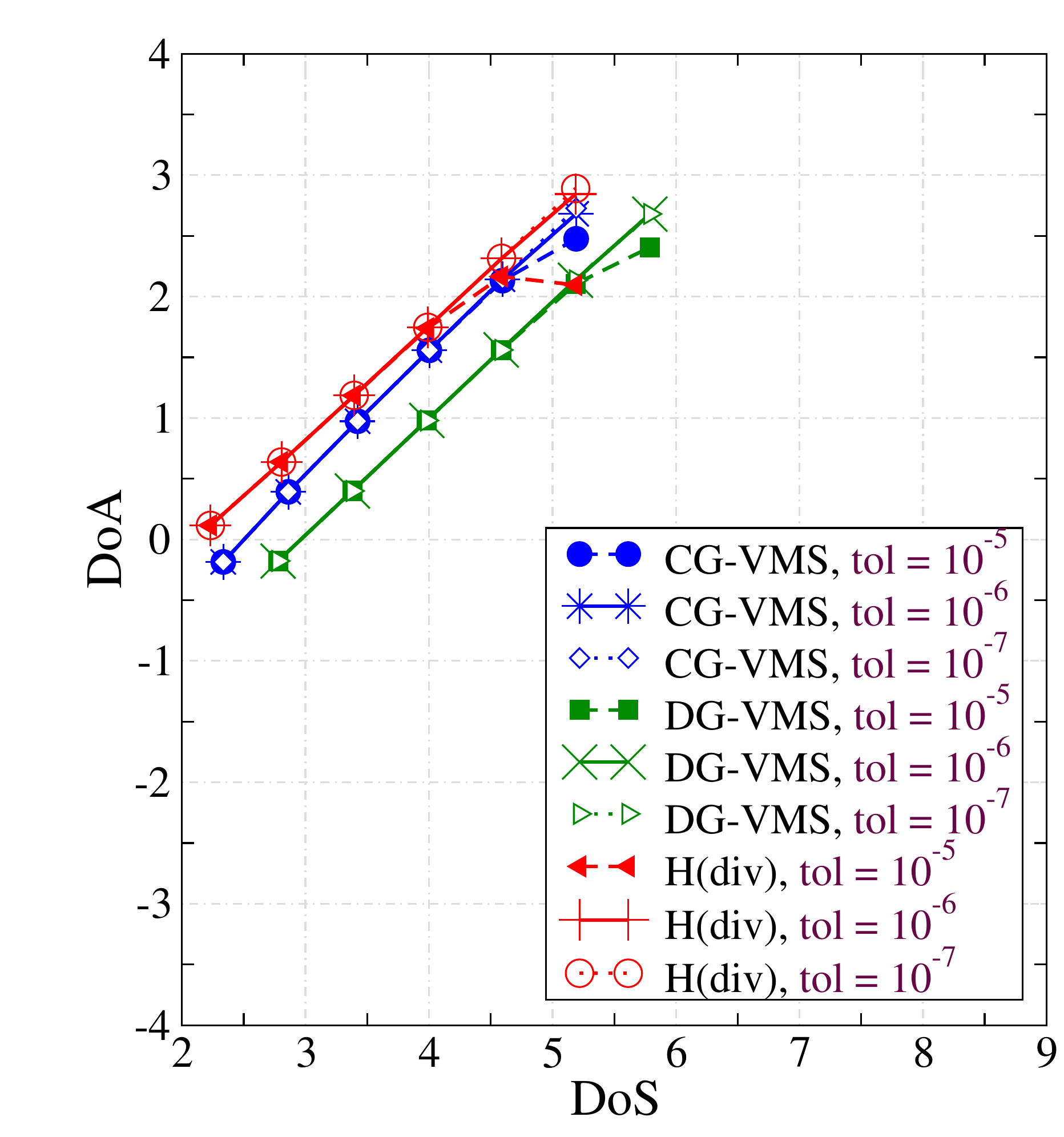}}
	\hspace{1 cm}
	\subfigure[Micro-velocity \label{v2_convergence_2D_Q4}]{
		\includegraphics[clip,scale=0.37,trim=0 1.25cm 9cm 0]{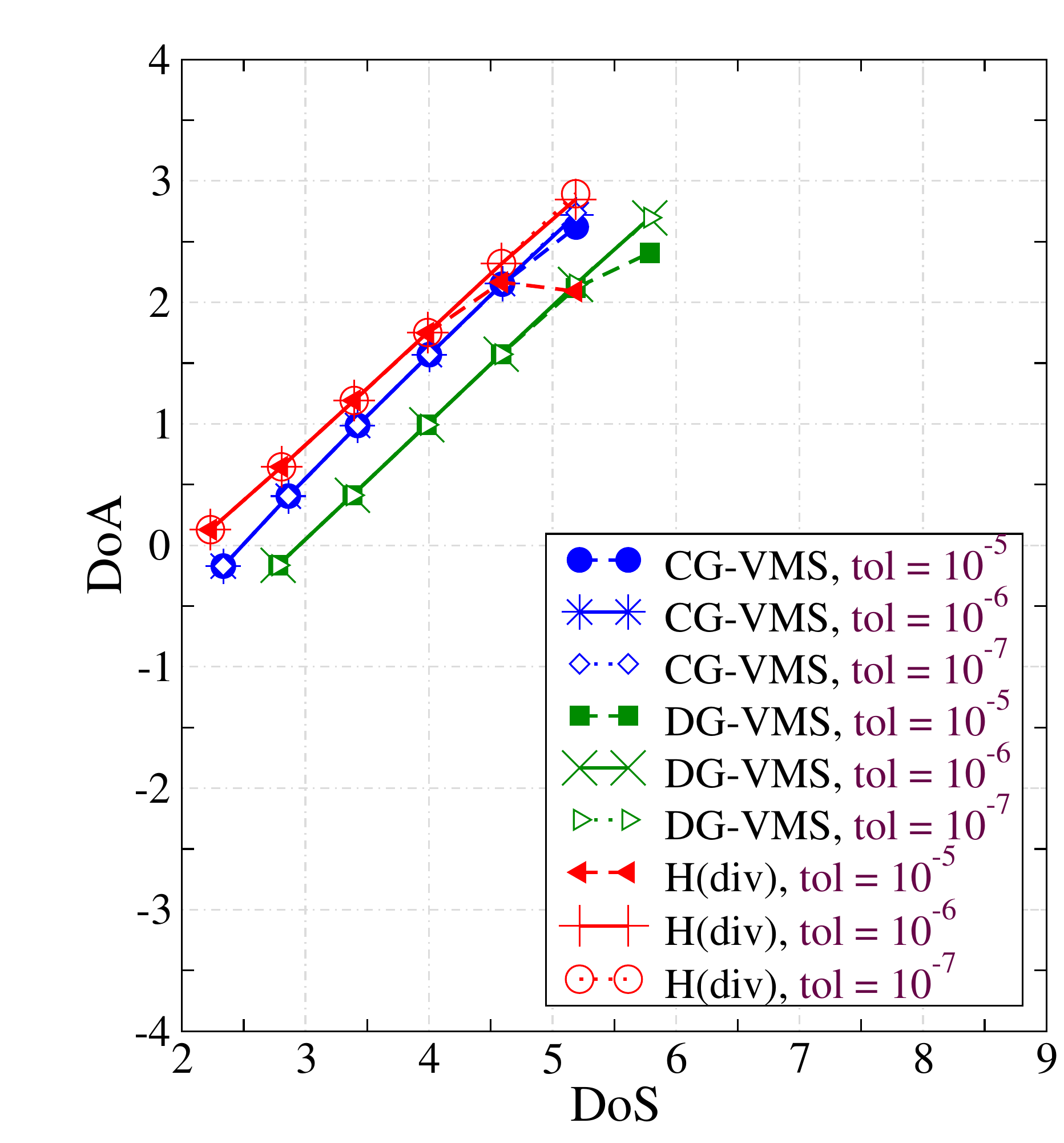}}
	
	\caption{\textsf{Two-dimensional problem using \textbf{QUAD} mesh:} 
		This figure compares the mesh convergence results
		for the chosen finite element formulations under
		various solver tolerances. The results are shown
		for field-splitting block solver methodology, but very
		similar results are also obtained under the
		scale-splitting solver (which are not shown
                for brevity). 
		The two main inferences are:
		(i) Similar to the results under TRI element, VMS
                formulations yield a slope of approximately $1$,
                whereas the H(div) formulations exhibit a superlinear
                convergence close to $1$.
		Nevertheless, the VMS formulations still outperform
                the H(div) formulation in terms of accuracy, though
                with lower margins.
		(ii) If the solver tolerances are not selected tight
                enough, the mesh convergence lines will flatten out.
	}
	\label{Fig:2D_convergence_Q4}
\end{figure}
%
\begin{figure}
	\subfigure[Assembly time {[TRI mesh]} \label{2D_SS_assembly_T3_e-7}]{
		\includegraphics[clip,width=0.485\linewidth,trim=0 0 0 3cm]{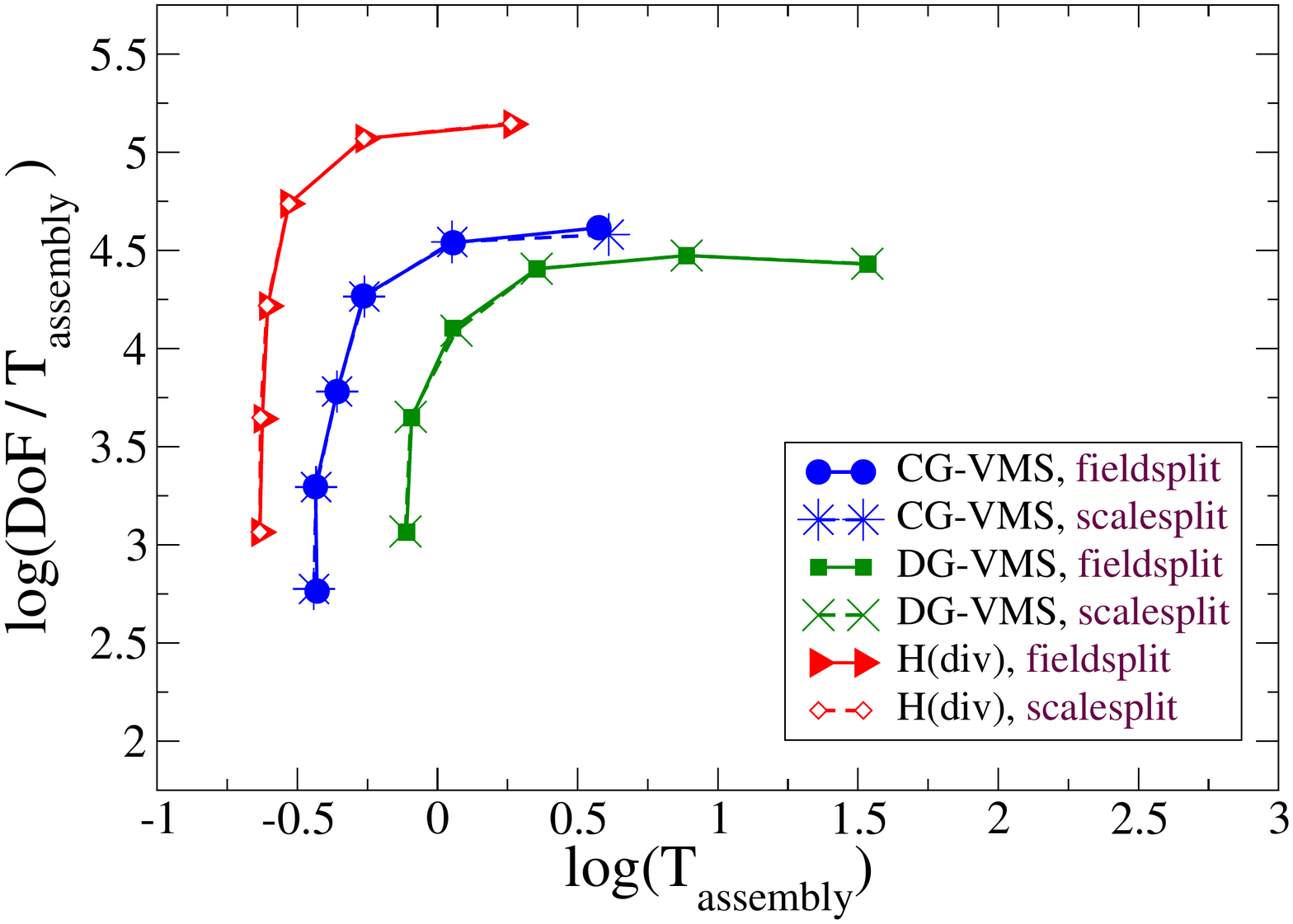}}
	\subfigure[Assembly time {[QUAD mesh]} \label{2D_SS_assembly_Q4_e-7}]{
		\includegraphics[clip,width=0.485\linewidth,trim=0 0 0 3cm]{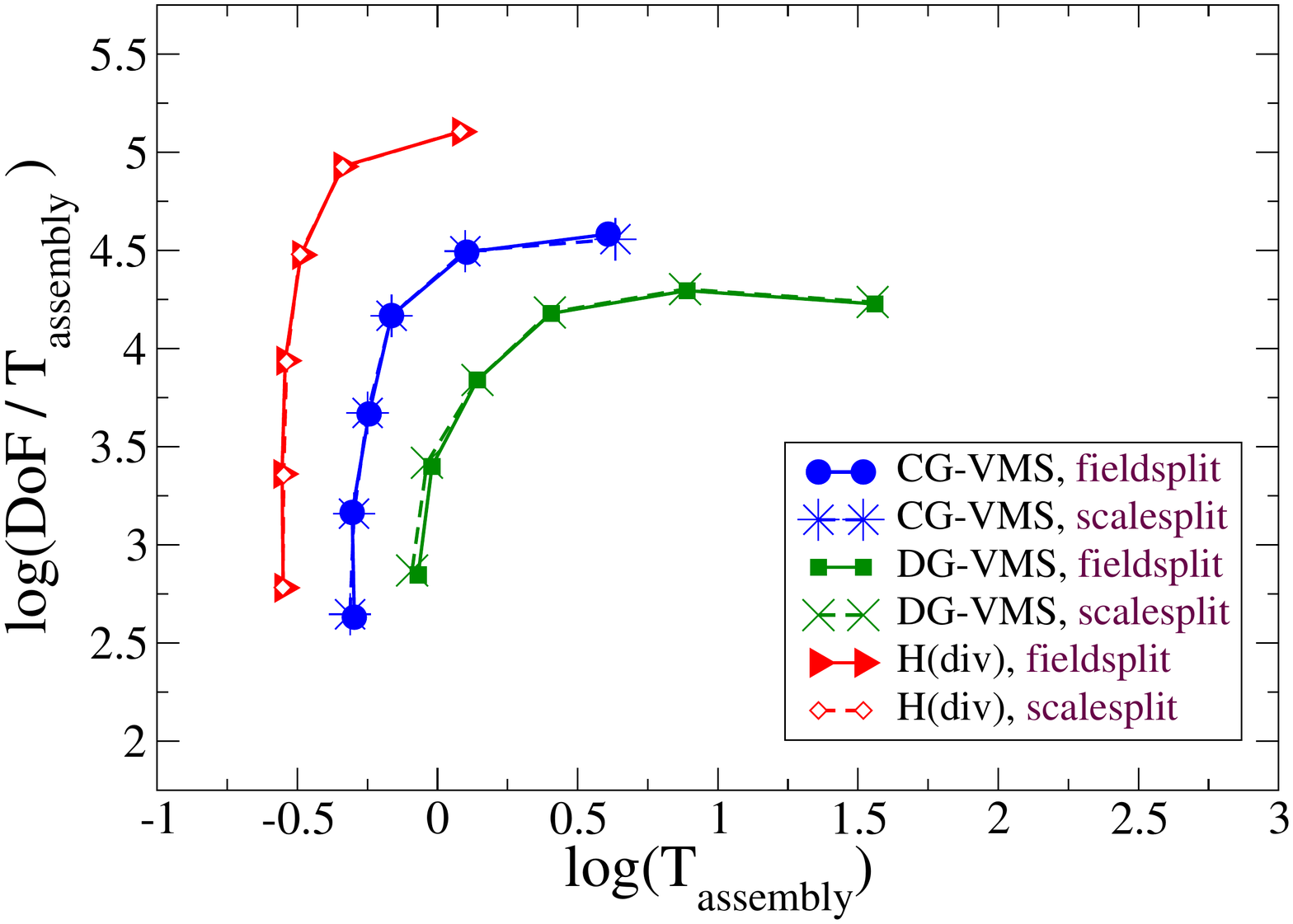}}
	\vspace{0cm}
	\subfigure[Solver time {[TRI mesh]} \label{2D_SS_solver_T3_e-7}]{
		\includegraphics[clip,width=0.485\linewidth,trim=0 0 0 3.0cm]{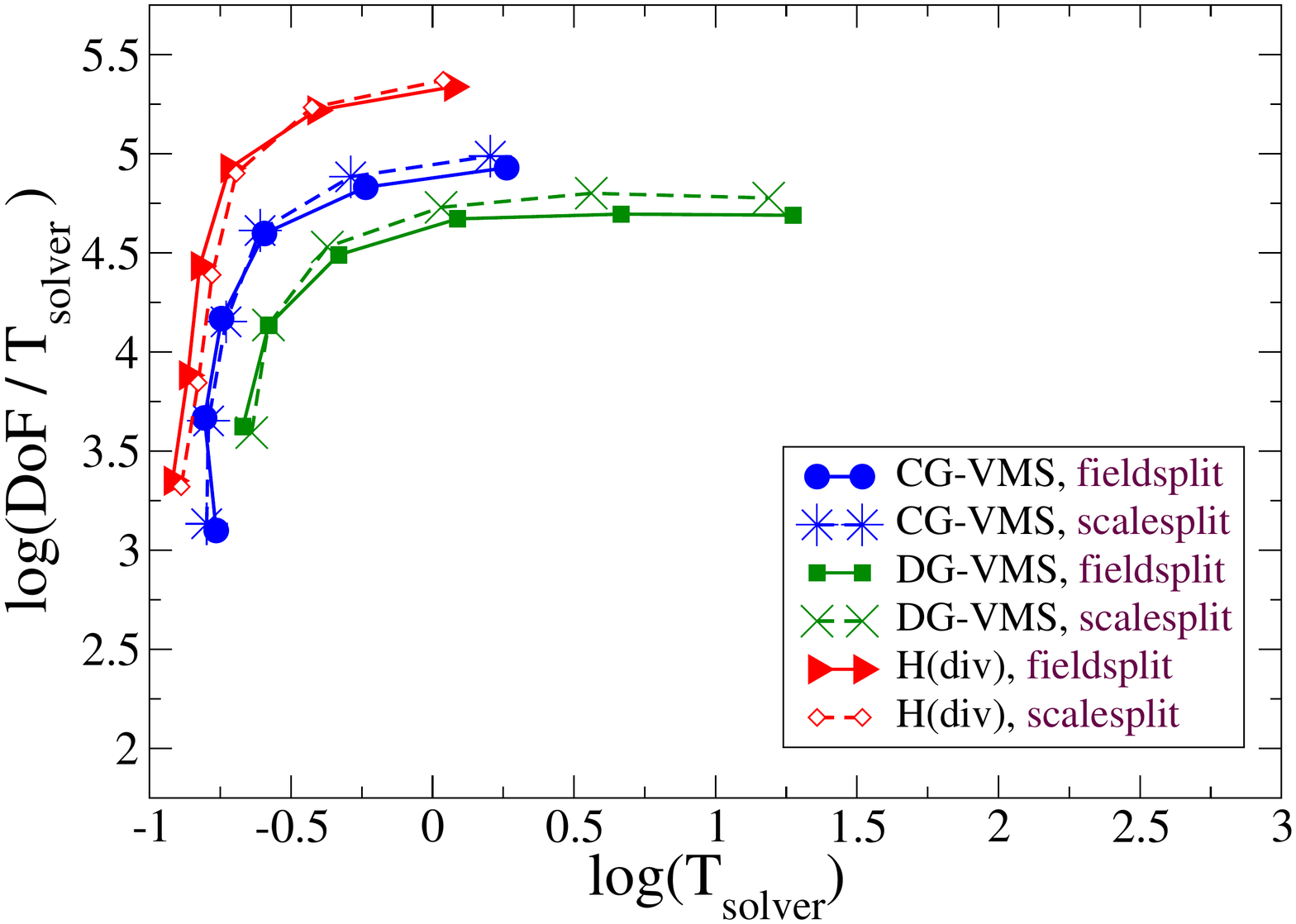}}
	\subfigure[Solver time {[QUAD mesh]} \label{2D_SS_solver_Q4_e-7}]{
		\includegraphics[clip,width=0.485\linewidth,trim=0 0 0 3.0cm]{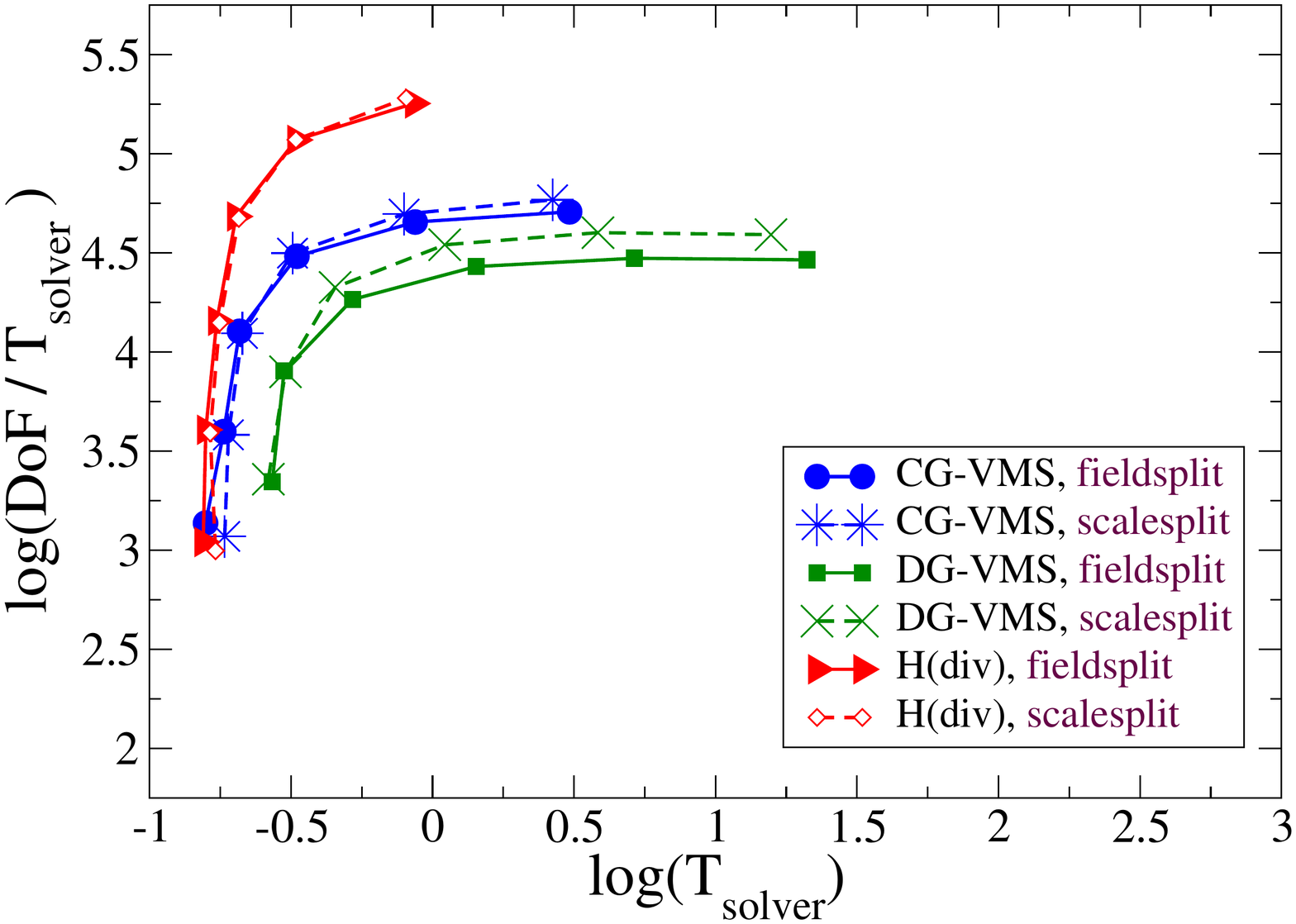}}
	\vspace{0cm}
	\subfigure[Total time {[TRI mesh]} \label{2D_SS_total_T3_e-7}]{
		\includegraphics[clip,width=0.485\linewidth,trim=0 0 0 3.0cm]{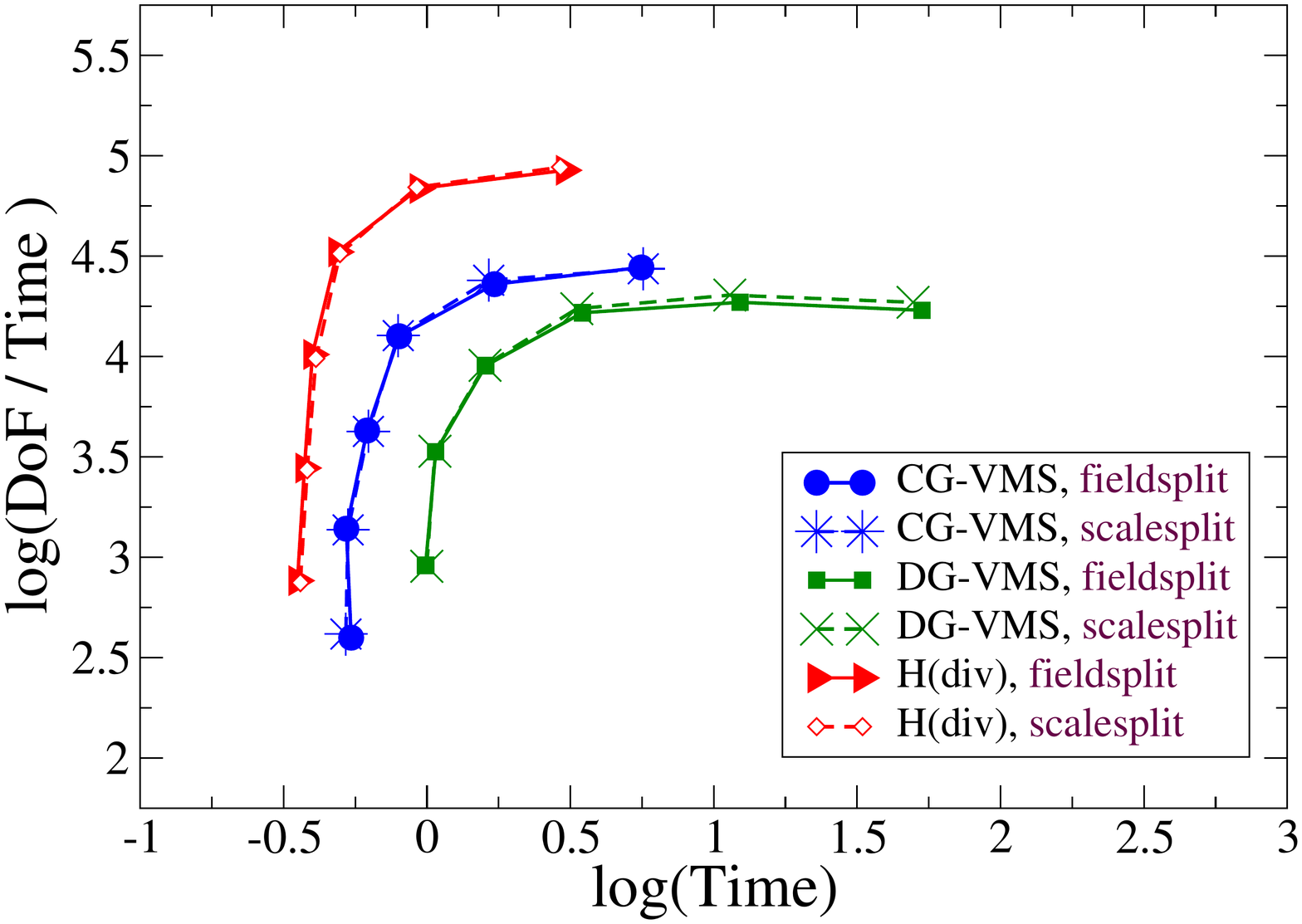}}
	\subfigure[Total time {[QUAD mesh]} \label{2D_SS_total_Q4_e-7}]{
		\includegraphics[clip,width=0.485\linewidth,trim=0 0 0 3.0cm]{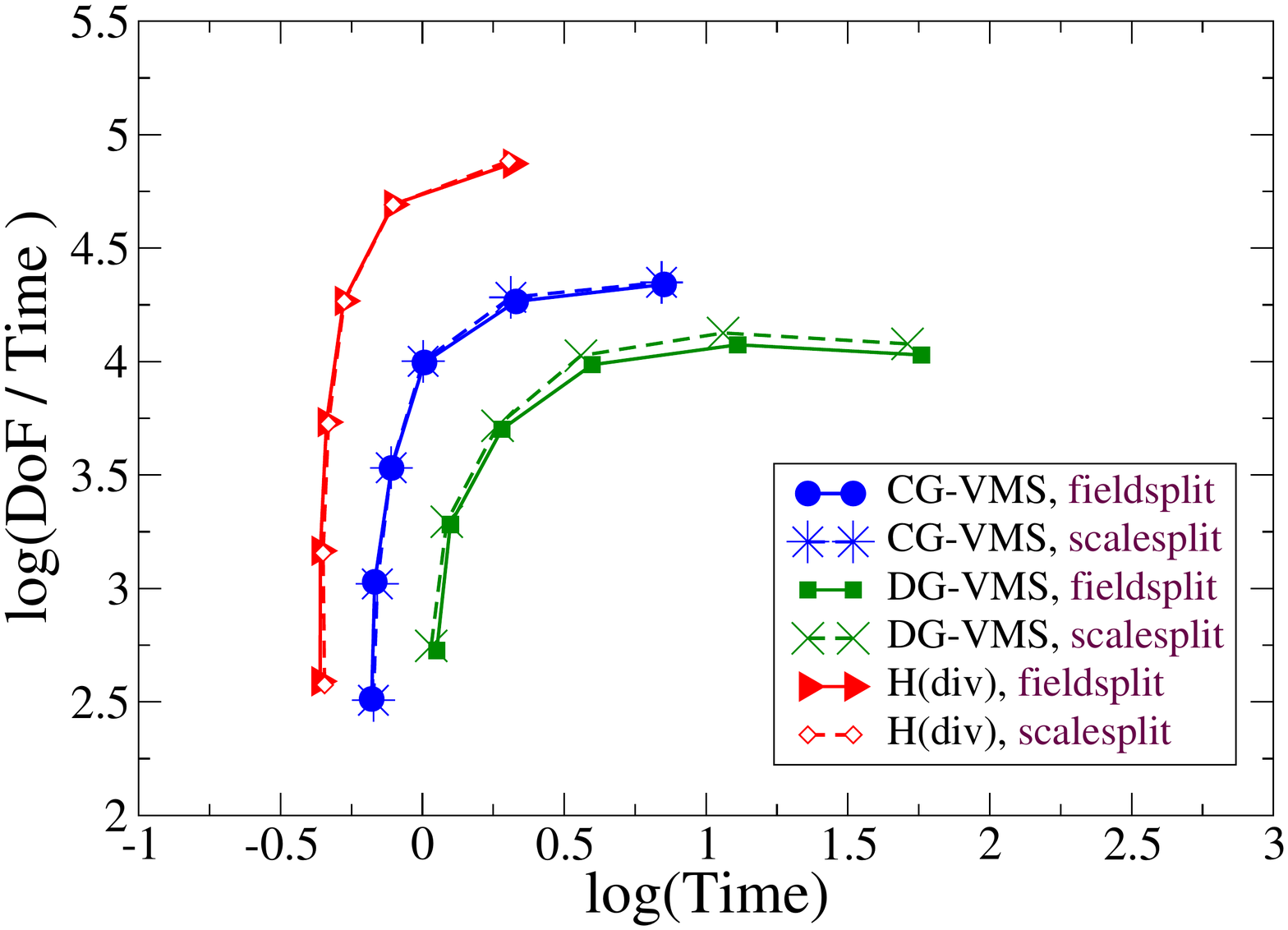}}
	\caption{\textsf{Two-dimensional problem:}~
		This figure compares the static-scaling results
		for the chosen finite element formulations
                using \textbf{TRI} and \textbf{QUAD} meshes.
		The results are shown for both field-splitting
                and scale-splitting block solver methodologies.
		The two main inferences are:
		(i) Under the VMS formulations, the
                field-splitting methodology is slightly
                worse compared to the scale-splitting
                with respect to solver time. However, the
                difference in performance is negligible
                with respect to the total time to the
                solution. 
		(ii) Regardless of the mesh type, the
                H(div) formulation processes its DoF count
                faster than either of the VMS formulations.
}
	\label{Fig:2D_Static_scaling}
\end{figure}

\begin{figure}
	\subfigure[Macro-pressure  \label{p1_DoE_2D_T3_e-7}]{
		\includegraphics[clip,scale=0.2875]{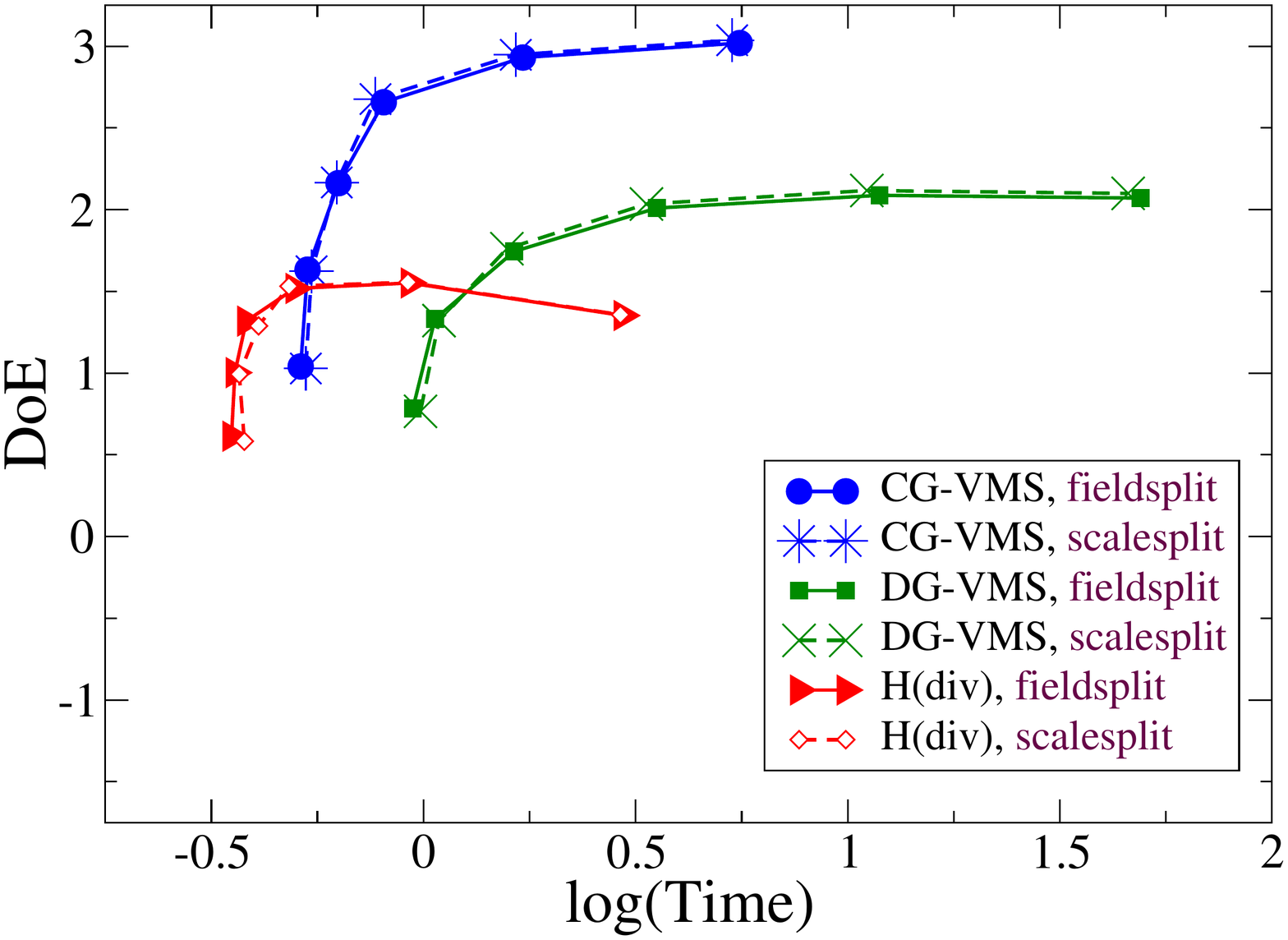}}
	%
	\subfigure[Micro-pressure \label{p2_DoE_2D_T3_e-7}]{
		\includegraphics[clip,scale=0.2875]{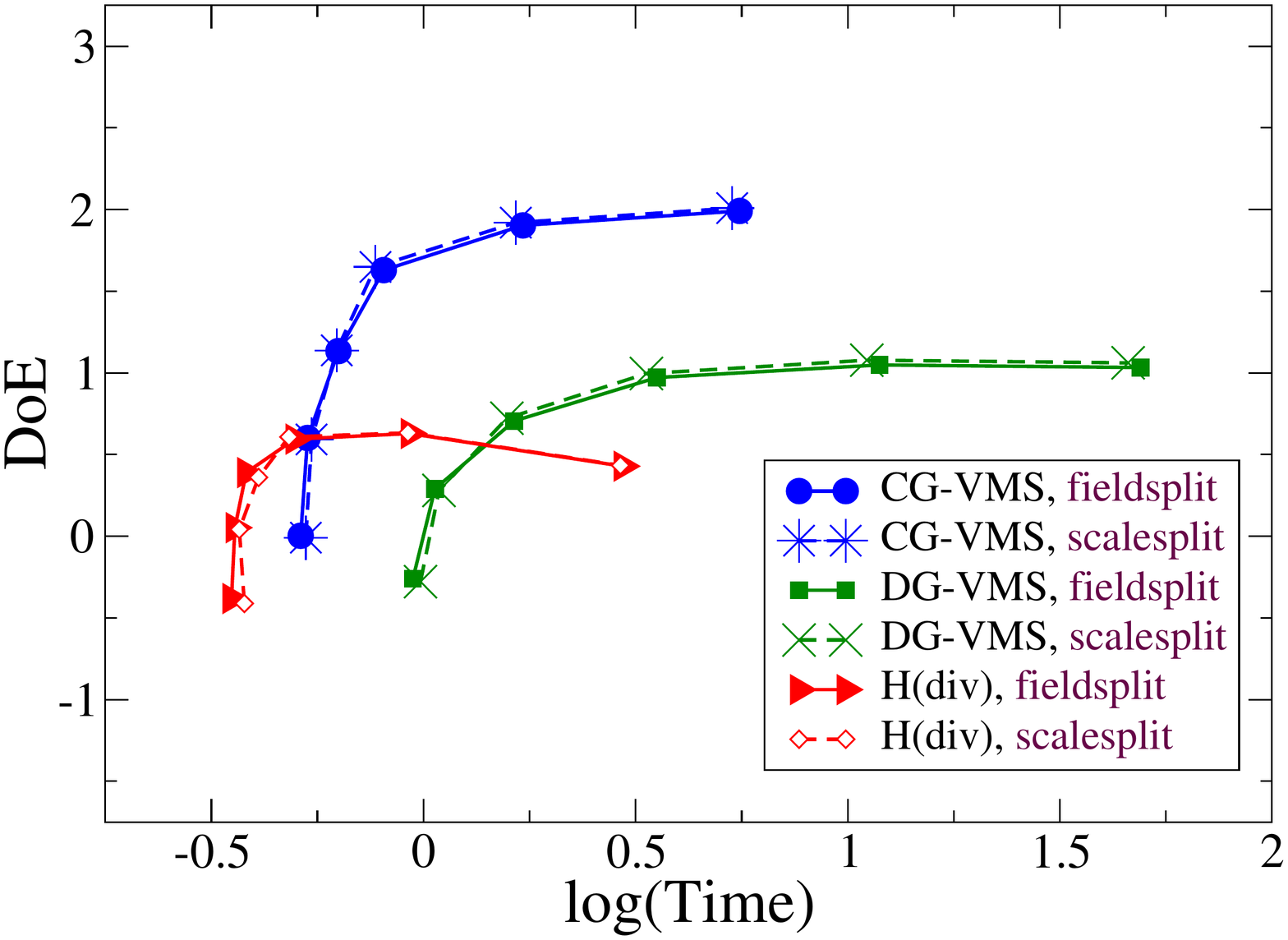}} \\
	%
	\subfigure[Macro-velocity \label{v1_DoE_2D_T3_e-7}]{
		\includegraphics[clip,scale=0.2875]{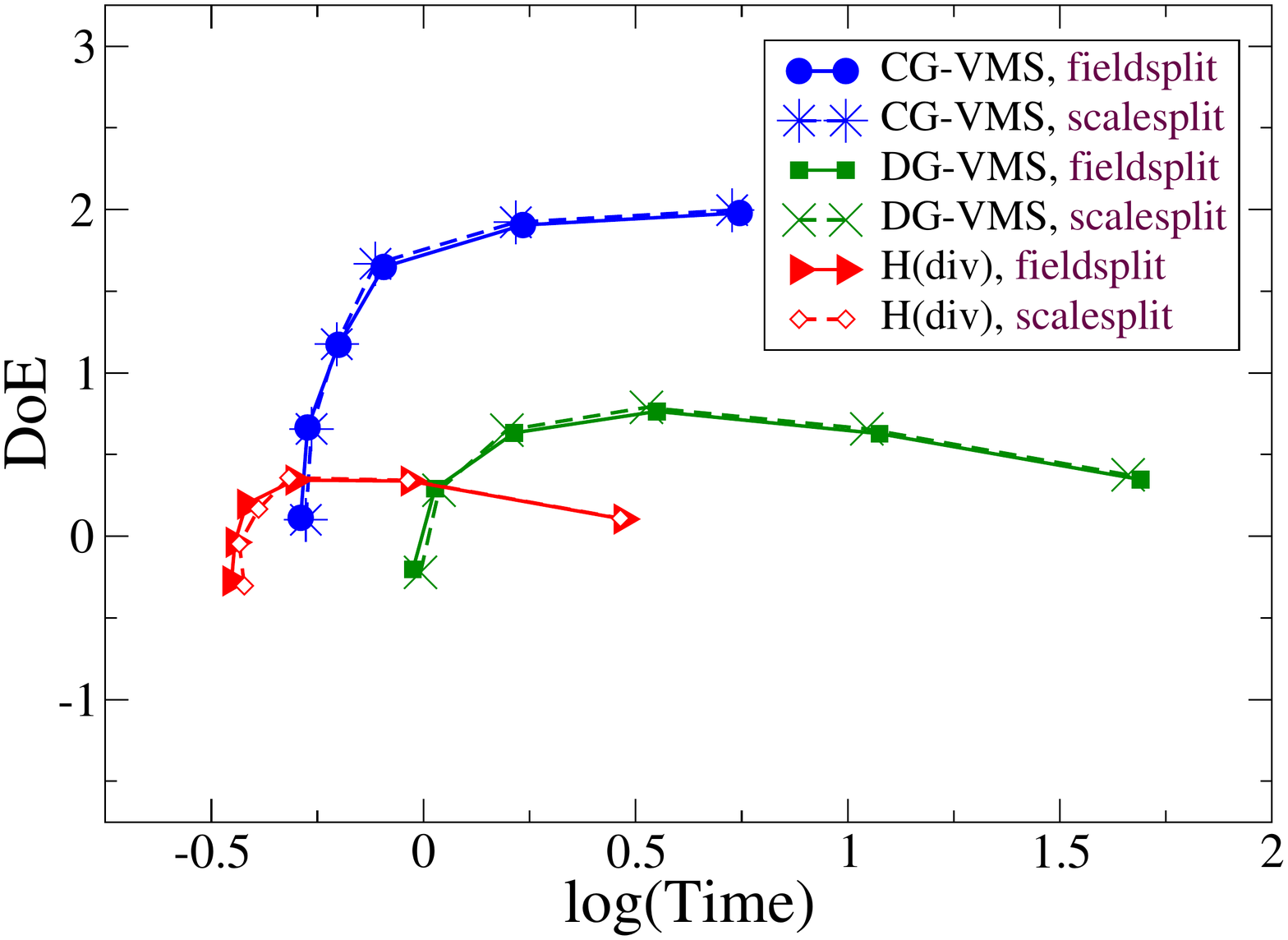}}
	%
	\subfigure[Micro-velocity \label{v2_DoE_2D_T3_e-7}]{
		\includegraphics[clip,scale=0.2875]{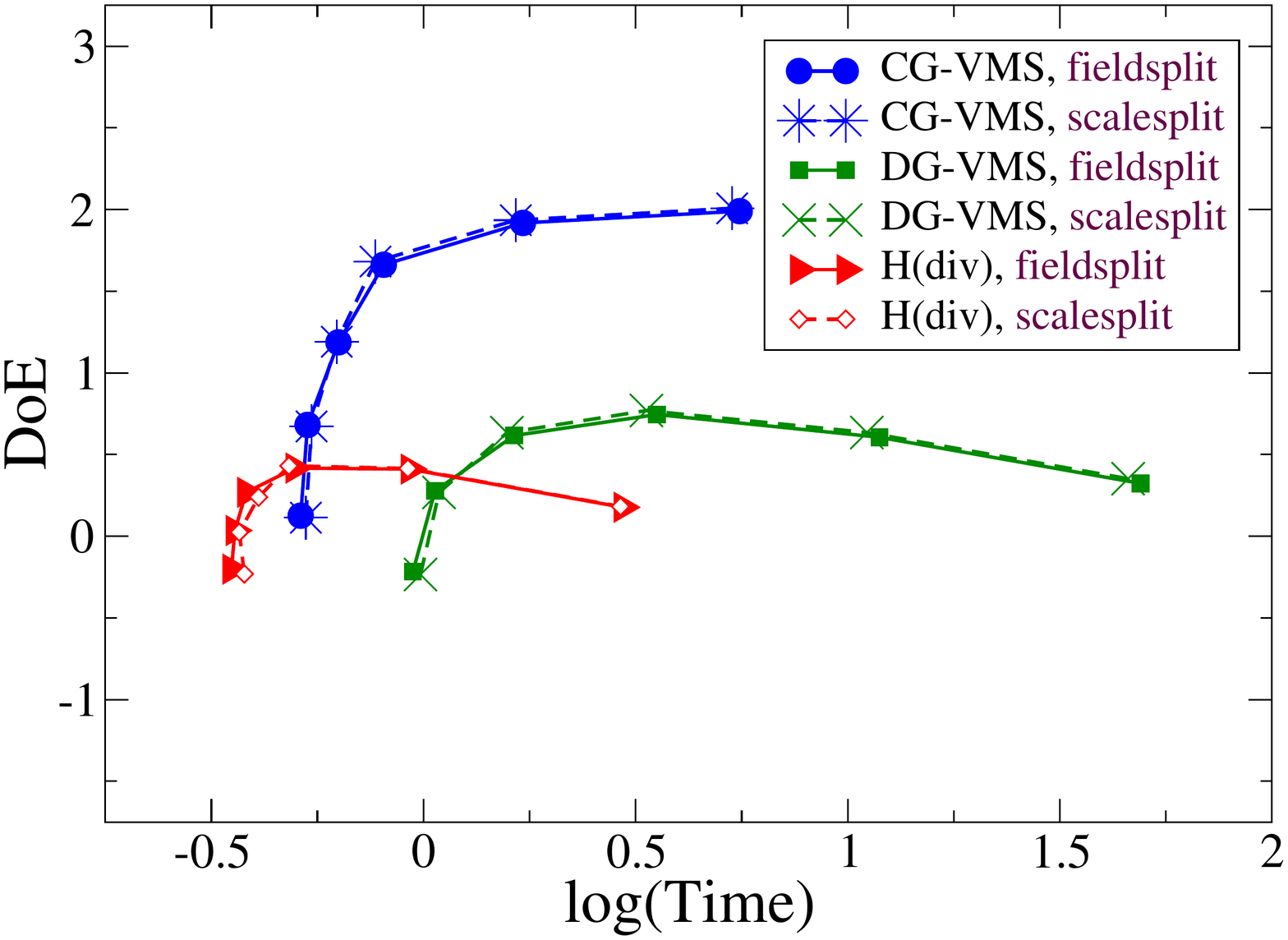}}
	
	\caption{\textsf{Two-dimensional problem for \textbf{TRI} mesh:} 
	  This figure compares the Digits of Efficacy (DoE)
          among the chosen finite element formulations. 
	  Results for both the composable block solver
	  methodologies with a tolerance of $10^{-7}$
	  are reported. The three main inferences are:
	  (i)The H(div) has much smaller DoE compared to
          either of the VMS formulations due to its
          lower DoA.
	  (ii) The CG-VMS formulation has a clear-cut
          advantage with respect to DoE compared to
          the other two formulations.
	  (iii) The choice of either field-splitting
          or scale-splitting has minimal effect on
          DoE for all the formulations and for all
          the fields.
        }
	\label{Fig:2D_DOE_T3}
\end{figure}
%
\begin{figure}
	\subfigure[Macro-pressure  \label{p1_DoE_2D_Q4_e-7}]{
		\includegraphics[clip,scale=0.2875]{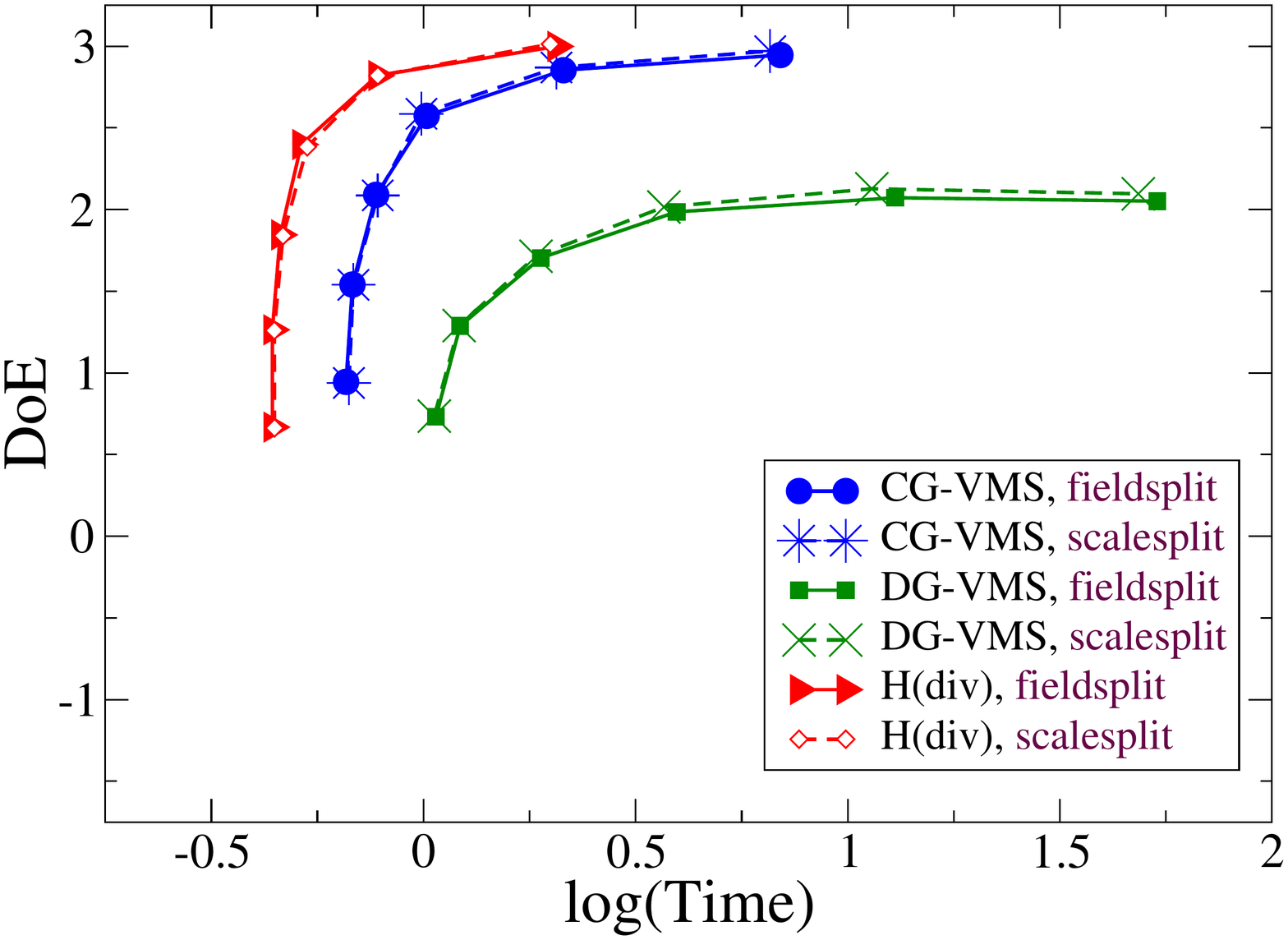}}
	%
	\subfigure[Micro-pressure \label{p2_DoE_2D_Q4_e-7}]{
		\includegraphics[clip,scale=0.2875]{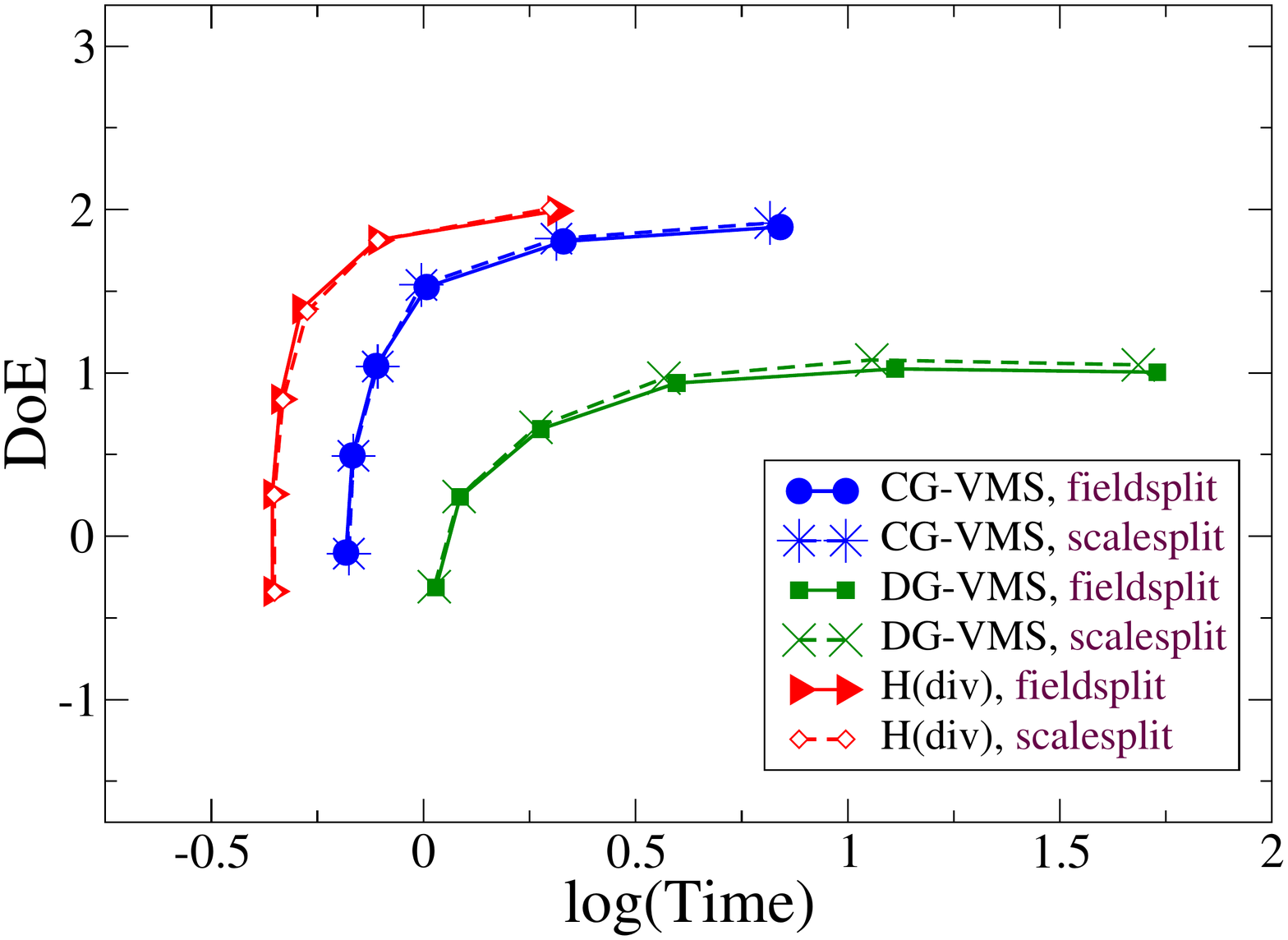}} \\
	%
	\subfigure[Macro-velocity \label{v1_DoE_2D_Q4_e-7}]{
		\includegraphics[clip,scale=0.2875]{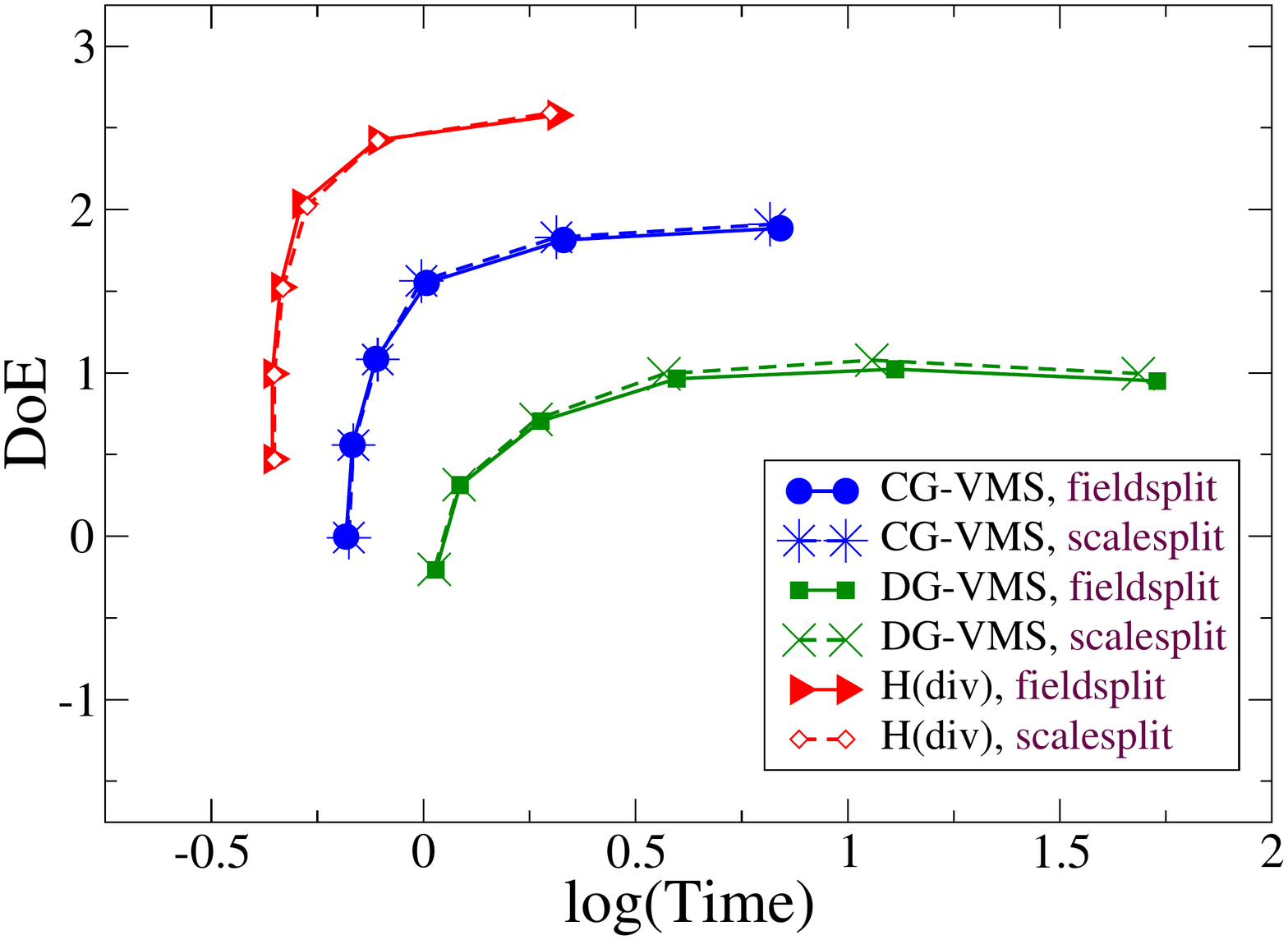}}
	%
	\subfigure[Micro-velocity \label{v2_DoE_2D_Q4_e-7}]{
		\includegraphics[clip,scale=0.2875]{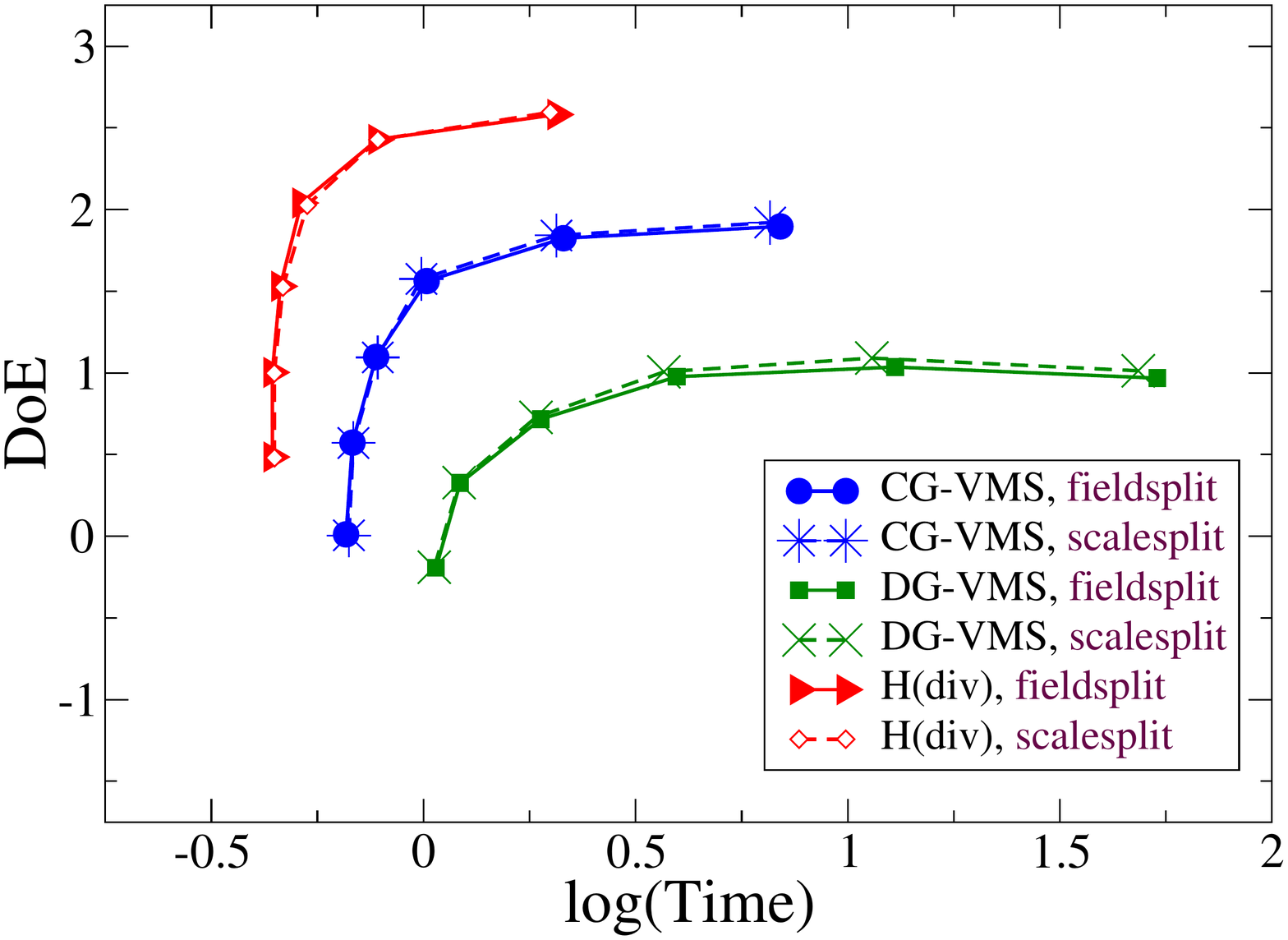}}
	
	\caption{\textsf{Two-dimensional problem for \textbf{QUAD} element:} 
	  This figure compares the Digits of Efficacy (DoE)
          among the chosen finite element formulations. 
	  Results for both the composable block solver
	  methodologies with a tolerance of $10^{-7}$
	  are reported. The main two inferences are:
	  (i) The H(div) formulation beats its VMS counterparts
          for all the fields. This is by virtue of
	  the H(div) formulation having the shortest
          total time to solution among all the formulations
          and at the same time maintaining relatively high DoA. 
	  (ii) The choice of either field-splitting or
          scale-splitting has minimal effect on DoE
          for all the formulations and for all the fields.
}
	\label{Fig:2D_DOE_Q4}
\end{figure}

\begin{figure}
	\centering
	\sbox{\measurebox}{%
		\begin{minipage}[b]{.55\textwidth}
			\subfigure
			[Schematic]
			{\label{Fig9:2D_schematic}\includegraphics[scale=0.55]{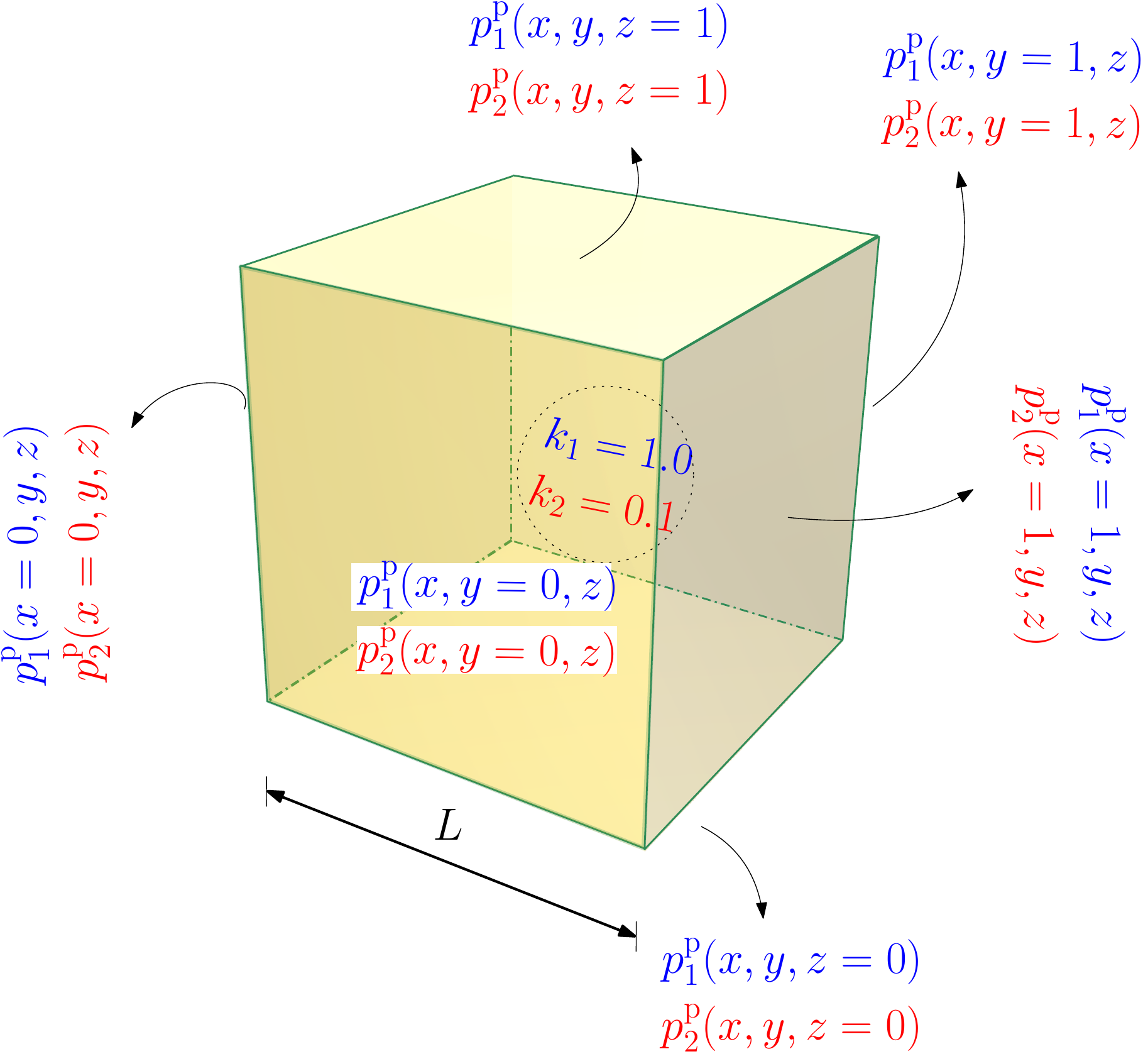}}
	\end{minipage}}
	\usebox{\measurebox}\qquad
	\begin{minipage}[b][\ht\measurebox][s]{.4\textwidth}
		\centering
		\subfigure
		[TET mesh]
		{\label{Fig1:Mesh_TET}\includegraphics[scale=0.33]{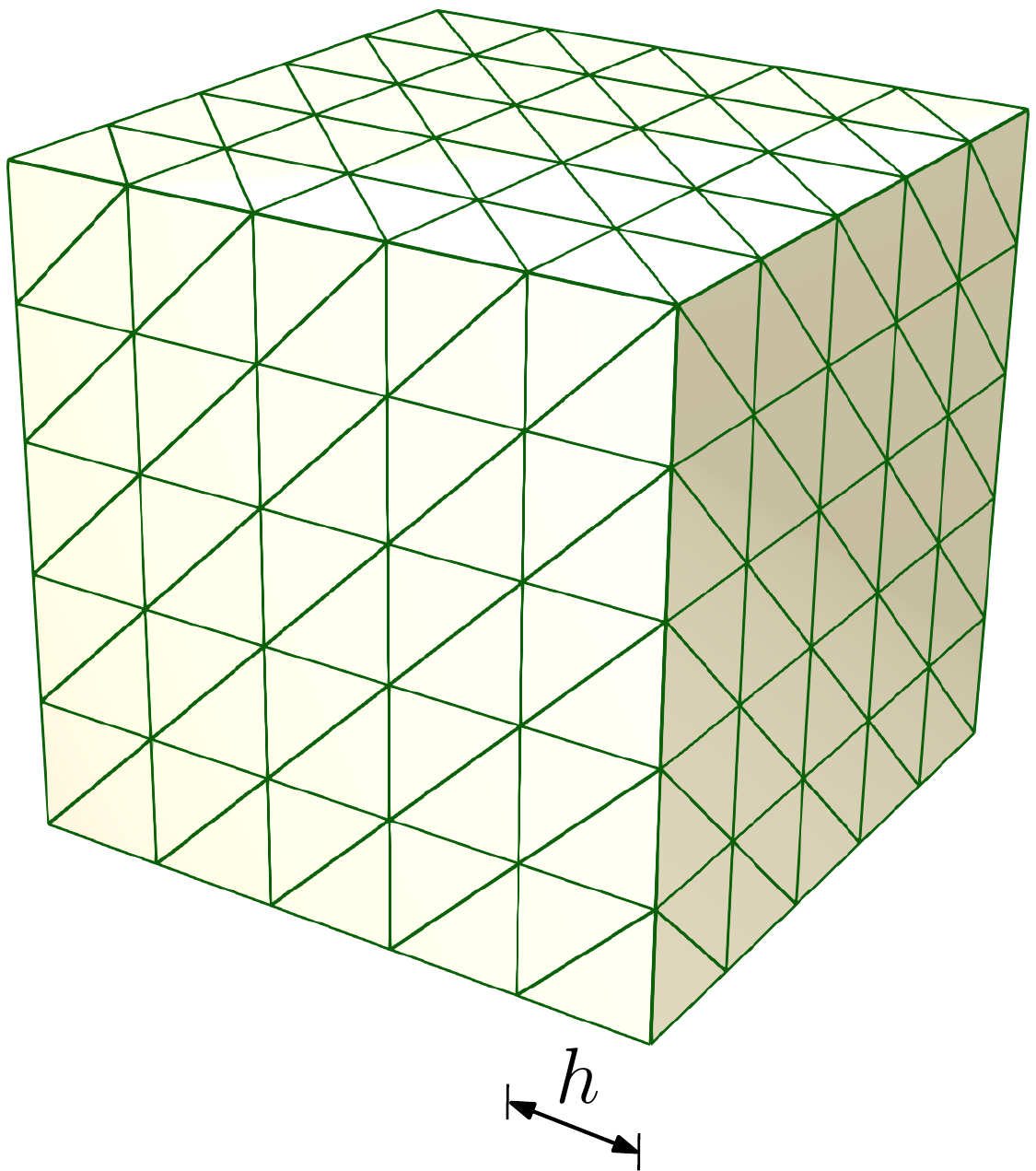}}
		
		\vfill
		
		\subfigure
		[HEX mesh]
		{\label{Fig1:Mesh_HEX}\includegraphics[scale=0.33]{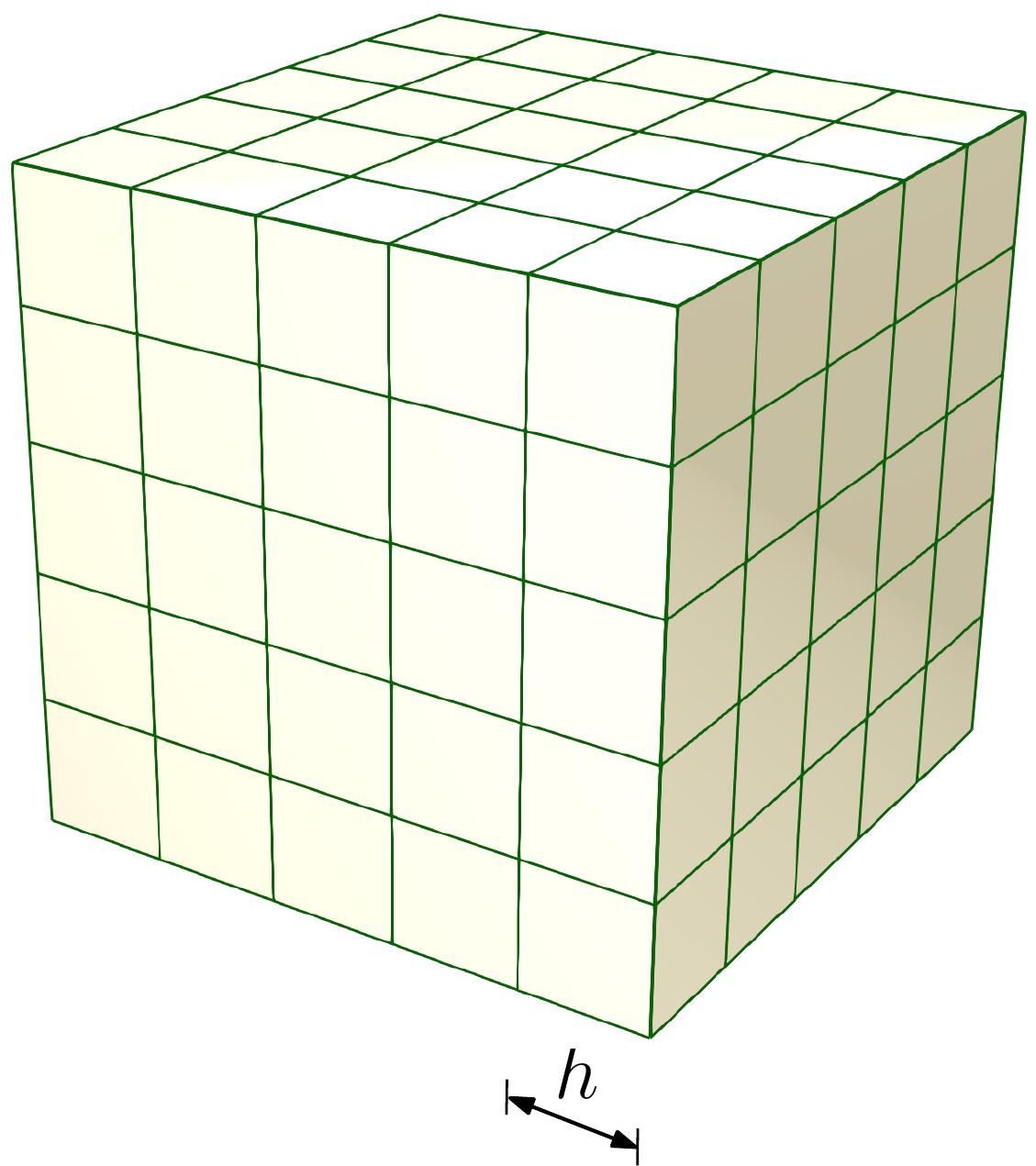}}
	\end{minipage}
	\caption{\textsf{Three-dimensional domain:} 
	  This figure provides a pictorial description
          of the boundary value problem and shows the
          typical meshes employed in our numerical
          simulations.\label{fig9:2D}}
\end{figure}

\begin{figure}
  \subfigure[Macro-pressure  \label{p1_convergence_3D_e-7}]{
    \includegraphics[clip,scale=0.37,trim=0 1.25cm 9cm 03]{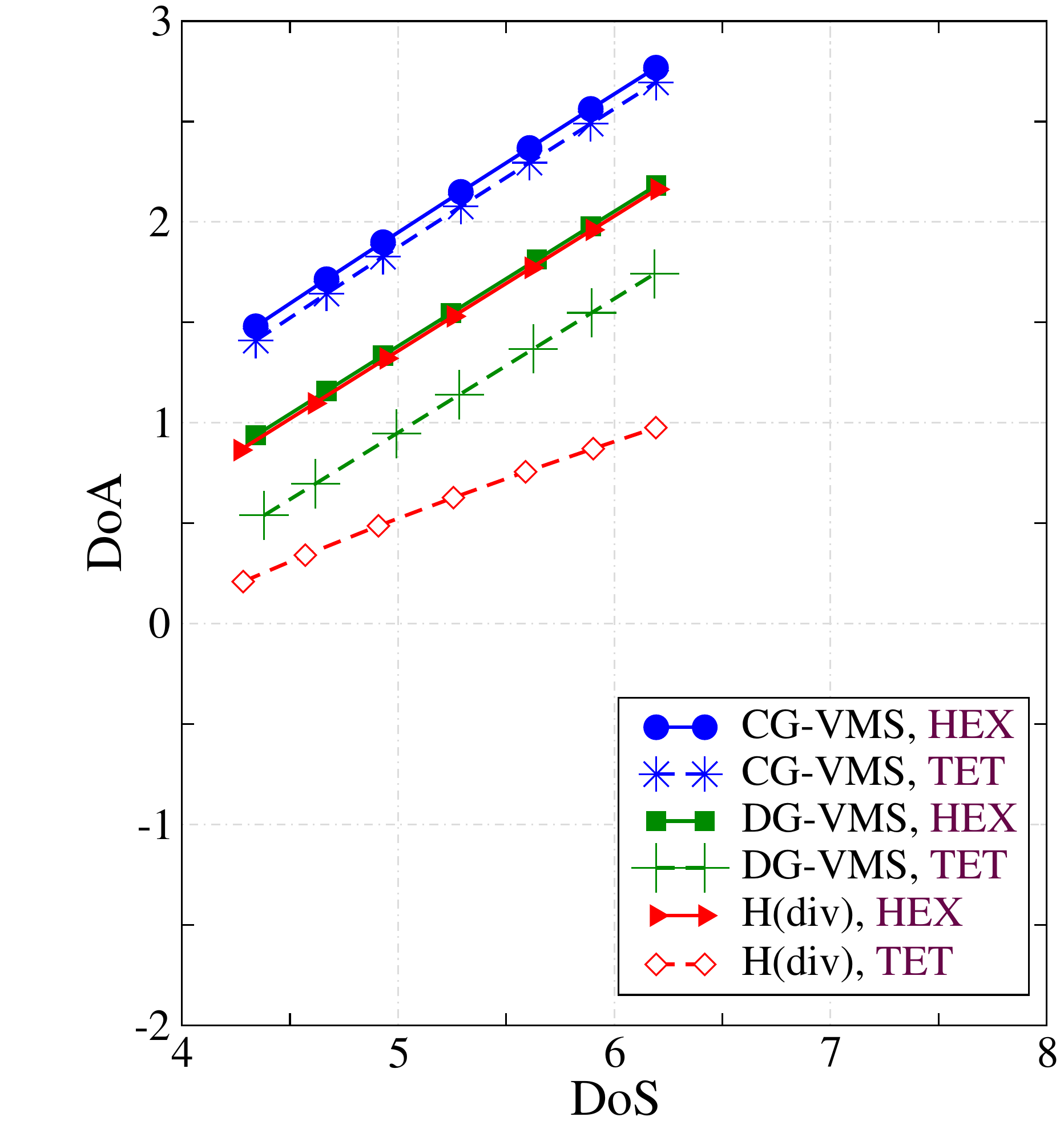}}
  \hspace{1 cm}
  \subfigure[Micro-pressure \label{p2_convergence_3D_e-7}]{
    \includegraphics[clip,scale=0.37,trim=0 1.25cm 9cm 03]{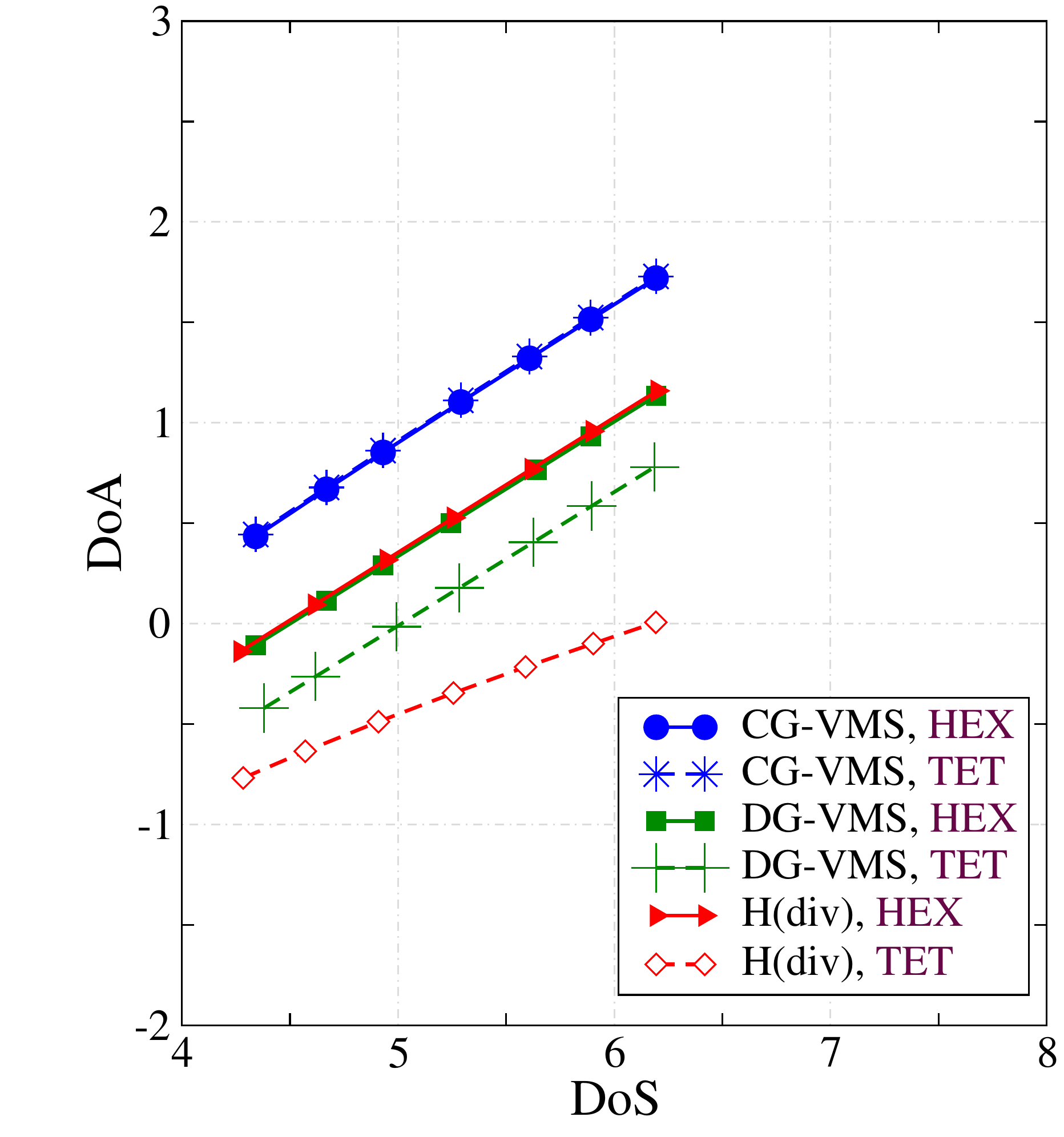}} \\
  \vspace{0.75 cm}
  \subfigure[Macro-velocity \label{v1_convergence_3D_e-7}]{
    
    \includegraphics[clip,scale=0.37,trim=0 1.25cm 9cm 03]{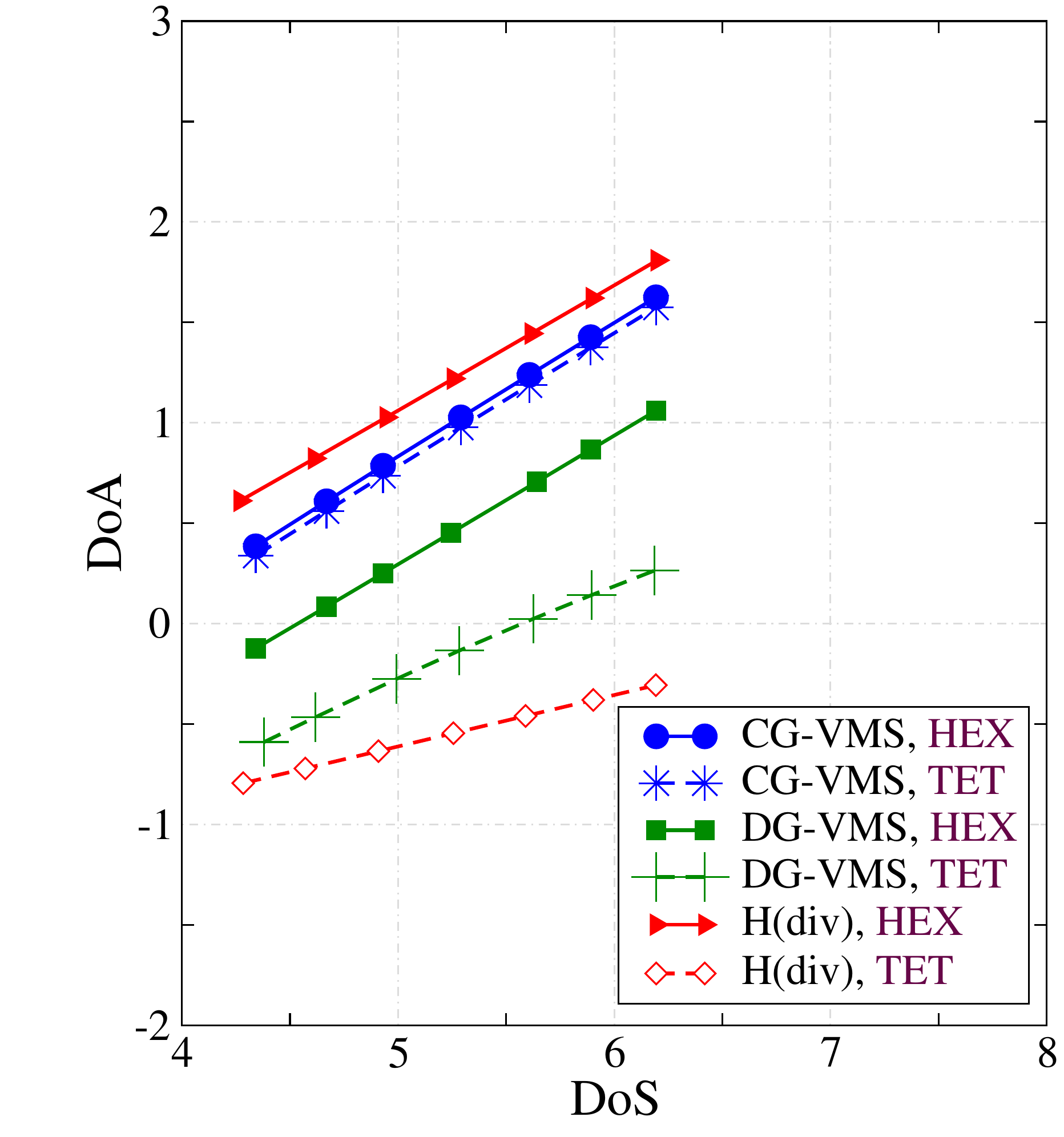}}
  \hspace{1 cm}
  \subfigure[Micro-velocity \label{v2_convergence_3D_e-7}]{
    \includegraphics[clip,scale=0.37,trim=0 1.25cm 9cm 03]{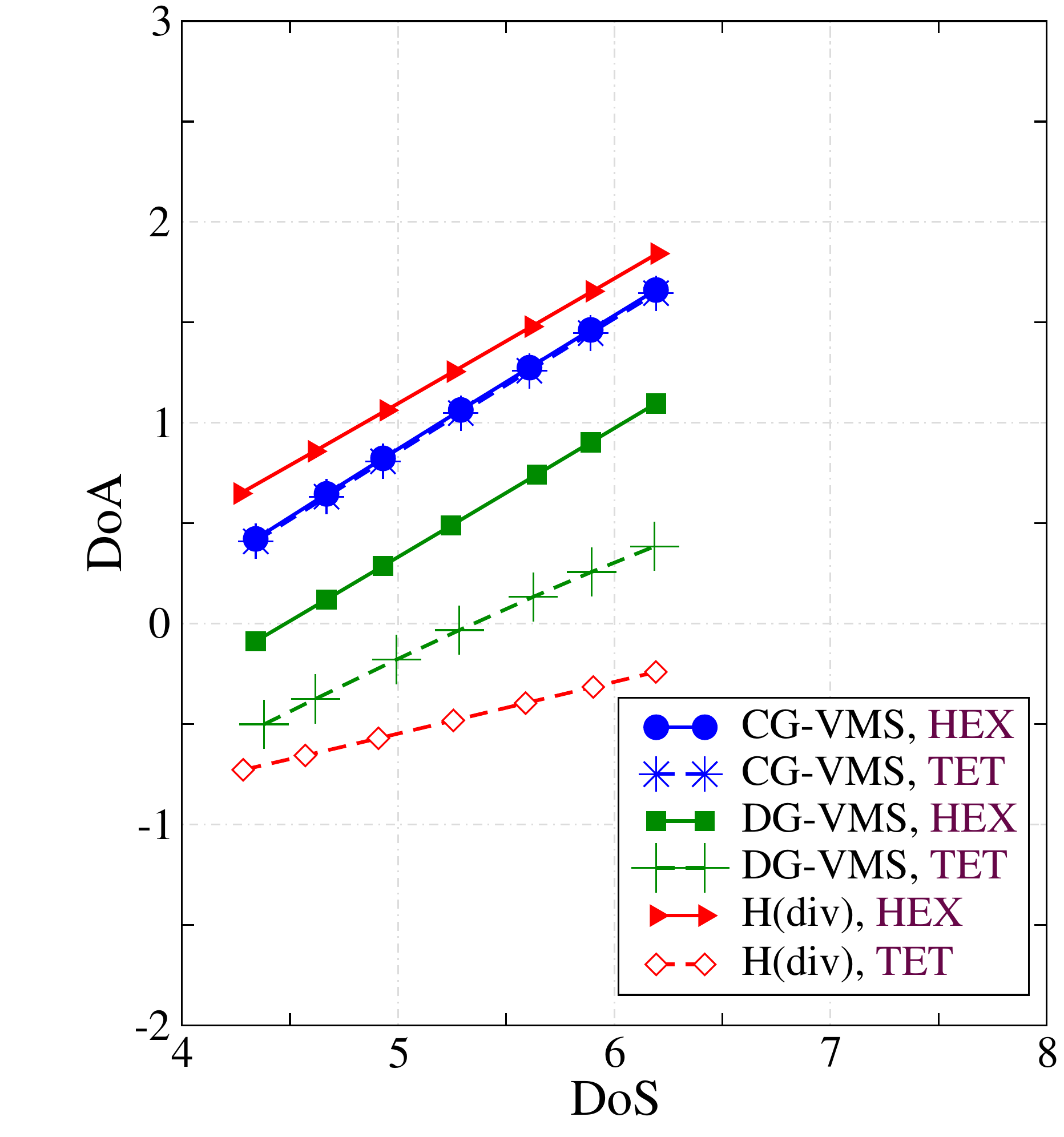}}
  
  \caption{\textsf{Three-dimensional problem:} 
    This figure compares the mesh convergence results
    for the chosen finite element formulations using
    \textbf{TET} and \textbf{HEX} meshes. The solver
    tolerance is taken to be $10^{-7}$. The results
    are shown for field-splitting block solver methodology,
    but very similar results are obtained under the
    scale-splitting solver. 
    Refinements are tuned in such a way that all the
    chosen finite element discretizations 
    maintain roughly the same DoF count at each stage.
    The two main inferences are:
    (i) The VMS formulations mark a slope of $\frac{2}{3}$,
    while the H(div) formulation yields a 
    slope of $\frac{1}{3}$. This indicates that the VMS
    formulations process Digits of Accuracy twice as fast
    as that of the H(div) formulation on a simplicial mesh. 
    (i) However, for the case of the non-simplicial element,
    similar to two-dimensional problem, we observe superlinear
    convergence under the RT0 formulation. 
  }
	\label{Fig:3D_convergence_e-7}
\end{figure}

\begin{figure}
  \subfigure[Assembly time {[TET mesh]} \label{3D_SS_assembly_T3_e-7}]{
    \includegraphics[clip,width=0.485\linewidth, trim= 0 0.5cm 0 3cm]{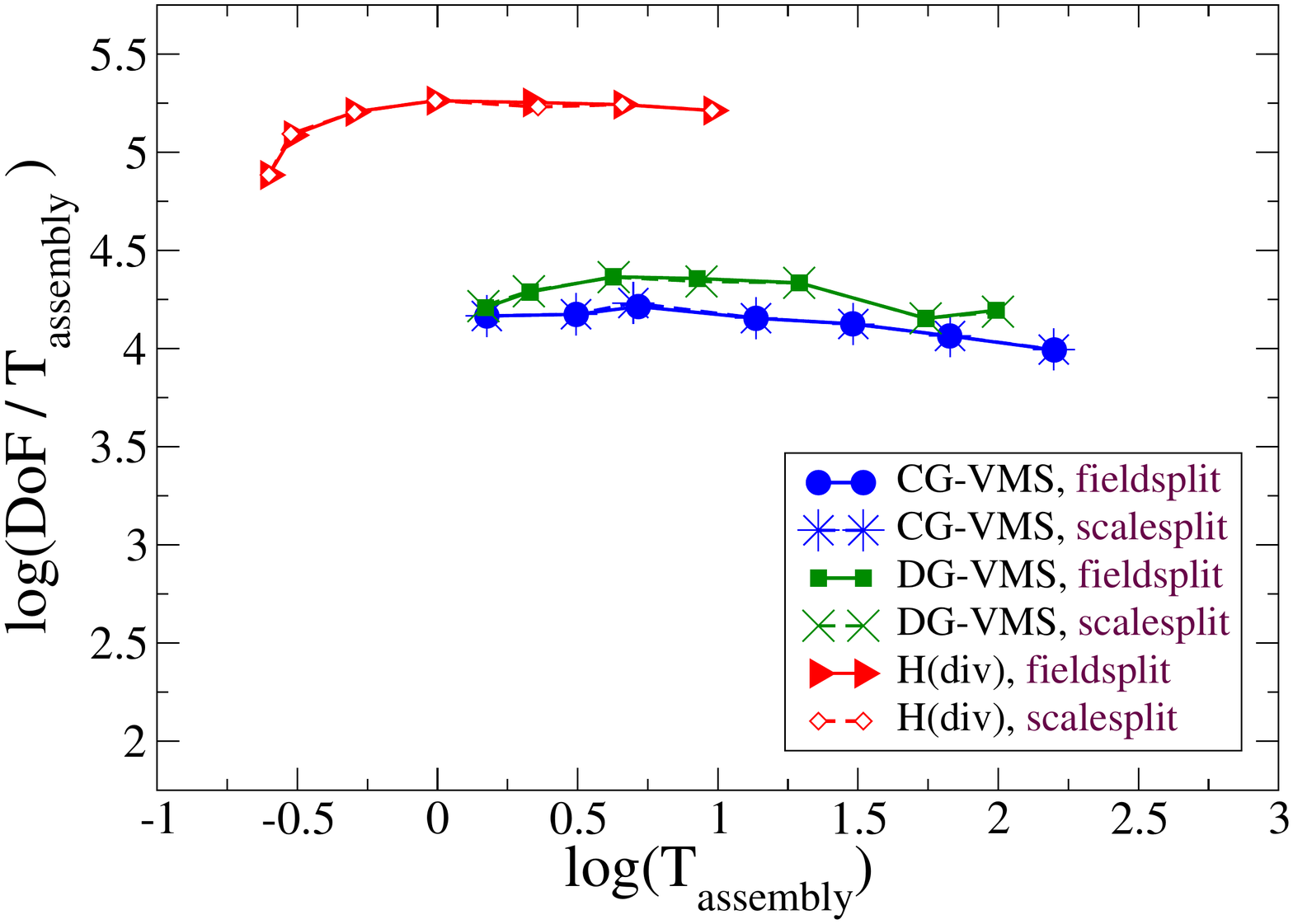}}
  \subfigure[Assembly time {[HEX mesh]} \label{3D_SS_assembly_Q4_e-7}]{
    \includegraphics[clip,width=0.485\linewidth, trim= 0 0.5cm 0 3cm]{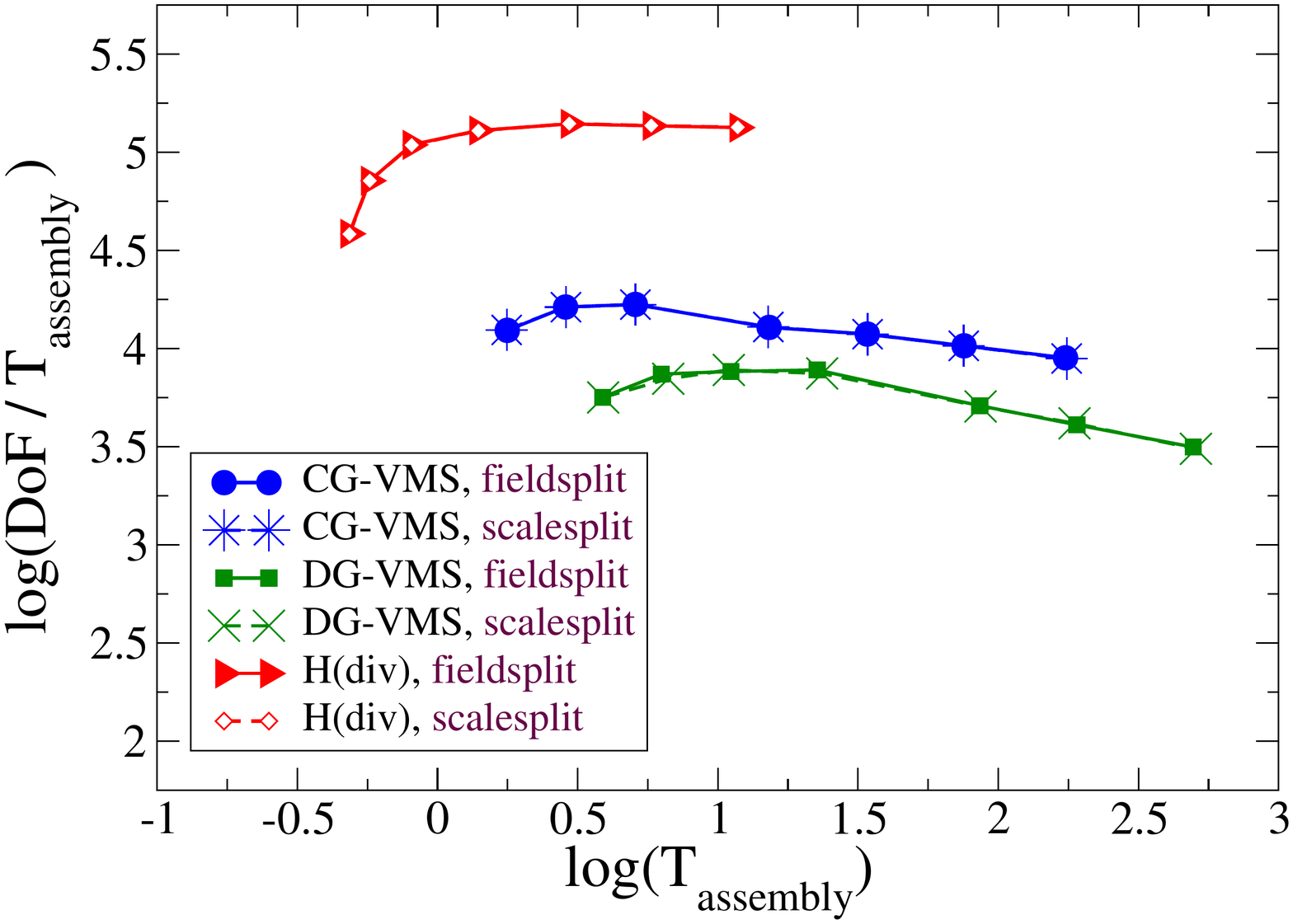}}
  \subfigure[Solver time {[TET mesh]} \label{3D_SS_solver_TET_e-7}]{
    \includegraphics[clip,width=0.485\linewidth, trim= 0 0.5cm 0 3cm]{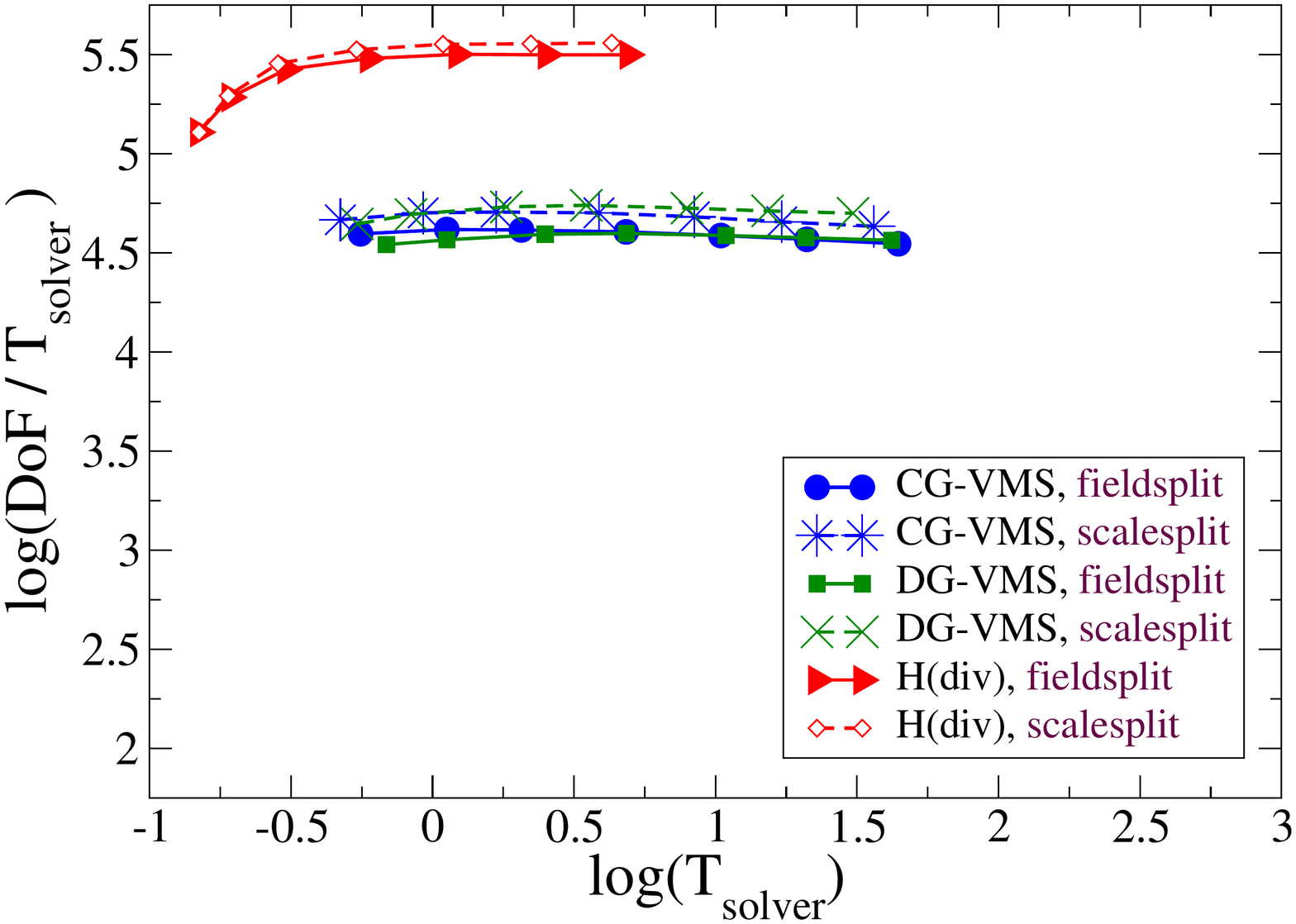}}
  \subfigure[Solver time {[HEX mesh]} \label{3D_SS_solver_HEX_e-7}]{
    \includegraphics[clip,width=0.485\linewidth, trim= 0 0.5cm 0 3cm]{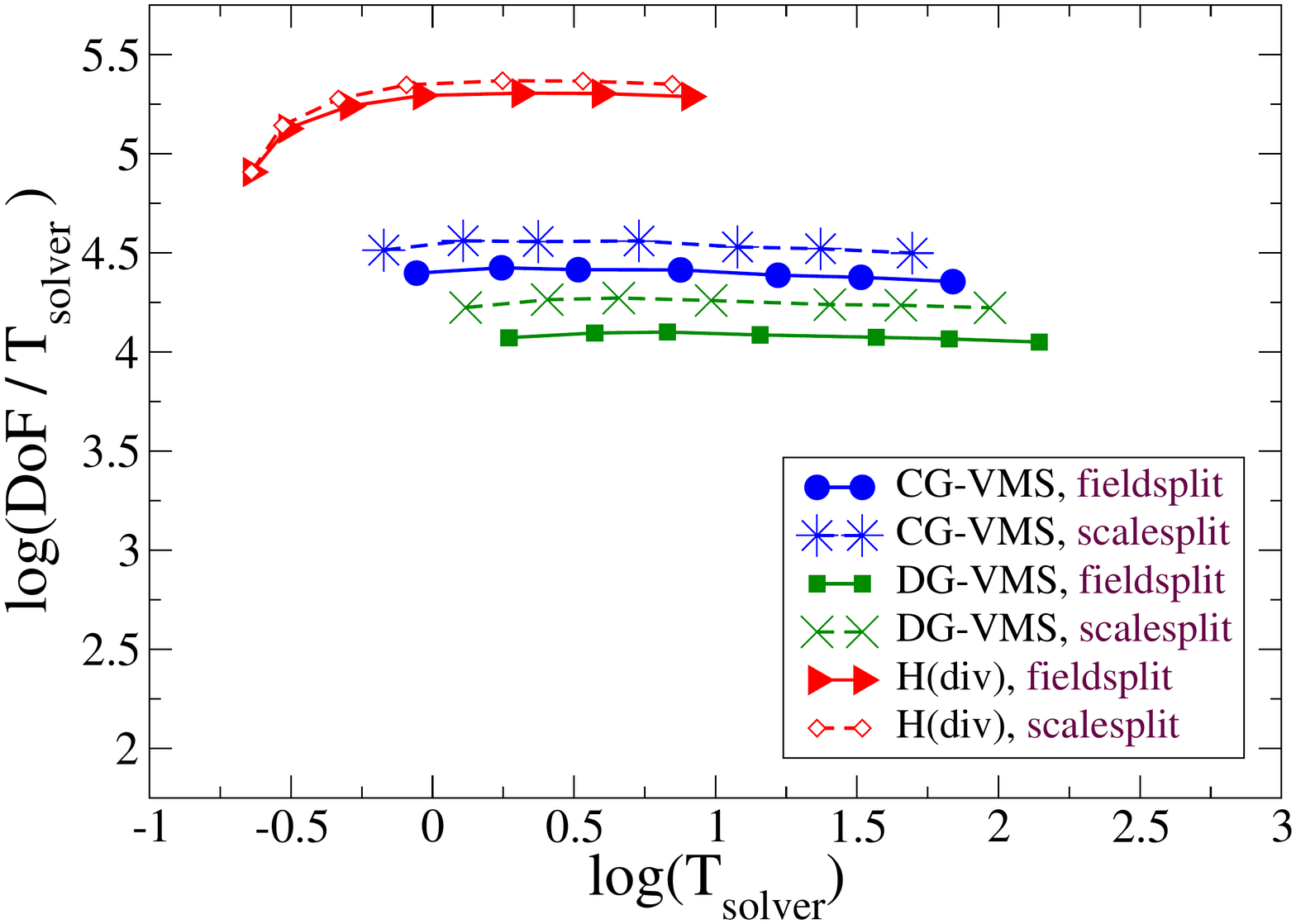}}
  \subfigure[Total time {[TET mesh]} \label{3D_SS_total_TET_e-7}]{
    \includegraphics[clip,width=0.485\linewidth, trim= 0 0.75cm 0 3cm]{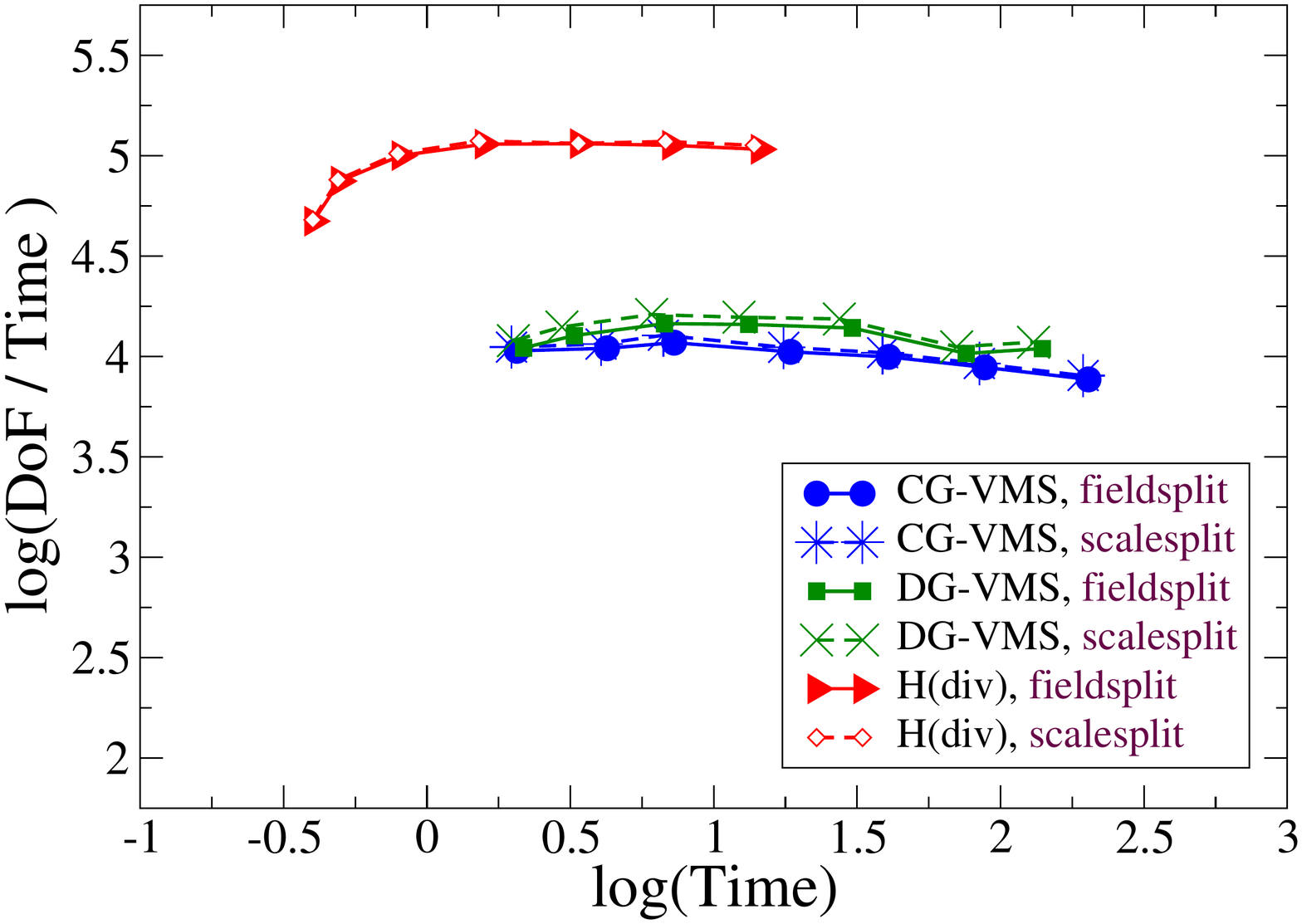}}
  \subfigure[Total time {[HEX mesh]} \label{3D_SS_total_HEX_e-7}]{
    \includegraphics[clip,width=0.485\linewidth, trim= 0 0.75cm 0 3cm]{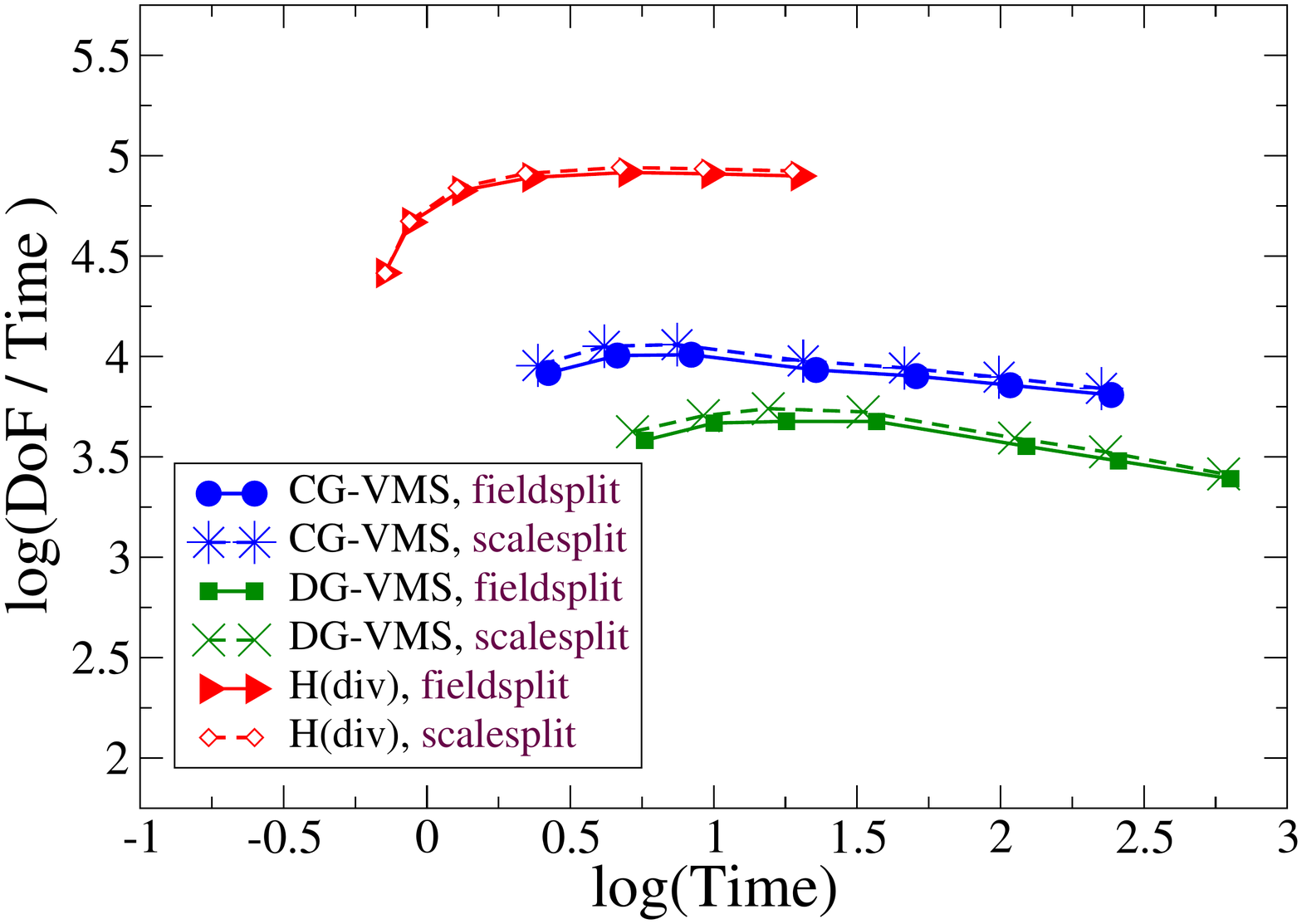}}
  \caption{\textsf{Three-dimensional problem:}
    This figure compares the static-scaling results
    for the chosen finite element formulations using
    \textbf{TET} and \textbf{HEX} meshes.
    The main inferences are:
    (i) Flat lines show the optimal scalability region of
    the proposed block solver methodologies for the
    chosen problem sizes and for the chosen
    hardware environment.
    (ii) For all the formulations, the
    scale-splitting processed more DoFs
    per unit solver time than the
    field-splitting. However, the gap
    between the two solvers is negligible
    in terms of DoFs processes per unit
    total time to solution.
    (iii) For both simplicial and non-simplicial elements, the
    H(div) formulation processes its DoF count faster than either
    of the VMS formulations.	
}
	\label{Fig:3D_Static_scaling}
\end{figure}

\begin{figure}
  \subfigure[Macro-pressure  \label{p1_DoE_TET_e-7}]{
    \includegraphics[clip,scale=0.2875]{Figures/figure_11a.pdf}}
  %
  \subfigure[Micro-pressure \label{p2_DoE_TET_e-7}]{
    \includegraphics[clip,scale=0.2875]{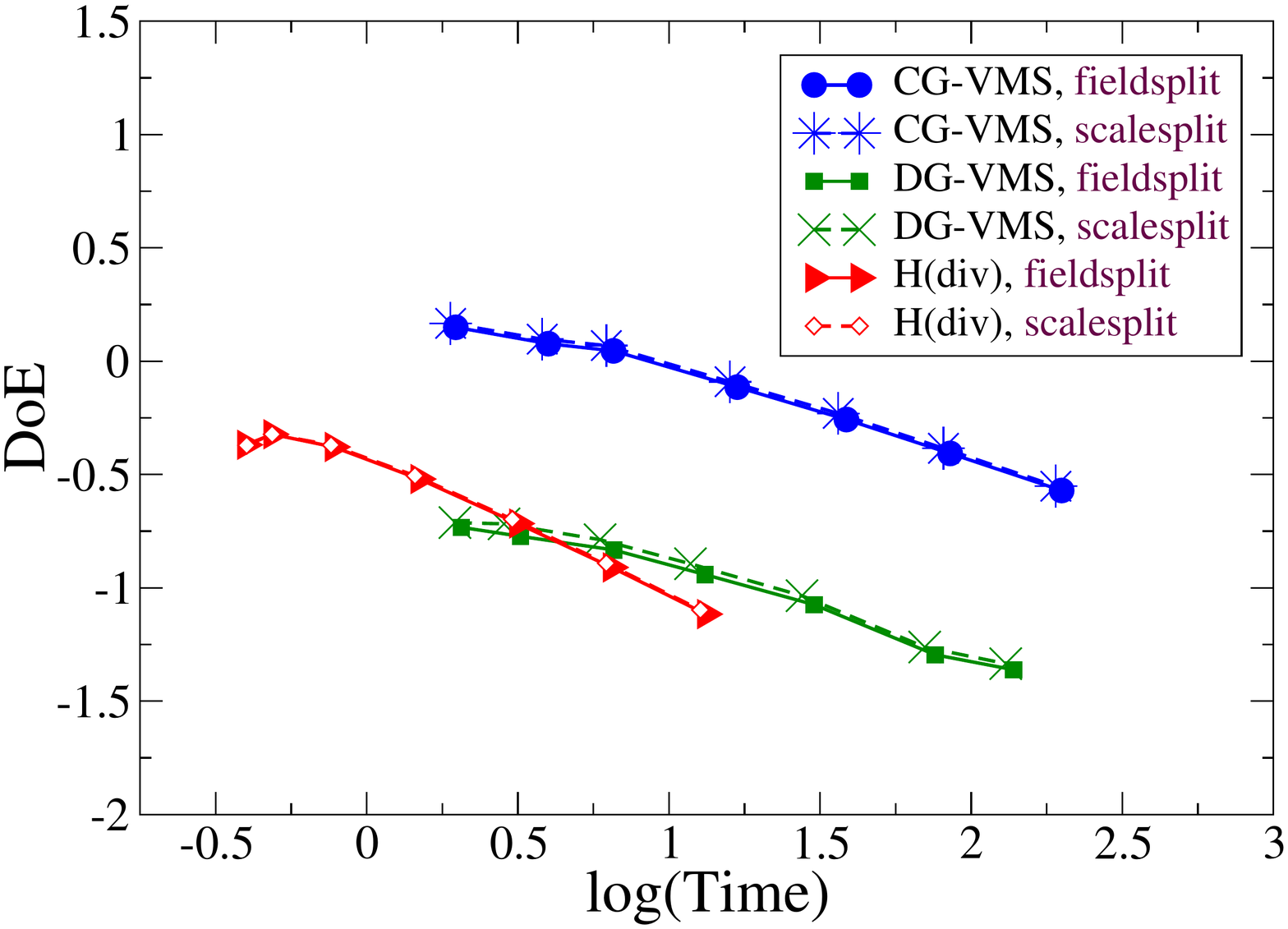}} \\
  %
  \subfigure[Macro-velocity \label{v_DoE_TET_e-7}]{
    \includegraphics[clip,scale=0.2875]{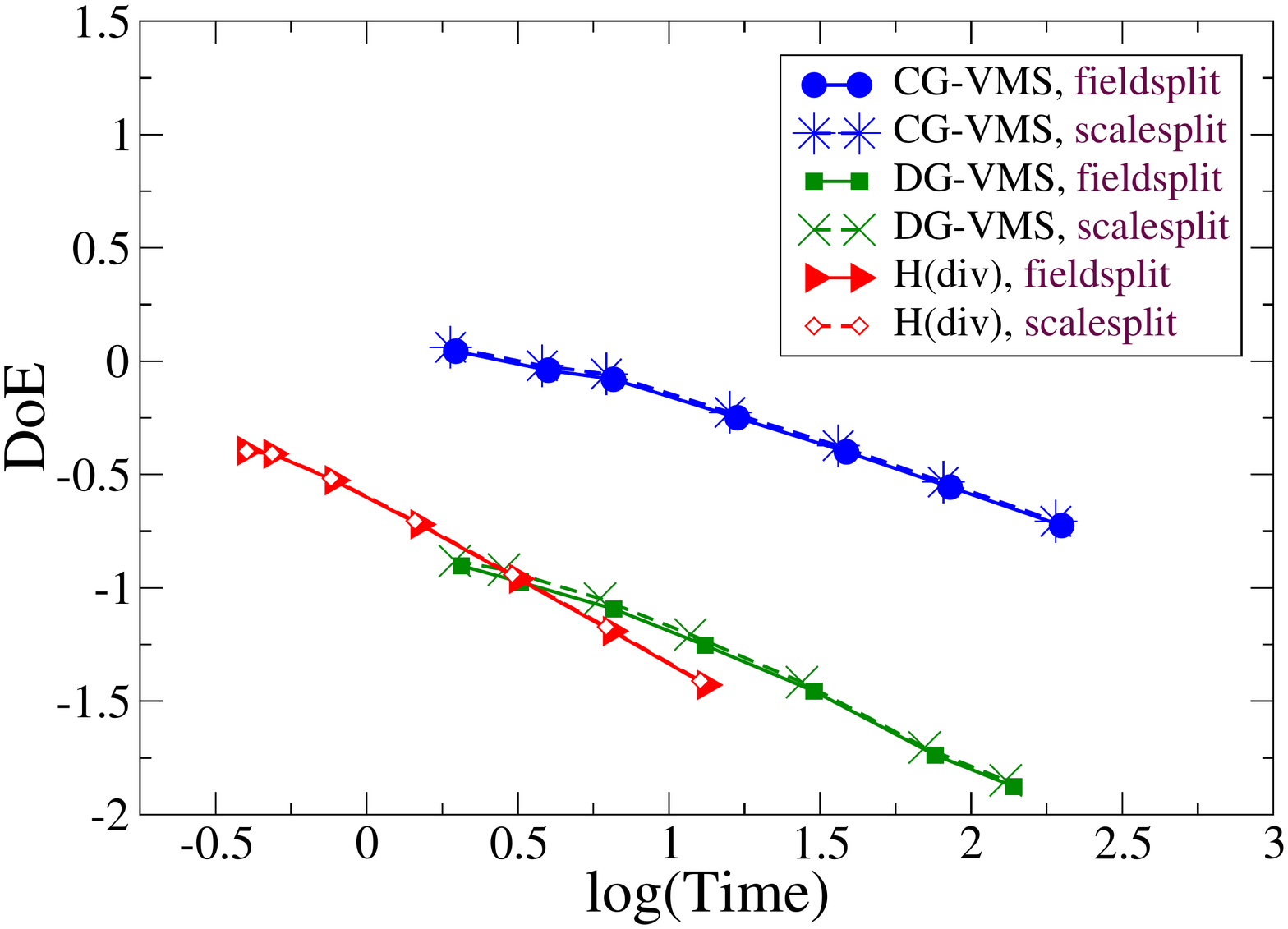}}
  %
  \subfigure[Micro-velocity \label{v2_DoE_TET_e-7}]{
    \includegraphics[clip,scale=0.2875]{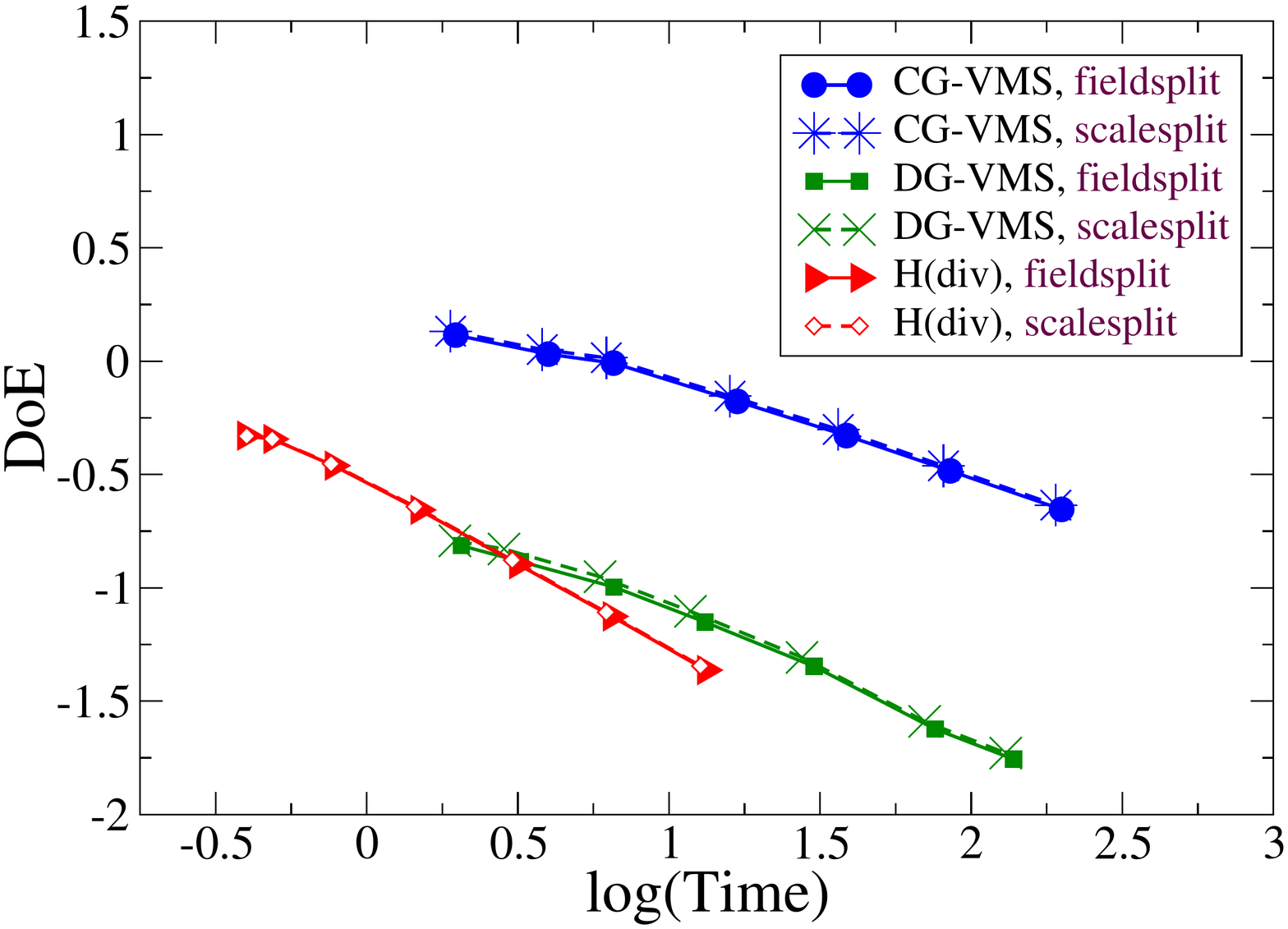}}

  \caption{\textsf{Three-dimensional problem using
      \textbf{TET} mesh:} 
    This figure compares the Digits of Efficacy (DoE) 
    for the chosen finite element formulations. 
    Results for both composable block solver
    methodologies with a tolerance of $10^{-7}$
    are reported. The two main inferences are:
    (i) The H(div) formulation has the smallest
    DoE with a steep declining curve. (ii) Similar
    to the two-dimensional problem, the CG-VMS
    formulation has the highest DoE, regardless
    of the block solver methodology.
    \label{Fig:3D_DoE_TET}}
\end{figure}

\begin{figure}
  \subfigure[Macro-pressure  \label{p1_DoE_HEX_e-7}]{
    \includegraphics[clip,scale=0.2875]{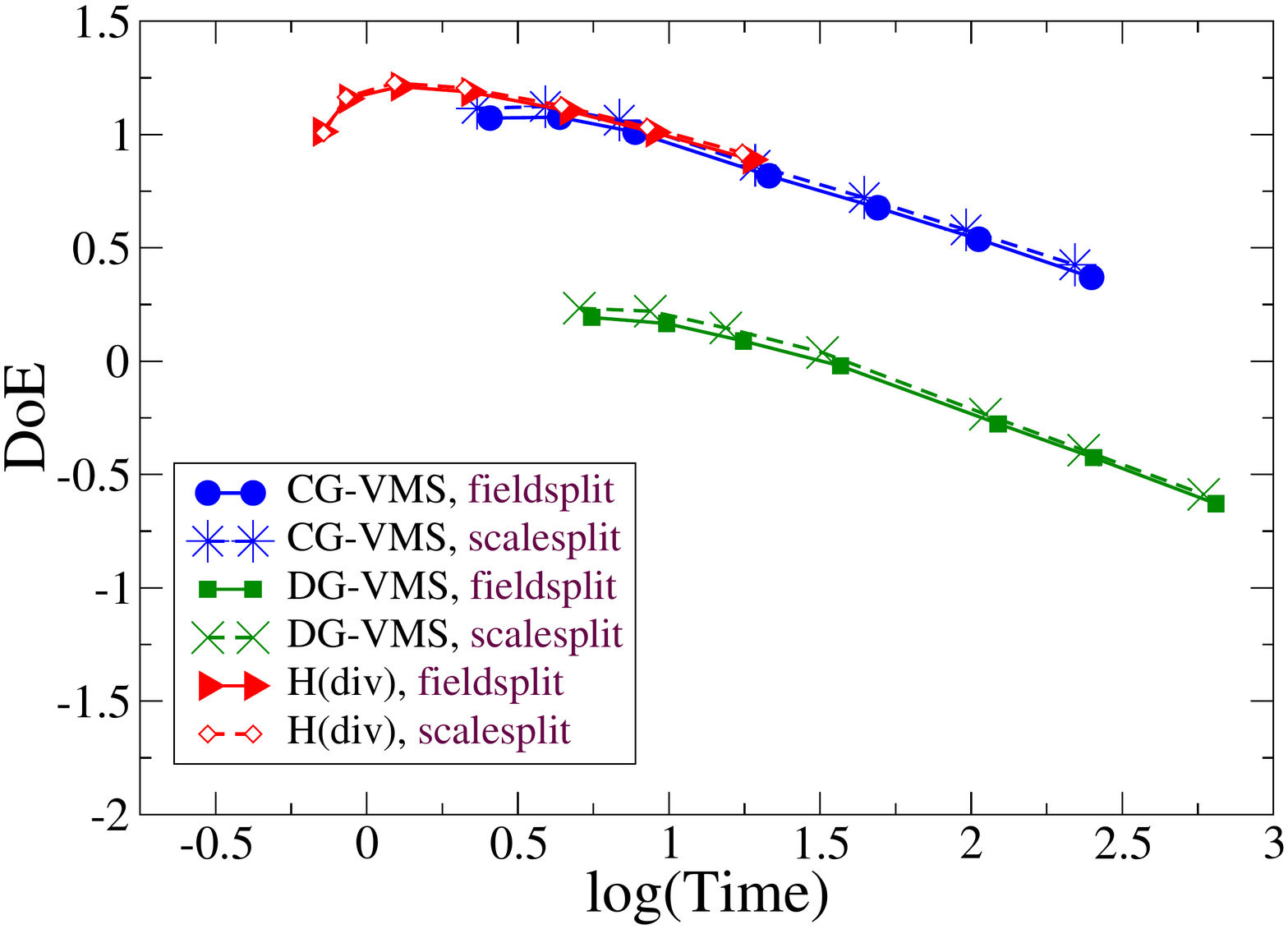}}
  %
  \subfigure[Micro-pressure \label{p2_DoE_HEX_e-7}]{
    \includegraphics[clip,scale=0.2875]{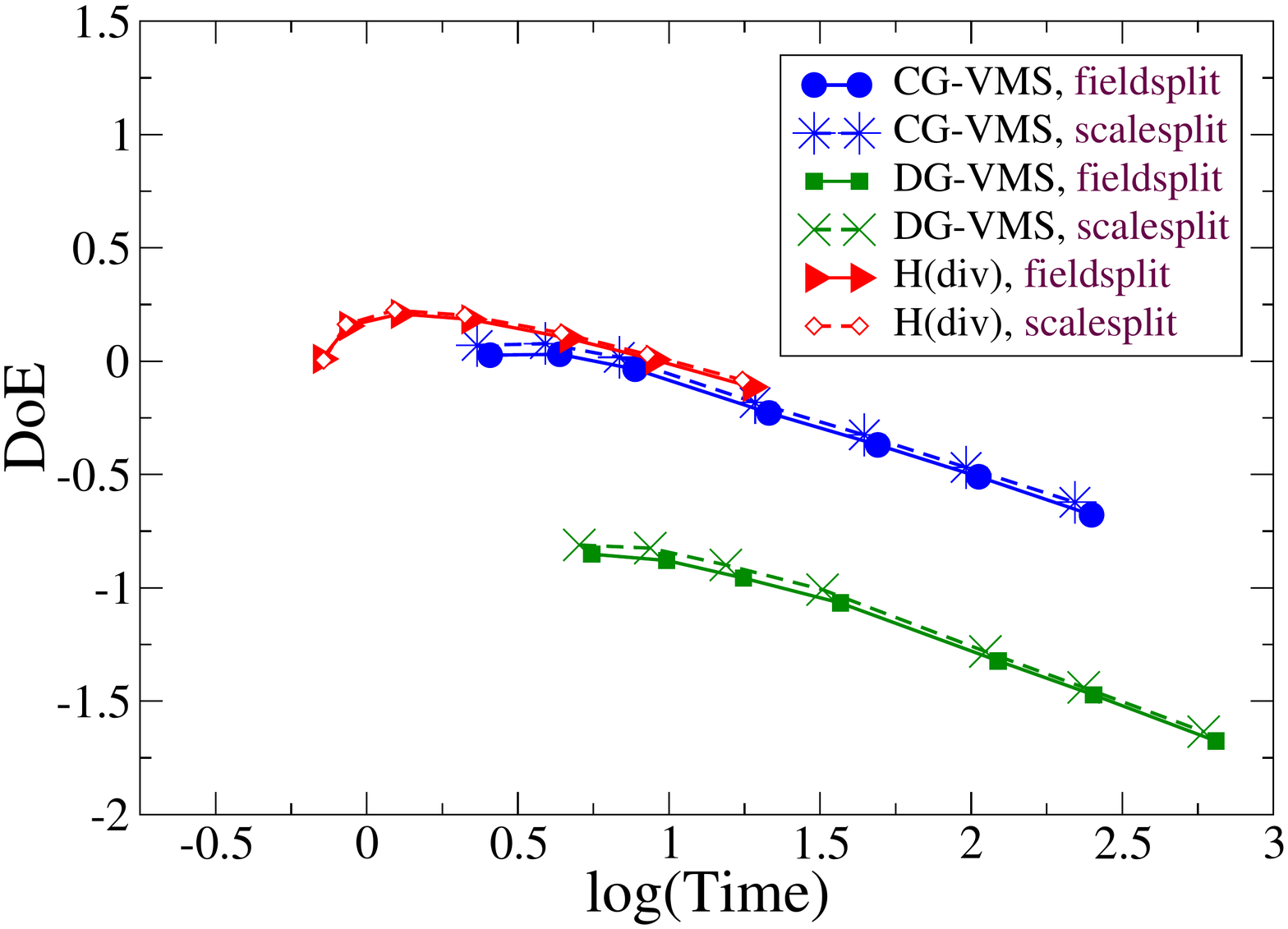}} \\
  %
  \subfigure[Macro-velocity \label{v_DoE_HEX_e-7}]{
    \includegraphics[clip,scale=0.2875]{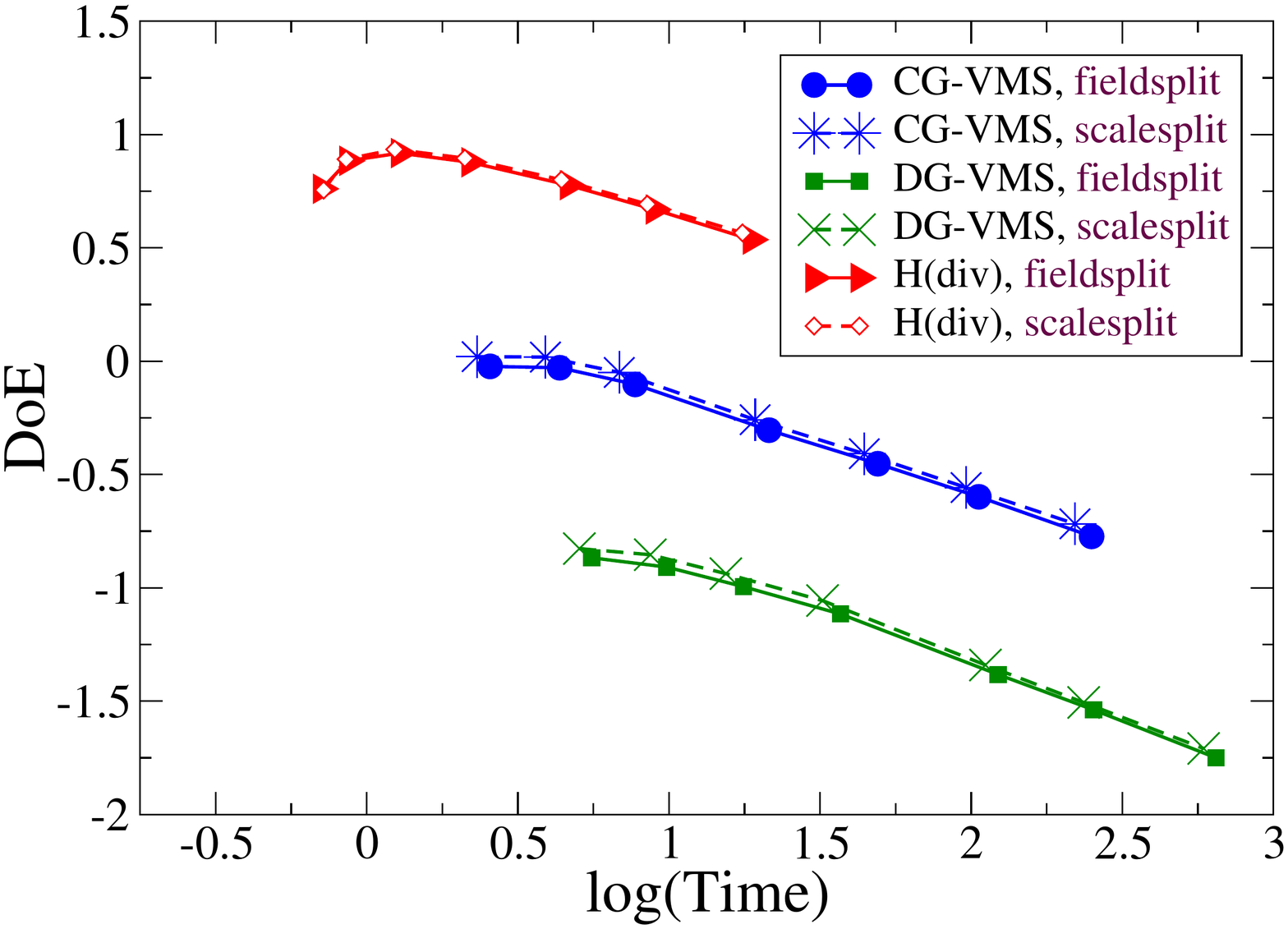}}
  %
  \subfigure[Micro-velocity \label{v2_DoE_HEX_e-7}]{
    \includegraphics[clip,scale=0.28755]{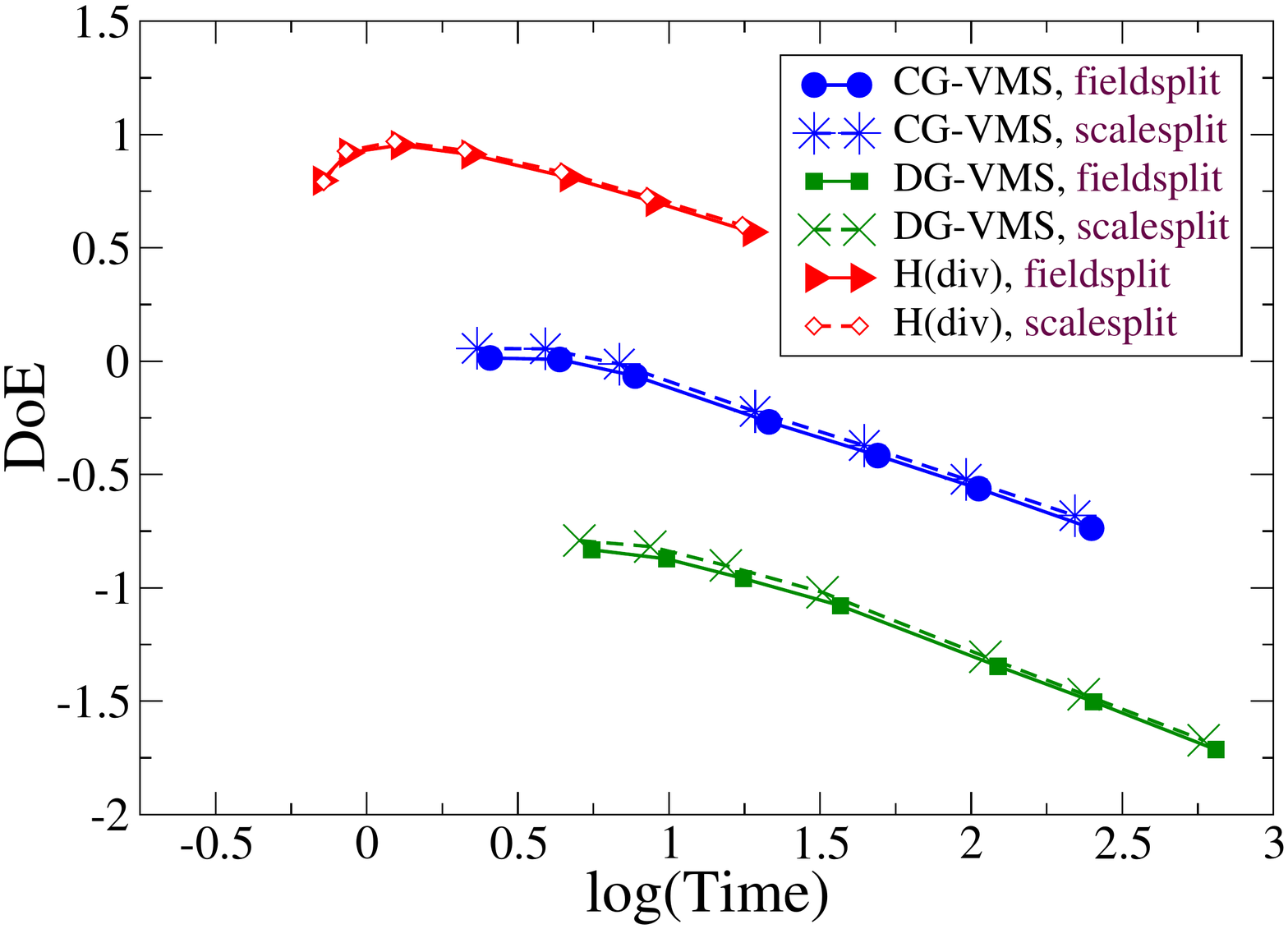}}

  \caption{\textsf{Three-dimensional problem
      using \textbf{HEX} mesh:} This figure
    compares the Digits of Efficacy (DoE) for the
    chosen formulations. Results for both
    composable block solver methodologies
    with a tolerance of $10^{-7}$ are reported.
    The main inference is that, unlike the
    simplicial mesh (i.e., TET mesh), the
    H(div) formulation has the highest DoE. In
    particular, this is more obvious with respect to
    the velocity solution fields.
    \label{Fig:3D_DoE_HEX}}
\end{figure}

\end{document}